\documentclass[12pt,preprint]{aastex}

\shorttitle{Morphological Annotations for Groups in the FIRST Database}
\shortauthors{Proctor}

\begin{document}



\title{Morphological Annotations for Groups in the FIRST Database}







\author{D. D. Proctor}
\affil{Institute of Geophysics and Planetary Physics, L-413, Lawrence Livermore National Laboratory, \\
7000 East Avenue, Livermore, CA, 94550; proctor1@llnl.gov}



\begin{abstract}

The morphology of selected groups of sources in the FIRST (Faint Images of the Radio Sky at Twenty Centimeters) survey and catalog is examined.
Sources in the FIRST catalog (April 2003 release, 811117 entries) 
were sorted into singles, doubles, triples and groups of higher-count membership based on a proximity criteria.   The 7106 groups with four or more components were examined individually for bent types including, but not limited to, wide-angle tail (WAT) and narrow-angle tail (NAT) types.  In the process of this examination, ring, double-double (DD), X-shaped, hybrid morphology (HYMOR), giant radio sources (GRS), and the herein described W-shaped and tri-axial morphology systems were also identified.  For the convenience of the reader separate tables for distinctive types were generated.  A few curiosities were found.  For the 16,950 three-component groups and 74,788 two-component groups, catalogs with probability estimates for bent classification, as determined by pattern recognition techniques, are presented.

\end{abstract}





\section{INTRODUCTION}

The FIRST survey \citep{BBWH} and catalog \citep{BWBHG} have proved to be a rich source of data for astrophysical investigations.  
However, no larger scale attempts on mining the information on radio galaxy morphology provided by the FIRST maps have been reported.  
The FIRST catalog is a catalog generated by fitting flux densities with elliptical Gaussians to provide representation of source morphology.  It is a catalog of components.  Thus a single physical system may have multiple catalog entries.
Considered here is the April 2003 release of the FIRST catalog with 811117 entries from the FIRST web site http://sundog.stsci.edu.  


With over 800,000 catalog entries, individual examination of sources is not feasible and another approach must be devised.
The approach taken here is to group sources by membership count,  
examining higher count groups individually and applying pattern recognition to the three- and two-component groups.
The single-component groups can be examined for statistical outliers, making use of morphological operators.
Finally groups can be examined for clustering.
Reported here are results of the visual examination of the higher-count groups and initial results for application of pattern recognition to the two- and three-component groups.   

The distribution of nearest neighbor distances for all catalog sources is shown in Figure 1.  It was used to determine a threshold distance for group construction.  This figure shows a  bimodal distribution, one sharp peak at approximately 0.16', an ill-defined minimum between approximately 0.7' and 1.5', and another broader peak at approximately 2.4'.  The first peak is attributed to single physical sources, mostly nearby, being resolved into their various radio components such as lobes, hot spots and cores, whereas the second peak is attributed to the distribution of nearest neighbors for physically unrelated systems.  A rough visual extrapolation of dropoff from the first peak suggests that for separations greater than about 0.96', the nearest neighbor of a source would more likely be a chance projection rather than a resolved component of the same physical system.  Choice of the 0.96' distance was compromise based on desire to identify most of the physically associated groups, yet have a manageable number of groups to examine.  
(See \citet{Cress96}, for discussion of the angular two-point correlation function for the FIRST survey.)



Using this 0.96' separation definition, the sources were sorted into singles, doubles, triples, and groups of higher count membership.
  This was done strictly on basis of proximity (projected sky distance) as determined from the catalog positions.  A group would grow until
no new source was found within 0.96' of any current group member.
The group number given in the tables and mosaics is simply that sequentially defined by processing software.
Of course, some groups defined in this manner are expected to have chance projection components and some may be the components of more widely spaced systems.
With this proximity definition, at the 5" resolution of the FIRST survey, $\sim$71\% of catalog sources
are single, $\sim$18\% of sources
are members of a double, $\sim$6\% are members of a triple, and the remaining $\sim$~5\% are in higher count groups.
Hereafter, groups of four or more components will be referred to as higher count groups.  Initial examination of some higher count groups suggested visual examination of the entire set of 7106 cutouts would be worthwhile.
Groups determined by visual examination to be of interest were assigned an IAU compliant designation beginning with the acronym 'FCG' (FIRST catalog radio galaxy or component group) and coordinate based sequence, format=JHHMMSS.s+DDMMSS, truncated, not rounded.
Individual examination of the 16,950 three-component groups and 74,788 two-component groups was considered impractical and automated pattern recognition techniques were employed to generate probability estimates for the group being bent.
Details of the pattern recognition process have been previously reported \citep{Proctor02, Proctor06}.

\subsection{\it Taxonomy Background}

\citet{Kempner03} in a conference note proposed a taxonomy for extended radio sources in clusters of galaxies.  
Morphologies discussed there and relevant to this work are Radio Galaxy (FR I, WAT, or NAT), and Classical Double (FRII), these sources being associated with active AGN.  For the reader unfamiliar with FR class, \citet{Fanaroff74} examined a sub-sample of the 3CR complete sample of \citet{Mackay71}, consisting of those sources clearly resolved into two or more components and found a division in luminosity between those sources depending on their high brightness regions being closer to (FRI type) or farther from (FRII type) the central galaxy or quasar than the low brightness regions.
FRIs are generally described as having well defined symmetric jets, possibly bent, expanding into diffuse edge-dimmed lobes, and usually prominent cores.  WATs (Wide Angle Tail) and NATs (Narrow Angle Tail) are radio galaxies in which the jets deviate from alignment.  Possible causes of this deviation are discussed in Section~\ref{watandnat}.  
FRIIs typically have weak jets, if any (often only one visible).  The lobes are edge-brightened with brightness peaks near the outer edges and the cores are weak.  They have been associated with giant ellipticals and most are not found in rich clusters.  Their luminosities are higher than other types of sources described in the conference note.  

Other morphologies described in the conference note include VLBI core and confined cluster core, associated with active AGN, and  AGN relic, radio phoenix, radio gischt, mini-halo and radio halos, associated with extinct or dying AGN.  These generally  have non-distinctive morphologies and require multi-spectral information for identification.

In early studies of parent galaxies of radio sources, 47 FRI and FRII mainly non-cluster sources were compared \citep{Owen89} with 52 sources in rich clusters \citep{Owen91}).  These samples were not intended to be complete but rather to explore the dependence of optical surface brightness as fully as possible in a small sample and give a first look at properties of several types of cluster sources.  
In these studies they proposed five morphological classes which they related to FRI and FRII types using plots of absolute radio luminosity at 1400 MHz versus absolute magnitude.  Their classes were as follows:

  (i) Classical Doubles (CD).  Double sources dominated by compact outer hotspots, found generally to be classified as FRII.

 (ii) Twin Jet sources (TJ).  Sources with smooth slowly widening jet on each side of the galaxy, required to be larger than 100 kpc and may be either straight or bent.  Includes WATs.  Generally FRI.

(iii) Fat Doubles (FD).  Sources with bright outer rims of radio emission and roundish diffuse radio lobes.
At the time called FRII or FRI/II, but proposed in \citet{Owen91} to really belong with FRI.

 (iv) Narrow Angle Tails (NAT).  Cluster sources with bright 'head' and diffuse 'tail' trailing to one side of galaxy, the tail often split into two structures at higher resolution.  Also includes Intermediate Tails, which could also be called Bent Twin Jet sources.  Considered probably FRI.

  (v) Small Twin Jets (STJ).  All jet sources smaller than 100 kpc, generally much smaller. Includes radio galaxies at the centers of cooling flow clusters and lower luminosity radio galaxies.  Again they considered them probably FRI.  Since these sources were not well represented in their data set, STJ were not included in most of their analysis.

The distinction between STJ and TJ requires distance information, considered beyond the scope of this paper. 
The distinction between CD, TJ/STJ, and FD could perhaps be better made analytically, especially for the smaller systems.  Such analysis could be made from the information in Table~\ref{masterlist} and the FIRST catalog, but will not be considered further here.

They concluded FRI sources, both in and out of clusters, show a correlation between radio and optical luminosity, the radio luminosity being mainly a function of the luminosity of the parent galaxy and the radio structure mainly a function of the properties of the external environment.  Further they found the FRII sources do not appear to fit this correlation, the parent galaxies for the CD (FRII) sources appear generally to be giant elliptical galaxies, normally found in regions of poor clustering.

\subsection{\it FIRST Taxonomy}
While in terms of radio surveys FIRST has relatively high resolution (5"), it is insufficient to discriminate between all of the above types.  The annotations here are generally limited to the visual morphology noted in the FIRST images, though occasional reference is made to other citations.
In particular while larger angular size and intensity WATs or NATs are distinctive, for the smaller angular diameter radio galaxies, it is seldom possible in general to distinguish between FRI and FRII from FIRST morphology alone.
Also distinguishable in FIRST are larger X-shape, W-shape, ring, and double-double (DD) morphologies, in addition to the common S-shape.  
As the angular diameter of the systems approach the survey resolution limit it becomes difficult to distinguish between bent
and S-shape morphologies.  It is also difficult to distinguish between double lobe, core-jet and head-tail types for the smaller angular diameter systems.  In some core-jet types the components have clearly different silhouette sizes or shapes, however, when the silhouettes are similar in size they become difficult to distinguish from double lobe morphology.  An analytical approach using the FIRST catalog and images might be a more appropriate for selection of core-jet candidates, as well as FRI and FRII discrimination.  

While some FIRST groups may eventually be identified as relics, phoenix, gischt, halo, or mini-halo, due to their more ambiguous morphologies and need for multi-spectral information such identification cannot be made here, though possible candidates for such types might be selected from groups annotated with particular characteristics.

\subsection{\it Note on Galactic sources: globular cluster sources, stellar sources}
Considering Milky Way galactic sources, \citet{Sun02} searched for radio emission of Milky Way globular clusters and found only two likely FIRST coincidences.  These were a two-component group of a few mJ peak flux, largest component major axis about 12" in cluster NGC 5634 and a possible planetary nebula in cluster NGC 6341, also only a few mJ peak flux, size about 13".  Radio stars are expected to be a rarity at flux densities above 1 mJy at 20 cm. \citep{BWBHG}.

\subsection{\it Organization}
This report is organized as follows:  The higher count groups and certain distinctive morphologies are discussed in Section 2.  Tables and mosaics are presented.  
  A brief review of possible mechanisms responsible for the various morphologies is included.  Section 3 reviews pattern recognition procedure and describes the three-component group catalog and Section 4 presents details of the two-component group pattern recognition and catalog.  Final comments are presented in Section 5.  

It should be noted the intensity scaling of the figures shown in this paper have often been chosen to enhance features of interest.  Unless otherwise noted the cutouts shown in the mosaics are 3.93' in size.  The choice of cutout size is a compromise between the desire to include enough of the background to recognize giant radio sources, but still be able to manage image display and storage efficiently.  The cutouts were generated to examine the groups as constructed by the grouping software and thus may not be centered on the coordinates noted in the tables.  
The significance of the system coordinate needs to be considered in the context of the type of group and the estimated size.  
Note that only the tabulated coordinates that are associated with a FIRST catalog source have the
significant figures of the FIRST catalog itself.  
The remainder are given to the same number of digits as the FIRST catalog for tabular consistency and convenience.
The key to the annotations used in the tables is given in the appendix.


\section{HIGHER-COUNT GROUPS}

This section discusses the construction of higher count groups and their annotations and includes subsections and tables for certain distinctive morphological types, with one subsection presenting a few curiosities.  The key for the annotations used in the tables is given in Table~\ref{masterkey}.  

The list of higher count groups was constructed as described above and approximately 4' square cutouts
were generated and visually examined as grayscale and contour plots.  An arbitrary selection of cutouts is shown in Figure~\ref{typical_ex}.
The individual cutouts are labeled with a group identification number and correspond to the sample entries of Table~\ref{masterlist}.  Those of special interest were given IAU compliant designations, details of which are discussed below.  The complete annotated list is given in the online data.
Discussion of the annotations is given in the appendix. 
For approximately 16\% of the higher count groups,  examination of the image suggested the desirability of a larger cutout, due to possibility of a larger angular diameter group or indications that the group was a sidelobe or artifact. For these an approximately 12' image was also extracted.  If there was still uncertainty, the FIRST web site (http://sundog.stsci.edu) was used to examine 30' square maps.  
 
In Table~\ref{masterlist}, the group identification number, group coordinate, and group size (diameter, excluding apparent chance projections) are for the groups as generated by the grouping software.  
Further processing is required since often these software generated groups were incomplete or appeared to contain multiple physical systems.  
Tables for specific morphologies were thus generated to provide coordinates and sizes for the apparent physical systems.  The sizes and coordinates are visual estimates generated using automated software.  
For the systems given in the morphologically distinct tables an IAU compliant designation was generated using the acronym FCG (FIRST catalog radio galaxy or component group) and coordinate based sequence, format=JHHMMSS.s+DDMMSS, truncated, not rounded.
The tables for the less common morphological types include sources of more tentative classification (those appended with '?').
For the more common types only the classifications deemed more reliable were included in the tables.
If a group appeared to contain multiple physical systems, there may be multiple entries for a given group number in the morphologically  specific tables, as for example in Table~\ref{watnattable}, FCG J010236.5-005007 and FCG J010242.4-005032, are both from Group 007807.
In the morphologically specific tables the system coordinates are for a visually estimated center for the system.  For those tables that include a core component coordinate, a 'c' type indicates a FIRST catalog entry and a 'v' type indicates a visual estimate from symmetry considerations.  If a component was deemed a likely core its coordinate was chosen for the designation, otherwise the system coordinate was used.

While an extensive search of the literature has been made for references to examples of the morphological types discussed and previous references to sources listed, no claims are made for it being complete.


\subsection{\it Bent types}
Discussed in this section are the morphologies variously described in the literature as
wide-angled tail (WAT), and narrow-angled tail (NAT) and the herein described W-shaped sources.
It is believed these bent radio galaxies can act as tracers of rich clusters and clusters at high redshift (\citet{Blanton00}, \citet{Blanton01}, and \citet{Blanton03}).  The sample selection for the Blanton papers was from the April 1997 release of the FIRST catalog.  They selected 384 objects as being bent by visually inspecting gray-scale plots of FIRST sources with more than one component within a circle of radius 60".  

Tables of apparent two-component bent (B) and three-component bent (TB) sources (found embedded in the higher-component groups) which have little extended structure are also presented, though their ultimate classification as NAT or WAT would require observations of higher resolution and/or sensitivity.

\subsubsection{\it WAT and NAT} \label{watandnat}
For these morphologies the jets or lobes appear as swept back from the core or center of the radio galaxy as if
by a wind.  They are also described as C-shaped or head-tail (HT) type in the literature.
\citet{Douglass08} give a review of recent views of formation and environment of these
interesting objects.  Briefly, the general view of WATs is that they reside at or near the center of clusters with low peculiar velocities relative to the intracluster medium (ICM), whereas a NAT's shape is due to the host galaxy's rapid motion through the ICM.
A radio source that is physically a WAT could appear as NAT given particular orientations to the plane of the sky.  Recently \citet{Mao09} studied five galaxies in the central region of the Horologium-Reticulum Supercluster they identified as HT type and suggested high densities allowed ram pressure to bend the jets as opposed to exceptionally high peculiar velocities.  They suggested a combination of close companion and cluster environment contributed to bent morphology.

Figure~\ref{typical_natwat} is a mosaic of WAT and NAT examples.
The full list of over 400 of what were deemed the more reliably classified WAT and NAT sources is given in Table~\ref{watnattable}.  The first right ascension and declination entry, labeled system, provides a position for the approximate visual center of the group, the second gives a presumed core position.  The type column indicates if the presumed core location is associated with a catalog entry (c) or is a visual estimate (v) from symmetry considerations.  The '?' suffix indicates less certainty in the identification.  A visual estimate of the size (diameter) is given in the next column, followed by the group identification number and description. 
Sources with opening angle greater than about 90 degrees were designated WAT and
sources with smaller opening angle were designated NAT.  Those for which the distinction was difficult to make
were designated NAT/WAT or WAT/NAT depending on which appeared the higher probability classification.

\subsubsection{\it W-shaped}
The list of W and possibly W shaped sources is given in Table~\ref{wtable}.
See Figure~\ref{typical_ws} for corresponding images of those sources deemed more reliably classified W-shaped.  These show wiggled structure.
They bear resemblance to radio images of  PKS 1246-410 (1.565, 4.76, 4.9, 5, 8.4 GHz, outside FIRST coverage) at the center of the Centaurus cluster shown in \citet{Taylor02}, though their 8.3 GHz BnA configuration image might be considered more X-like.   There, this PKS system was studied as one of a sample of 14 cooling flow clusters each containing at least one embedded radio galaxy stronger than 100mJy at 5Ghz \citep{Taylor94}.  The unusual morphology was suggested to perhaps result from interactions with or confinement by the hot, X-ray-emitting gas of the cluster.  The radio galaxy 3C 338, (four-component FCG J162838.2+393304, Group 611594, Figure~\ref{typical_ws}), a member of cluster A2199, also of the Taylor sample, also has a W-shaped radio morphology in FIRST.  \citet{Burns83} discussed two models for 3C 338.  One model involved ram pressure produced by a cooling accretion flow and the other invoked motion of an intermittent radio nucleus about the barycenter of the cluster/galaxy potential leaving behind a steep spectrum jet.   Another member of the Taylor sample 1508+059 in cluster A2029 (five-component FCG J151056.2+054441, Group 531415) was classified as WAT/NAT in FIRST.  A further look suggests some indication of W-shape morphology there also.  2MASX J03272476-5325179 (outside FIRST coverage), one of the galaxies from \citet{Mao09}, discussed in Section 2.1.1, also has this morphology.  These galaxies were found in regions of enhanced X-ray emission.

Faraday rotation measure (FRM) and projected magnetic fields calculated by a 3-dimensional simulation of MHD jets was used by \citet{Kigure04} to investigate radio observations of large scale wiggled AGN jets on kpc scales and showed a good match with a part of the 3C 449 jet (no FIRST coverage).  They urged high quality observations of the FRM distribution to investigate the mechanism of jet formation.

\subsubsection{\it TB and B types}
Examples of sources designated TB and B in this paper are given in Figures~\ref{typical_tb} and ~\ref{typical_bs} respectively.  Some may be ultimately identified as WAT or NAT.  
Here, the TB type consists of three components, more point-like in morphology, which could be identified with a core and two similarly sized lobes that are not aligned and the
B type consists of two close or overlapping components for which the major axes are not aligned.  
Lists of the more prototypical are given in Table~\ref{tbtable} and Table~\ref{btable} respectively.
The identifications of TB and B sources as bent (WAT or NAT) is much more tentative.  The B-type are typically of smaller size or in groups with possible chance projections, as for example FCG J124209.6+243658 (Group 366839) and FCG J131139.2+325235 (Group 399294) in Figure~\ref{typical_bs}.


\subsection{\it Ring and Ring-like Morphologies} \label{ringandringlike}

Examples of ring and ring-like structure are shown in Figures~\ref{typical_ring} and ~\ref{typical_ring-lobes}.
There is considerable variation in ring morphology in these figures.  The rings considered here could be classified  into three types, 'embedded rings', and 'whole-lobe rings', and 'system rings', depending on if the ring appears embedded in a lobe, appears to form the lobe boundary, or appears around the core external to the lobes, though possibly through them also.

  Many of the systems in Figure~\ref{typical_ring} appear to be extreme cases of NAT radio galaxies.  Some may be edge-brightened lobes (designated ring-lobes) for which only a single lobe is apparent.  Some rings would perhaps be more accurately describes as loops.  The intended distinction between the sources in Figure~\ref{typical_ring} and Figure~\ref{typical_ring-lobes} is that in the former figure the ring appears to form the entire radio system at the given wavelength, while in the latter the ring constitutes or is embedded in a single lobe.
FRIIs and FDs are described as having edge or rim brightened lobes.  With further information, some of the sources in Figure~\ref{typical_ring} may be determined to be the lobes of FRIIs or FDs.

The table corresponding to Figure~\ref{typical_ring}, the ring mosaic, is Table~\ref{ringtable} and the table corresponding to Figure~\ref{typical_ring-lobes}, the ring-lobes mosaic, is Table~\ref{ringlobetable}.

\citet{Buta96} provide a comprehensive review paper on ring classification, properties, formation theory, and references for optical, IR, and radio.  They state that about one fifth of all spiral disk galaxies include a ring-shaped pattern in the light distribution and an additional third appear to have partial rings made up of spiral arms.  They attribute the vast majority of rings to resonance phenomena and a small fraction to galaxy merger or accretion of intergalactic gas, with photometric data showing that most nuclear rings are the site of current active star formation.  Star formation is a known source of such structures in the radio regime.  Also hypothesized are shocks and bubbles in thermal gas.  More rarely nova and supernova remnants, masers and Einstein rings (gravitational bending) are seen.  Various mechanisms are discussed in more detail in the following subsections.

\subsubsection{\it Star Formation Rings} \label{sfsection}
\citet{Chyzy08} obtained 8.46 and 4.86 GHz data on NGC 4736 that showed a distinct ring of total radio emission precisely corresponding to the the bright inner pseudoring visible in other wavelengths.
They described the approximately  47" radius ring as a well-defined zone of intense active star formation visible in optical, $H_{\alpha}$, UV emission \citep{Waller01} and infrared and CO maps \citep{Wong00} and
found  the observed distribution of Faraday rotation suggested the possible action of large-scale MHD dynamo.  
 They suggest galaxies of ringed morphology and resonant dynamics can be excellent targets to investigate questions of galactic magnetism.  NGC 4736 is shown as FCG J125053.0+410713 (Group 376229) in Figure~\ref{typical_ring}.
\citet{Sofue91} has proposed molecular rings as a possible cosmic distance indicator.

In a search for structure that could be associated with a spatial or causal connection between the nuclear activity and circumnuclear starburst rings \citet{Laine06} compared five Seyfert and five starburst galaxies.  They found four with ringlike structures in their 20 cm. observations, evenly divided between Seyfert and starburst types.  (These were outside FIRST coverage.)  They believe high-resolution radio continuum observations are a powerful way to study the  properties of circumnuclear rings since radio wavelengths are free of extinction effects of dust in optical and NIR.  They also state the caveat that radio continuum, especially at 6 and 20 cm. is not a reliable tracer of the location of star formation activity, but instead traces the overall magnetic fields of the galaxy \citep{Beck05}.  Further they state the radio emission from the rings consists of two components, thermal emission that traces bremsstrahlung from free electrons spiraling in magnetic fields that are thermalized by optically thick medium and nonthermal emission from electrons spiraling in magnetic fields.

FCG J122232.0+295343 (Group 345577) in Figure~\ref{typical_ring} is identified with NGC 4314.  \citet{Garcia91} and  \citet{Combes92} have reported a small nuclear ring of star formation of 7" radius for NGC 4314.
FCG J104634.9+134502 (Group 239008), identified with an SDSS galaxy at z=0.00989702 (NGC 3367, a barred spiral) and
FCG J140708.9+045257 (Group 460487) coincident with an SDSS galaxy at z=0.133602 appear similar.

\subsubsection{\it Shock and Bubble Rings} \label{shocks_and_bubbles}

{\it Embedded rings.}  The ring (FCG J150458.8+255940) and nearby point-like source (FIRST 150457.107+260058.30) in the image of FCG J150458.8+255940 (Group 524877) in Figure~\ref{typical_ring} was found coincident with the lobe and core of 3C 310.  Van Breugel \& Fomalont (1984) suggest the rings of this system are edge-brightened boundaries of bubbles of hot gas blown up by weak jets and are surrounded by magnetic fields, while the filamentary structures are remnants of burst bubbles. 
The system 3C 310 was classified as FD (fat double) and FRI/II by \citet{Owen89}.   

\citet{Kundt87}, considering the fine structure of the lobes Cyg A (outside FIRST coverage), suggested jets ram channels through thin massive shells surrounding AGN, the jets being temporarily choked and blowing radio bubbles.  The warm shell matter in the cocoon shows up radio-dark through electron scattering.  They noted about a dozen other systems with lobes showing considerable structure.

The radio lobes of Hercules A (outside FIRST coverage), also contain ringlike embedded structures concentric with the jet axis \citep{Gizani02}.  \citet{Saxton02} in hydrodynamic simulations of Hercules A found the observed ringlike structures well explained as nearly annular shocks propagating in the backflow surrounding the jet.  \citet{Sadun02} also proposed explanations for morphology of these structures.

While there are few systems presenting such a nearly circular ring as 3C 310, FCG J093750.7+013445 (Group 162607), FCG J105539.8+143351 (Group 249174), FCG J134038.3+555020 (Group 431064), FCG J142440.5+263730 (Group 479753), FCG J150721.7+200414 (Group 527485), and FCG J162757.5+334548 (Group 610998) in Figure~\ref{typical_ring-lobes} show some similarity.

{\it Whole-lobe rings.}  \citet{Conner98} considered several interpretations for the lobe of MG 0248+0641 (no FIRST coverage).  Its western radio lobe appears as a nearly continuous ring.  Chance interposition of Galactic supernova remnant, nova, planetary nebula or HII region was ruled out.  They also considered lensing, however no lensing object was found and high linear polarization seen around the ring then could not be easily explained.  Suggested causes were bubbles resulting from instabilities in the energy flow down the jet triggered by infall of gas via tidal interaction with nearby galaxy.  FCG J030041.7-075334 (Group 024123), FCG J114546.5-041449 (Group 304503), FCG J134811.5+071640 (Group 439402), FCG J145941.9+290333 (Group 518886), and J162757.5+334548 (Group 610998), among others, in Figure~\ref{typical_ring-lobes} appear similar.  
Some ring type sources may be HYMORS (HYbrid MOrphology Radio Sources).
HYMORS have an FR I-type lobe (edge dimmed) on one side of the core and an FR II-type lobe (edge brightened) on the other.  A mosaic of proposed HYMOR candidates is given in Figure~\ref{hymors} along with FIRST cutouts of some previously identified HYMORS.
Several of the HYMOR candidate edge-brightened lobes also appear to be complete rings.
\citet{Gopal-Krishna00} introduced HYMORS as support of an explanation for the FR dichotomy based upon jet interaction with the medium external to the central engine as opposed to there being fundamental differences in the central engine.
Of the six HYMORS discussed there, one 1004+130 (4C+13.41), was within FIRST coverage.  Its FIRST cutout is shown in the last row in Figure~\ref{hymors} as FCG J100726.1+124856 (Group 195513), a 14 component group.
\citet{Gawronski06} visually examined 1700 sources with flux densities $F_{1.4Ghz} \geq$ 20 mJy and angular size greater than 8" from five subareas of FIRST to select 21 sources they felt to be hybrid morphology.  Of those candidates, they verified three to be actual HYMORS using spectral index maps and considered two more were likely HYMORS, though good spectral index maps could not be made for those two systems given the available data.  These five are shown as the final five cutouts of Figure~\ref{hymors}.  For those five, three were higher count groups, one was three-component group (FCG J134751.5+283630, Group 439024), and one was a two-component group (FCG 120622.0+501748, Group 327615).  It is noted the Gawronski et al. selection criteria generally resulted in smaller angular diameter systems than those candidates found during examination of the higher count groups.

The list corresponding to the mosaic is given in Table~\ref{hymortable}.  The new candidate HYMORS in this list should be considered incidental as opposed to comprehensive, as the morphology was not considered in the original classification pass.

{\it System rings.}  The FIRST image of M87, a three-component group, Figure~\ref{M87}, is similar in morphology to the 6 cm VLA radio image from \citet{Hines89} showing the radio jet and synchrotron emission from the cocoon.
They suggested M87 provides a unique laboratory for investigating the interaction of a SMBH and surrounding intracluster medium.
\citet{Forman07} in a study of M87 reported a nearly circular ring of outer radius 2.8' at hard energies (3.5-7.5 keV) and at ~0.6' and ~1' (the approximate radius of the FIRST ring), surface brightness edges and pressure enhancements.
\citet{Forman07} suggest this cocoon is the "piston" that mediates outbursts from the central MSBH and drives shocks into the surrounding X-ray-emitting, thermal gas.
Other systems that may be similar are FCG J112838.4+360144 (Group 285515), FCG J114546.5-041449 (Group 304503), FCG J131812.9+512638 (Group 406496), FCG J153642.7-020132 (Group 559796), and FCG J162757.5+334548 (Group 610998) in Figure~\ref{typical_ring-lobes}.

\subsubsection{\it Nova, Supernova Remnant, and Maser Rings}
As the coverage of the FIRST survey was chosen to avoid the galactic plane, it is less likely for galactic nova or supernova remnants or masers to found in FIRST.  However, since on order of a half dozen members of the local group are within FIRST coverage, and a shell shaped SNR, 10" in diameter, located at the center of dwarf irregular galaxy NGC 6822 (outside of FIRST coverage), was observed by \citet{Kong04}, extra-galactic SNR are deemed possible but not likely.

\subsubsection{\it Einstein Rings, Gravitational Bending} \label{Einstein_rings}

Apart from scale, a few of the images in Figures~\ref{typical_ring} and ~\ref{typical_ring-lobes} are suggestive of the images and models of gravitational lens systems in the SLACS survey, the VLA B-configuration observations of the CO($J = 2 \to 1$) transition made by \citet{Riechers08}, or the CO($J = 2 \to 1$) and 1.4 GHz radio continuum emission reported by \citet{Carilli03}.  
FCG J073929.8+394711 (Group 045877) and FCG J140708.9+045257 (Group 460487) are examples.
\citet{Riechers08} found the molecular gas in host galaxy of quasar PSSJ2322+1944 (outside FIRST coverage) lensed into a full Einstein ring of diameter ~1.5".
\citet{Carilli03} modeled their observations as a starforming disk surrounding a QSO nucleus (PSS J2322+1944).
However, the great majority of the lenses reported in the SLACS \citep{Bolton08} survey are on order a few arc seconds or less in radius and would not be resolved by FIRST.  
While \citet{Ghosh09} reports a 6" ring in the optical, not apparent in FIRST, and \citet{Belokurov07} reports an almost complete Einstein ring of diameter 10" in the Sloan Digital Sky Survey, few complete large optical rings have been found.   Thus, while possible, given its 5" resolution, FIRST would not appear to be a rich source of gravitational lens candidates.

\subsection{\it X-shaped} 
For X-shaped sources the classical double lobe structure is augmented with emission along a second axis of symmetry.  They have also been described in the literature as 'winged'. Also included in this morphology, in this compilation, are those sources exhibiting emission transverse to the major axis of the system, if it was not at the extremum of the major axis, i.e. dog-leg morphologies were not included as X-shaped.  
See Figure~\ref{typical_xs} for examples of X-shaped sources.
The complete list of the 155 X-shaped sources deemed more reliably classified is given Table~\ref{xtable}.

Van Breugel et al. (1983), in an early study of the X-shaped source 3C 433 (outside FIRST coverage) confirmed it to be a member of a close pair and considered models based on tidal torquing and precessing models of double galaxies.  They presented a scenario based on tidal torquing they felt least contrived.

In another early study, \citet{Duric83} observed the edge-on spiral galaxy NGC 3079.  While cataloged as a single-component source in FIRST, its X-shaped morphology is quite apparent.  It is noted that the Duric observations showed readily apparent ring shaped eastern lobe and a partial ring in the western lobe, suggesting a possible connection between some cases of these morphologies.  

\citet{Chen07} have discussed the relationship between X-shaped radio sources and the double-double morphology for a small sample of Xs and DDs.
\citet{Lal07} in a study of twelve X-shaped sources, almost all then known, grouped them into three categories based on spectral indices of wings and active lobes.  

X-shaped radio sources have attracted interest due to their possible association with binary black hole mergers.  \citet{Cheung07} has presented an initial sample of 100 candidate objects of this type from FIRST using a selection criteria of fields with image peaks of 5mJy/bm and greater and fitted major axis greater than 15", and visually examined 1648 source fields to find candidate sources.  Of his 100 candidates, fifty were higher count FIRST sources and thus would be expected to be included in this section.  Of those 50, 24 were designated X or W-shaped and 12 designated possible X-shaped by this author.  Differences can be attributed to Cheung's initial size and dynamic range requirements and inclusion of sources that here were considered resolved cores or S/Z-shaped.

\citet{Zhang07} studied SDSS J1130+0058, an object with X-shaped radio structure, and favored the binary black hole model on the basis of information available at the time.   Here it is FCG J113021.3+005824 (Group 287448) in Figure~\ref{typical_xs}.

Recently \citet{Lal08} made a comparison of a sample of eight nearby FRII radio galaxies, matched in size, morphological properties and redshift, with a sample of known X-shaped radio sources.  They review the formation models and investigate whether the standard spectral aging models adequately describe the normal sources and consider implications for the X-shaped sources.

It is of note that of the seventeen FRII sources listed in \citet{Owen89} six appear to have X morphology in FIRST.  Their X-morphology varies.  The systems 3C 223.1 (J2000 09h41m24.0s +39d44m42s, lower count group) and 1205+341 (J2000 12h07m32.9s +33d52m40s, FCG J120732.9+335240 (Group 328869) are very suggestive of concomitant emission from binary black hole systems, whereas 3C 227 (J2000 09h47m45.1s +07d25m21s, FCG J094745.2+072518 (Group 173559 and links))  has a more irregular X shape, though the arms do appear to cross at the core.  3C 234 (J2000 10h01m49.5s +28d47m09s, lower count groups) and 3C 319 (J2000 15h24m05.5s +54d28m15s, FCG J152404.8+542811, Group 545833) the transverse emission does not appear to cross at the core.  3C 315 (J2000 15h13m40.1s +26d07m31s, FCG J151340.0+260730, Group 534378) is more suggestive of edge brightened lobes for which only the sides/edges near the core are apparent.  It is however associated with a galaxy pair in NED (NASA Extragalactic Database).

\subsection{\it Double-Doubles}
The definition adopted here is that a double-double radio galaxy (DDRG) consists of a pair of double radio sources with a common center.  On order of a few dozen of these have been described in the literature.  
Double-doubles are of interest because it is believed they can provide information for understanding episodic jet activity and studying jet propagation in different media as well as investigation of the relationship between star formation and AGN fueling. 
In an early series \citet{Schoenmakers00a}, \citet{Kaiser00}, and \citet{Schoenmakers00b} studied seven
such systems, all of which were megaparsec sized.  The luminosity ratios of the lobe pairs and its relation to projected linear size of the pairs was examined. 
 \citet{Liu03} and \citet{Liu04} relate DDRGs to binary black hole mergers and suggest all X-shaped radio sources evolve into DDRGs.  \citet{Marecki06} in a study of the cores of more compact sources with symmetric relic lobes found that four of fifteen were doubles or core-jets on the subarcsecond scale.  The morphology of three of those four was attributed to a rapid repositioning of the central engine, most likely from a merger.  \citet{Saikia06} studied twelve DDRGs, examining the relationship between the luminosities and symmetry parameters of the inner and outer lobe pairs.  While trends were found they concluded larger samples were needed.  \citet{Chen07} further discuss the proposed relationship between X-shaped radio sources and DDRGs.

While \citet{Sirothia09} did not find any unambiguous evidence of episodic activity in a list of 374 sources from an 18' square field using a multifrequency search, they did note that the median size of their objects was expected to be less than about a 100 kpc, which could be partly responsible for their non-detection of fossil lobes.

Table~\ref{ddtable} contains a list of 242 systems and includes triple-double as well as double-double candidates.  \citet{Brocksopp07} presented an interesting case of a triple-double which they felt increased the likelihood that these lobes represent multiple episodes of jet activity as opposed to knots in an underlying jet.  On order of two dozen possible triple-doubles are included in Table~\ref{ddtable}. 
A mosaic of examples of double-doubles candidates are presented in Figure~\ref{typical_dds}.  Over 600 more candidates, designated 'dd?' or 'td?', are contained in Table~\ref{masterlist}.  Of course some may be chance projections of similar doubles, two singles and a central double or possible lensing.  Further observations are necessary to distinguish between knots in the jets and true restart activity.

\subsection{\it Core-jet}
A group was annotated 'cj' (core-jet) in the group annotation table, Table~\ref{masterlist}, if components had significantly different silhouette sizes and the morphology suggested physical association.  It might be expected these are star forming regions of spiral galaxies.
Typical core-jet sources are shown in Figure~\ref{typical_cjs}, and the table is Table~\ref{cjtable}.  As noted in the introduction, analytical selection using the FIRST catalog might be a more appropriate approach to core-jet selection.

\subsection{\it S or Z-shape}

See Figure~\ref{typical_sps} for typical examples of S or Z-shaped morphology. 
In an early compilation \citet{Florido90} found approximately 11\% of radio sources with two jets showed
S or Z-shaped morphology while approximately 9\% showed both jets curving in same direction.  The sample was 365 sources from published observations.  For the higher count groups in Table~\ref{masterlist},  11\% are designated S or Z-shaped, with another approximately 30\% showing suggestions of that morphology.  In the tables, the 'sz' annotation refers to either S or Z shapes, with Z-shapes, those with abrupt direction change, also annotated dogleg.  It should be noted that some number of S-shaped groups could be FRIIs for which only part of the edge-brightened lobe is apparent.
The S- or Z-shape morphology is generally believed due to precession in the orientation of the radio jet, due to
precession of the central black hole.  \citet{Begelman80} proposed the precession being due to the presence of another hole in the same nucleus while \citet{Lu90} suggested a tilted accretion disk as a cause. 
\citet{Lu05} suggest observational evidence likely in favor of precession caused by disk.  They present an updated P-$M_{abs}$ relation, where P is the period of precession and $M_{abs}$ is the absolute magnitude of the nucleus.  \citet{Kurosawa08} present discussion of a three-dimensional model for formation of outflows by an accretion disk, but note it is difficult to determine the exact cause of disk/jet precession for a given AGN system due to large uncertainties in model parameters and observed precession periods, also often model dependent.  As noted in Section~\ref{sfsection}, the S or Z shape of some radio galaxies may be star forming regions.  

Table~\ref{sptable} gives a list of 399 of the more distinctive S or Z-shaped systems.

\subsection{\it Giant Radio Sources} \label{GiantRadioSources}
Here we consider Giant Radio Sources (GRSs) as those sources with projected angular size $\geq$ 4' and 
and Giant Radio Galaxies (GRGs) to be those sources known to have overall projected size of $\geq$ 1Mpc.
These giant sources are of interest in the study of evolution of galaxies and testing consistency with the unified scheme for radio galaxies and quasars.

A list of 53 known Giant Radio Galaxies was compiled from the literature by \citet{Ishwara99}. Of these they found 48 were associated with galaxies and 5 with quasars.  The radio sources ranged in size from about 2' to 59' diameter and up to 5.6 Mpc projected linear size.  They used the sample to investigate the evolution of giant sources and found consistency with the unified scheme.
 
\citet{Lara01} constructed a sample of 84 galaxies of angular size $\geq$ 4' above declination $+60^o$ from
the NRAO VLA Sky Survey (NVSS: \citet{Condon98}, angular resolution 45") and found 37 giants.  
\citet{Lara04} noted just 3 of the 46 FRIIs were quasars.

\citet{Machalski01} selected three dozen FRII or FRI/II sources using NVSS and FIRST with angular size between brightest regions $\gtrsim 3'$ and a flux density limit (one a possible quasar) and in \citet{Machalski06b} concluded GRGs do not form a separate class of radio sources but rather most likely evolve with time from smaller sources under specific circumstances.  

\citet{Schoenmakers00} and \citet{Schoenmakers01} visually inspected WENSS radio maps and selected sources $\geq$ 5' and 
more than $12.5^o$ from the galactic plane and identified 47 GRGs including 19 previously known.  The 1.4-GHz NVSS and FIRST surveys were also used to aid in removal of confused sources.

\citet{Saripalli05} constructed a sample of megaparsec-size double radio sources in the southern hemisphere with angular size greater than 5'.  Two of the 18 cores were identified with quasars, the remainder with galaxies.

Many of the GRGs discussed above are present in FIRST and provide morphological information, but would be difficult to identify as systems from the FIRST image alone.  However the reverse also applies.  There are GRSs in FIRST that were not identified in the other overlapping samples, suggesting a systematic approach could prove fruitful.  Given the different resolution and sensitivity, FIRST may well provide GRGs in an expanded region of parameter space.
A table of candidates for large angular size radio galaxies was constructed from linked groups of Table~\ref{masterlist} and is given in Table~\ref{grgtable}.   It should be considered an incidental as opposed to a comprehensive list.
Table~\ref{grgtable} gives a core position (visually estimated or catalog source) for the system, estimated diameter, constituent groups, and annotations.  The corresponding images are shown in Figure~\ref{typical_grg}.  Some of these GRSs may have linear sizes exceeding 1 Mpc and be GRGs, some are double radio sources with smaller linear size and some may be composed of unrelated components.

\subsection{\it The Curiosities} 

A few curiosities are included in this section.

\subsubsection{\it STARWARS (Fig.~\ref{Cur_starwars})}
FCG J010236.5-005007 and FCG J010242.4-005032 (Group 007807) consists of a pair of WAT/NAT morphology galaxies.  This variant, with 'lobes' appearing approximately transverse to the jets, was denoted the 'starwars' morphology.  This pair appear to be associated with SDSS giant ellipticals at z=0.243214 and 0.242836.  Only a few other sources were noted to have approximately the same morphology in FIRST.

\subsubsection{\it PRETZEL (Fig.~\ref{Cur_pretzel})}
FCG J162304.4+375523 (Group 606579), a four-component group was designated 'pretzel'.  It may be a superposition of multiple systems.
 This group might also be considered 'W-shaped'.
There appear to be several candidates for optical counterparts.
FCG J162317.9+370529/FCG J162317.9+370535 (Group 606754) shown in Figure~\ref{typical_ring} and FCG J014519.7-015951 (Group 012983) in Figure~\ref{typical_int} also show hints of this pretzel morphology.

\subsubsection{\it PRETZEL-WITH-HANDLE (Fig.~\ref{Cur_pretzelwh})}
FCG J113222.5+555821 (Group 289726) is another pretzel or perhaps 'figure 8' shape, this time with a 'handle'.
Optical counterparts for components at (J2000) 11h32m22.754s +55d58m20.55s and 11h32m23.219s +55d58m06.52s appear to be SDSS galaxies at about z=0.05.  The other components have apparent counterparts in SDSS labeled star and galaxy, though neither of these components has a spectra available.

\subsubsection{\it PRETZEL LOBE (Fig.~\ref{Cur_pretzellobe})}
FCG J105420.0+022700 (Group 247661) shows another pretzel, in this case it appears that the pretzel may be a lobe of core at J(2000) 10 54 21.167  +02 27 55.34.

\subsubsection{\it RING-WITH-PLUME (Fig.~\ref{Cur_ring_and_jet})}
FCG J133724.5+455142, a three-component group, appears to be a ring with plume emerging from center of ring.  Optical counterparts were not clear.  It does appear similar to the lower portion of FCG J150721.7+200414 (Group 527485) in Figure~\ref{typical_ring} which would seem to be an edge-brightened lobe.
NED shows two galaxies and a star within 20" of the approximate center of the ring. 

\subsubsection{\it BUG (Fig.~\ref{Cur_bug})}
FCG J074331.0+375908 (Group 048758), a four-component group shown in Figure~\ref{Cur_bug}, has an interesting bug-like appearance. The wings have SDSS galaxy counterparts and while the head has a galaxy counterpart, the larger body component is an SDSS labeled star.

\subsubsection{\it NOVA-LIKE RING (Fig.~\ref{Cur_nova-like})}
FCG J112555.0+200501 (Group 282486) shown in Figure~\ref{Cur_nova-like} is unique 'nova-like' ring.  
The source may be a lobe of the bright point-like source in the image.  That source has as an apparent optical counterpart an SDSS galaxy (QSO) at z=0.132317.  Alternatively, the SDSS galaxy J112555.72+200516.2 appears to be at center of 'ring'.

\subsubsection{\it CORE GAP (Fig.~\ref{Cur_core_gap})}
FCG J132309.9+415318 (Group 411941) shown in Figure~\ref{Cur_core_gap} shows an interesting gap around presumed core.
It is similar in morphology to FCG J091735.0+425906 (Group 140244) shown in Figure~\ref{typical_ring}.

\subsubsection{\it PERPENDICULAR RING (Fig.~\ref{Cur_ppd_ring})}
FCG J162824.8+230105 (Group 611411) shows a ring perpendicular to axis of triple.
The ring in FCG J122749.4+215520 (Group 351269) also appears perpendicular to the system axis, although in this case it seems more centered on the core.
Both of these rings are more teardrop in outline.
The system 3C 234 (J2000 10h01m49.5s +28d47m09s, lower count groups) may also have a ring perpendicular to major axis of system.

\subsection{\it Additional Tables}
Three additional tables with corresponding mosaics are presented.
Table~\ref{tritable} is a list of sources exhibiting possible tri-axial morphology.  The corresponding cutouts are shown in Figure~\ref{typical_tri}.  They may be chance projections, interacting systems, lobes, distorted NATs or WATs, or lens candidates.  Especially interesting is FCG J165132.8+422925 (Group 630062) with arcs or a ring as well as the three more point-like sources.
Table~\ref{quadtable} is a list of groups of four or more point-like sources.  The corresponding cutouts are shown in Figure~\ref{typical_quad}.  While most of these are probably chance projections, it is possible there are lens candidates.
Table~\ref{inttable} is a list of interesting sources not otherwise included in previous tables.  The corresponding cutouts are shown in Figure~\ref{typical_int}.  Especially interesting is FCG J075241.4+455628 (Group 055632), with a beaded-ring appearance and FCG J113830.7+600949 (Group 296465), with a 'gull-like' appearance.

\section{THREE COMPONENT GROUPS}
With nearly 17,000 three-component groups, individual visual classification nears impractical.  Thus
pattern recognition techniques were applied to the selection of a particular morphology from the three-component groups.  Bent morphology (WATs, NATs, TB and B) was chosen as the target morphology.  Supervised pattern recognition was used.  This involves construction of a training set.  This training set consists of typical examples of data with known classifications.  A random sample of 2823 three-component groups was taken from the entire population of three-component groups.   Each training-set group was examined and assigned a classification, bent, non-bent, or ambiguous.  Feature sets, the numerical values of variables of presumed relevance, were constructed and statistical methods employed to generate a classifier.  This classifier was then applied to process the entire population of three-component groups.
Details of these procedures applied to three-component bents can be found in \citet{Proctor02}, and \citet{Proctor06}.  Brief summaries are included here.  \citet{Proctor02} discussed the implications of low resolution for pattern recognition and methods for dealing with more ambiguous populations.  Two types of pattern recognition algorithms were studied, decision trees and artificial neural networks.  Results were compared and showed substantially equivalent results for the two types.  Since the decision tree algorithm was significantly faster, the use of artificial neural networks was discontinued.  \citet{Proctor06} presented significance tests on the sort-ordered, sample-size normalized vote distribution of an ensemble of decision trees as a method of evaluating the relative quality of feature sets.  Hereafter this distribution is designated vote curve.  The vote curve provides an estimate of the probability of the group being bent.

With $d_{min}$, $d_{mid}$, $d_{max}$, being respectively the short, intermediate, and long sides of the
projected triangle formed by the three-component sources, and $R_{ss}$ being defined as the ratio of silhouette sizes of assumed lobes (smaller to larger) calculated at 0.65 mJ level,
the four feature set, {$d_{mid}$, ($d_{mid}$+$d_{min}$)/$d_{max}$, $R_{ss}$}, {$d_{min}$/{$d_{mid}$, provided the most desirable vote distribution of those examined. However, the {$d_{min}$/{$d_{mid}$ feature is of arguable necessity.

A selection of groups of interest from the three-component-group training set is presented in Table~\ref{threecomptrsetint}.

The catalog of all three-component groups is given in Table~\ref{three_comp_vote}.  The vote for each group is followed by the right ascension and declination of the three components.
Mosaics were presented in \citet{Proctor02} and \citet{Proctor06}.

\section{TWO COMPONENT GROUPS}

A training set for two-component groups was constructed in a similar manner as that for the three-component groups.  
Table~\ref{twocomptrsetint} is a selection of a few of the more interesting groups from the two-component-groups training set.
A few different feature sets were evaluated, the best found being the feature set given in Table~\ref{two_comp_features}.

The catalog of all two-component groups is given in Table~\ref{two_comp_vote} and gives the probability estimate for the group being bent.

Figure~\ref{F_maxvex} and Figure~\ref{F_minvex} show mosaics of random selections of the highest and lowest ranked sources respectively.

Figure~\ref{F_2-3vcc} shows vote curve comparisons for three-component groups with that of
the two-component groups.  The two curves in (a) were tested with Conover's distribution functions \citep{BCo} for Tsao's truncated Smirnov statistics \citep{BTs} with the null hypothesis $H_{o}$: no difference in distributions under consideration.  Similarly the two curves in (b) were tested  using the Kolmogorov-Smirnov test comparing two cumulative distribution functions, and the Wilcoxon signed rank test \citep{BOS}, comparing effects of two treatments on paired data.  Discussion of choice of these statistical tests is given in \citet{Proctor06}.  In both instances the null hypothesis was accepted.  This is somewhat surprising since it might be expected that the two-component sources would be be predominately FRII whereas the three-component sources would have higher FRI population.   

\section{DISCUSSION, SUMMARY, AND OPPORTUNITIES FOR FURTHER RESEARCH}

From the set of higher-count groups, extensive tables of certain distinctive radio morphologies have been presented, including WAT, NAT, W, X, DD and S/Z-shaped systems, as well as ring and ring-like systems.  Incidental lists of HYMOR and GRG candidates were also compiled. 
Probability estimates for three and two-component systems being bent were tabulated.
This set of tables should provide considerably expanded populations for the various morphologies.
A few curiosities were found.

Table~\ref{MorphTypesTable} gives estimated percentages of the higher-count groups for certain morphologies.  While these percentages can't be readily extrapolated to the entire population, they do give some sense of relative populations, at least for FIRST resolution, sensitivity, and wavelength.  

Concerning optical counterparts, \citet{deVries06} characterized the radio morphological makeup of a quasar sample selected from the Sloan Digital Sky Survey, regardless of whether the quasar (nucleus) itself was detected in radio.  The radio sources found using that method had sizes predominately under 100".  Combining this information with the predominance of galaxy optical counterparts noted in Section~\ref{GiantRadioSources} for Giant Radio Sources suggests that the higher count groups constructed here will generally have galaxies as optical counterparts.  This remains to be verified.  The counterparts at other wavelengths also need to be investigated.

Many opportunities exist for further mining of this rich database.
While for the vast majority of systems the FIRST source finding software appears to adequately represent the data, a few cases were found where the catalog seems to be missing sources.  For perhaps a few dozen of the higher count groups with complex morphology and/or large dynamic range, the catalog seems to have missed sources or components.
Preliminary examination of 1000 single-component groups suggested that approximately 13\% appeared to be core-jets or resolved cores of doubles not separated by the source finding routine.  Less than 4\% appeared to be widely ($ > 0.96' $) separated doubles.  There were only five examples of sources with some possibility of ring structure.  No WAT, NAT, DD or other unusual morphologies were found.
Application of image processing techniques to single-component sources to search for unusual morphology might prove fruitful.
The single component groups could be examined for statistical outliers from model residuals to identify interesting systems.  These topics are left for future work.

A further look at the  morphology of the 'T' and 'D' systems in Table ~\ref{masterlist}, combined with
component size information in the FIRST catalog, could be used to provide an expanded population of
CD, FD, and TJ/STJ systems.
Examination of the lower-count groups for nearby companions could provide more giant radio source candidates.
Combinations of groups could be examined for clustering.
Updates are needed for the sources added and revisions made in the latest version of the FIRST catalog.

Questions arise concerning S/Z-shaped systems.  Where do these systems fall in the Owen-Ludlow plane?
Why does it seem there are so few with larger lobes?  Are they strictly an early phase of the evolution
or is it an environmental or distance effect?


Successful pattern recognition procedures were developed for estimating probabilities for the three- and two-component groups being 'bent'.  It would be interesting to target other morphologies, e.g. S/Z morphology. 
The morphological tables need to be combined with optical/redshift information to determine positions of the various
morphologies in the Owen-Ludlow plane.
It would be interesting to apply pattern recognition techniques, including optical/redshift
information, to determine whether or not specific morphological types can be sorted automatically, and if so
determine the boundary planes and any implications for galaxy evolution.



\acknowledgments
Thanks are due the anonymous referee for suggestions that lead to improved clarity and detail of the paper.

The original FIRST data was taken by the NRAO Very Large Array from Spring 1993 through 2008.  The National Radio Astronomy Observatory is a facility of the National Science Foundation operated under cooperative agreement by Associated Universities, Inc.  The author appreciates publication charge support provided by the National Radio Astronomy Observatory.

R. Becker provided useful suggestions as well as computer resources.   Richard White provided software to access the FIRST images.  Extensive use was made of the IDL astro package (http://idlastro.gsfc.nasa.gov).  The free availability of the OC1 decision tree software (anonymous ftp from ftp.cs.jhu.edu directory pub/oc1) was greatly appreciated.
  The author is grateful for office space and computing facilities provided by the Institute of Geophysics and Planetary Physics (IGPP), John Bradley, director, and Kem Cook as well as publication charge support provided by Lawrence Livermore National Laboratory Physical and Life Sciences Directorate, William Goldstein, Associate Director.

This research has made extensive use of NASA's Astrophysics Data System (ADS) and also some use of the NASA/IPAC Extragalactic Database (NED) and the Sloan Digital Sky Survey (SDSS). 

This work was performed under the auspices of the U.S. Department of Energy by Lawrence Livermore National Laboratory in part under Contract W-7405-Eng-48 and in part under Contract DE-AC52-07NA27344.

\clearpage


\begin{appendix}
\section{APPENDIX MATERIAL}

Table~\ref{masterkey} provides the key for annotations in Table~\ref{masterlist}, the table of group annotations, as well as for Tables 1 through 16, 18, and 21 (primarily tables of individual morphological types).  This key is divided into three sections, (1) annotations for group visual morphological classifications, (2) annotations of additional comments, (3) annotations of individual component designations.  Each group was given at least one morphological classification.  Note the additional comments may not be comprehensive for the list, e.g. all HYMORS may not have been noted.  Individual source designations are discussed further below.  Table~\ref{masterlist} is an arbitrary segment of the online group annotation table containing 7106 higher count groups.
For each group in the list, the first line of the group includes the group identification number as generated by the grouping software, followed by a group position, group size, group source count, and an annotation comment as to the group morphology.  The group coordinate is a simple average of the component coordinates, excluding components designated single or sidelobe.  A range is given if there are linked groups to be considered.  Linked groups are discussed below.  The size given in this table is the greatest pairwise distance between group components, excluding components designated single sources or side-lobes.  This group line is followed by a line for each component in the group. These individual-member source lines contain the group identification number as well as the source identification number which was generated by the IDL astro database software (IDL Astronomy Users Library, http://idlastro.gsfc.nasa.gov/homepage.html) applied to the April 2003 FIRST catalog.   These group and component identification numbers are followed by the component's right ascension and declination and a component annotation.  The cutouts corresponding to this Table~\ref{masterlist} segment are presented in Figure~\ref{typical_ex}.
It should be noted the annotations are not necessarily mutually exclusive, for example an S or Z-shaped source may also be a double-double candidate, thus groups may have multiple morphological classification types listed.  If multiple morphological types are given, they generally are given in order of estimated probabilities.
Source annotations should be taken in the context of the group annotations.  In particular, uncertainty in visual group identification implies corresponding uncertainty in the source identifications.  Of course, core identifications should be regarded as apparent regardless, until verified by other evidence.  The individual source annotations should be taken as an initial attempt to
improve signal-to-noise ratio for statistics on the various components.  Obviously many of the identifications require further observation for accurate classification. 

Some group annotations include a list of linked groups, as for example Group 009409 in Table~\ref{masterlist} and Figure~\ref{typical_ex}. 
Generally, these indicate that the linked groups appear to be physically associated with the corresponding group due to symmetry or detail of morphology, i.e. the criteria for automatic group construction was not sufficiently large for the original grouping.  These linked groups provide candidate giant radio sources (GRSs).  However, for some links the relationship is source and sidelobe, as indicated.  Linked groups that are not in the four-or-more component group list have been appended to the online version of Table~\ref{masterlist}.  They generally have not been annotated.

 A '+' sign in the group annotation indicates a group was believed to consist of multiple systems, for example Group 009687 in Figure~\ref{typical_ex} is a possible double-double candidate or a triple (the catalog missing the core component) with nearby nearly aligned double and a sidelobe. 
Apparent chance-projection single sources were not necessarily included in the group comment, just indicated as 'S' in the individual source classifications.  The suffix '?' for an annotation suggests that while there were good indications of the associated type, it was felt there was not sufficient detail for unambiguous identification. The suffix '??' indicates some possibility of designated annotation.  Those groups consisting of probable sidelobes of a source outside the group were annotated 'sl ext source'.  The individual source classifications were left as classified before the originating source was determined. 

{\it Further Comments on Some Specific Annotations of Table~\ref{masterkey}:}
While most keyed items in Table~\ref{masterkey} are self-explanatory, a few comments are in order.

{\it Under Group Visual Morphological Classifications heading:} 

The annotation 'mg' (multiple possible groupings) was used to refer to groups for which multiple possibilities for physical associations appeared reasonable.  This annotation was followed by possible likely groupings.  These groups require higher resolution, sensitivity, and/or spectral data for definitive classification. 

The annotation 'rc' (resolved compact source) was used to describe a single contiguous non-pointlike, extended source that may or may not have associated jets or lobes apparent and may or may not have been resolved into multiple catalog components.  Their silhouette may be irregular. Early in the study the optical counterparts for a few of these sources were found to be the center of large galaxies.  
A brief discussion of \citet{Turner94} provides some context for resolved cores.   They studied ten nearby normal spiral galaxies, at 6 cm. and 2 cm. with ~1.5" beam size, chosen as likely to have bright continuum emission with the goal of determining the morphology of star-forming activity.  They found bright radio continuum within the inner arcminute of eight of the ten.  They were able in some cases to separate thermal and nonthermal emission and suggested the nonthermal component originates in a large population of young SNRs associated with recent star formation.  They identified a dozen SN or SN candidates.  Of the ten radio galaxies, three, Maffei 2, NGC 3031 (M81) and NGC 5236 (M83), have what is here denoted 'rc' morphology.  None of these three galaxies were covered by FIRST.
Two of the ten had indications of spiral morphology.  For NGC 4258, the Turner and Ho 6 cm. image showed S-shape in their 2 kpc map, while the FIRST image (three-component group) shows only possible faint tracings of spiral arms. 
In contrast, NGC 5194 (M51) shows better suggestions of spiral arms in FIRST (nine-component FCG J132952.8+471140 (Group 419136), annotated 'rc sz? pinwheel? cj?') than the 6 cm. map of Turner and Ho. 
For NGC 4826, the 6 cm. map would have been classified 'tri'  and the 2 cm. map 'mg'.  It appears as a single extended source in FIRST, with low level halo.
NGC 4736 (M94), with ring morphology around a central source, is discussed in Section~\ref{ringandringlike}.  Their remaining systems, NGC 2403, NGC 4236, and NGC 5457 (M101), were classified quiet types by Turner and Ho, having little radio continuum emission within the central arcminute.  


The distinction between double, resolved core, and core-jet morphology becomes rather arbitrary for smaller angular diameter groups.

The annotation 'unu' was generally applied to unusual configurations that could be explained by chance projection.  If they are not chance projections, they would be deemed interesting.
The annotation 'irr' denotes a group whose members appear to be physically associated but are not well described by another morphology class.   Such groups may provide candidate merger or relic type sources. 
The annotations 'quad' and 'quint' were used to describe more point-like sources with little or no low level extended structure.  Such groups may provide candidate core-type sources or possibly lens candidates.  They may of course be chance projections.

The designations 'tb','b','bs' also indicate bent types, 'tb' being a three-component group for which there was less indication of extended structure but symmetry suggested a physical association, 'b' being a group for which the core was not well defined but there was clear indication of bent structure, and 'bs' for groups for which the core was not well defined with only slight deviation in orientation.  In general they have less detail in morphology.

{\it Under Additional Comments:}

 Since the focus was on more morphologically definitive types, the annotations in this section should be considered as exemplary rather than being comprehensive for the list.

Concerning arcs, while it is expected that the great majority of groups with this annotation may be portions of edge-brightened lobes for which a core and/or other lobe are not apparent, a possible cause of arcs may be gravitational lensing.  
Recently \citet{Belokurov09} have reported two large-separation gravitational lenses (of radius ~4" and ~11") found by the Cambridge Sloan survey of Wide Arcs in the Sky.  There is further discussion on gravitational lensing in Section~\ref{Einstein_rings}.  Arcs may also be indicative of relics or shocks.  Some may be bent jets for which lobes and core are not apparent.

The comment 'distorted' was applied to groups for which there was a smeared appearance to the morphology.  These may provide possible merger candidates.

For the purposes of this paper, the annotation 'dl' (dogleg) is used to describe jets or lobe portions for which there is an abrupt change in direction. 

The annotation 'int' (interesting) is a subjective evaluation, which may well have varied over the course of the classifications.

The comment 'jet/tail' indicates a protrusion (normally low level) from lobe in direction away from core.  The corresponding component, if cataloged, was labeled jet.

The groups annotated 'low level' are examples of groups with extended low level flux, that may or may not have been resolved by the software.  Candidate mini-halos and halos are perhaps to be found in those with groups with the 'low level' comment.

The 'ss' (single sided) comment was applied to groups for which the morphology suggested a single sided jet or lobe.  They are also possible core-jet sources or Doppler shifted doubles for which only one lobe is apparent.

{\it Concerning Individual Source Classifications:}

An 'S' annotation for a source indicates the source appeared to be a source associated with the group by chance projection.  An 'S' annotation was also used to indicate a point-like source, other group members being sidelobes.  For some source components, the assignment of jet or lobe annotation was difficult and even arbitrary. 


\begin{deluxetable}{ll}
\tabletypesize{\scriptsize}
\tablenum{A-I}
\tablecaption{Keys for Group Classifications, Comments, and Source Designations \label{masterkey}}
\tablewidth{0pt}
\tablehead{ \colhead{Code} & \colhead{Classification/Comment/Designation} }
\startdata
\sidehead{ Group Visual Morphological Classifications:}
 a&   ambiguous\\
 b&   bent (B), little visual indication of core\\
 bs&  bent slightly\\
 cj&  core-jet\\
 d&   double lobe radio galaxy - may include core-jet sources\\
 dd&  double double radio galaxy (DD)\\
 irr& irregular, distorted\\
 lobe& lobe, resolved lobe\\
 mg&  multiple groupings possible\\
 nat& narrow angle tail (NAT)\\
 quad& group of four approximately point-like sources (QUAD)\\
 quint& group of five approximately point-like sources (QUINT)\\
 rc&  resolved compact source, non-pointlike\\
 ring& ring\\
 ring-lobe& edge brightened lobe or embedded ring\\
 s&   single source (often with sidelobes) or probable chance projection of point-like source into group\\
 sl&  sidelobe\\
 sz&  S-shaped or Z-shaped\\
 t&   triple, no bend, little extended structure\\
 tb&  bent triple (TB), little indication of extended structure\\
 tbs& slightly bent triple\\
 td&  triple double\\
 tri& 120 deg. rotational symmetry - may include chance projections\\
 unu& unusual, uncommon - may be due to chance projection\\
 w&   W-shape, wiggles\\
 wat& wide angle tail (WAT)\\
 x&   X-shape\\

\sidehead{ Additional Comments:}
 arc& arc or C, possible edge brightened lobe, bent jets, or ring fragment\\
 asym& asymmetric\\
 bifurcation&  a same side splitting of jet from apparent core\\
 butterfly& four lobe morphology\\
 cat& catalog\\
 cp&  chance projection\\
 diag& diagonal\\
 distorted& smearing of prototypical morphologies with low level emission\\
 dl&  dogleg, generally an abrupt change in direction\\
 pinwheel& spiral type similar to optical star forming regions of spiral galaxies\\
 fan& low level triangular shape lobe\\
 hook& hook, an apparent approximately 180 change in direction of jet\\
 hymor& hybrid FRI-FRII morphology - a few HYMOR candidates were specifically annotated\\ 
 id& identification\\
 int& interesting\\
 jet& jet or occasionally protrusion from lobe\\
 low level& extended low level flux, diffuse and/or filamentary\\
 ppd& perpendicular\\
 ss& apparent single sided jet and/or lobe\\

\sidehead{Individual Source Classifications:}
 A&  ambiguous\\
 CJC& core-jet core component\\
 CJJ& core-jet jet component\\
 CORE&  core component\\
 J&  jet\\
 LOBE& lobe\\
 RC& resolved core\\
 RJ& resolved jet\\
 RING&  ring component\\
 RL& component of resolved lobe\\
 S&   single source, no structure apparent, likely chance projection into group\\
  &   or single bright point-like source, other group members being sidelobes\\
 SL& sidelobe\\
 SZJ& jet component of S-shape or Z-shape\\
 SZC&   core component of S-shape or Z-shape\\
 X&  transverse component of X group\\
\enddata
\end{deluxetable}
\clearpage


\begin{deluxetable}{lllll}
\tabletypesize{\scriptsize}
\tablenum{A-II}
\tablecaption{Morphological Annotations and Components for Groups with Four or More Components \label{masterlist}}
\tablewidth{0pt}
\tablehead{
\colhead{group\tablenotemark{a}} & \colhead{source\tablenotemark{b}} &
                 \colhead{R.A.(2000)\tablenotemark{c}} & \colhead{Dec.(2000)\tablenotemark{d}} & \colhead{Source Morphology\tablenotemark{e}}
}
\startdata

\sidehead{\hspace{0.8em}009186\hspace{2.3em}- - - - -\hspace{2.65em}01 14 26.187 \hspace{.8em} +00 29 41.11 \hspace{1em} size=  1.05' \hspace{1em} count:  4  \hspace{1em}comment: wat}
009186& 011015&  01 14 24.099&  +00 29 37.18&  LOBE\\
009186& 011020&  01 14 25.594&  +00 29 32.59&  CORE\\
009186& 011024&  01 14 26.998&  +00 29 35.58&  RL\\
009186& 011028&  01 14 28.058&  +00 29 59.09&  RL\\
 
\sidehead{\hspace{0.8em}009304 \hspace{2.3em}- - - - -\hspace{2.65em}01 15 27.666 \hspace{.5em}  +00 00 03.56 \hspace{1em} size=  1.34' \hspace{1em}  count:  4  \hspace{1em}comment: x int}
009304& 011149&  01 15 27.281&  -00 00 41.85&  LOBE\\
009304& 011151&  01 15 27.463&  +00 00 05.65&  X\\
009304& 011152&  01 15 27.675&  +00 00 38.50&  LOBE\\
009304& 011154&  01 15 28.247&  +00 00 11.94&  X\\
 
\sidehead{\hspace{0.8em}009409 \hspace{2.3em}- - - - -\hspace{2.65em}01 16 16.777  \hspace{.5em} -10 41 20.50 \hspace{1em} size=  0.52'  \hspace{1em}  count:  5  \hspace{1em}comment: mg lobe of wat? sz? linking grp= 009430 }
009409& 011277&  01 16 15.869&  -10 41 18.97&  RL\\
009409& 011278&  01 16 16.235&  -10 41 23.26&  RL\\
009409& 011280&  01 16 17.011&  -10 41 20.03&  RL\\
009409& 011283&  01 16 17.993&  -10 41 19.74&  RL\\
009409& 011292&  01 16 20.138&  -10 41 09.56&  A\\
 
\sidehead{\hspace{0.8em}009441 \hspace{2.3em}- - - - -\hspace{2.65em}01 16 32.009  \hspace{.5em} -01 27 33.29 \hspace{1em} size=  0.37' \hspace{1em}  count:  4  \hspace{1em}comment: t sz?}
009441& 011317&  01 16 29.001&  -01 26 57.56&  S\\
009441& 011325&  01 16 31.854&  -01 27 44.31&  LOBE\\
009441& 011326&  01 16 31.983&  -01 27 32.98&  CORE\\
009441& 011328&  01 16 32.189&  -01 27 22.59&  LOBE\\
 
\sidehead{\hspace{0.8em}009687 \hspace{2.3em}- - - - -\hspace{2.65em}01 18 28.604  \hspace{.5em} +01 14 20.03 \hspace{1em} size=  0.69' \hspace{1em}  count:  5  \hspace{1em}comment: mg dd t+d+sl?}
009687& 011606&  01 18 27.311&  +01 14 18.17&  LOBE\\
009687& 011608&  01 18 27.896&  +01 14 17.35&  LOBE\\
009687& 011613&  01 18 29.170&  +01 14 21.92&  LOBE\\
009687& 011619&  01 18 30.039&  +01 14 22.66&  LOBE\\
009687& 011627&  01 18 31.550&  +01 14 27.08&  SL\\
 
\sidehead{\hspace{0.8em}009752 \hspace{2.3em}- - - - -\hspace{2.65em}01 19 01.044  \hspace{.5em} -10 03 23.44 \hspace{1em} size=  0.94' \hspace{1em}  count:  4  \hspace{1em}comment: nat sz?}
009752& 011688&  01 19 00.129&  -10 03 08.69&  A\\
009752& 011690&  01 19 00.162&  -10 03 46.55&  A\\
009752& 011691&  01 19 00.228&  -10 03 15.24&  A\\
009752& 011697&  01 19 03.659&  -10 03 23.27&  A\\
 
\sidehead{\hspace{0.8em}009861 \hspace{2.3em}- - - - -\hspace{2.65em}01 19 46.935  \hspace{.5em} -00 39 47.57 \hspace{1em} size=  1.06' \hspace{1em}  count:  5  \hspace{1em}comment: sl ext source}
009861& 011820&  01 19 45.134&  -00 39 57.92&  SL\\
009861& 011821&  01 19 45.266&  -00 39 43.06&  SL\\
009861& 011824&  01 19 46.688&  -00 39 51.43&  SL\\
009861& 011831&  01 19 48.395&  -00 39 45.41&  SL\\
009861& 011833&  01 19 49.191&  -00 39 40.02&  SL\\
 
\sidehead{\hspace{0.8em}009902 \hspace{2.3em}- - - - -\hspace{2.65em}01 20 12.538  \hspace{.5em} -00 38 41.08 \hspace{1em} size=  3.27' \hspace{1em}  count: 13  \hspace{1em}comment: t dd??}
009902& 011870&  01 20 06.194&  -00 39 04.48&  RL\\
009902& 011871&  01 20 06.785&  -00 39 04.52&  RL\\
009902& 011872&  01 20 07.306&  -00 38 58.48&  RL\\
009902& 011874&  01 20 08.637&  -00 38 55.63&  RL\\
009902& 011875&  01 20 09.111&  -00 39 01.08&  RL\\
009902& 011878&  01 20 09.534&  -00 38 48.17&  RL\\
009902& 011884&  01 20 12.510&  -00 38 37.71&  CORE\\
009902& 011888&  01 20 14.903&  -00 38 29.52&  RL\\
009902& 011890&  01 20 16.319&  -00 38 38.01&  RL\\
009902& 011891&  01 20 16.665&  -00 38 19.22&  RL\\
009902& 011894&  01 20 17.631&  -00 38 24.66&  RL\\
009902& 011895&  01 20 18.430&  -00 38 10.47&  RL\\
009902& 011896&  01 20 18.966&  -00 38 22.14&  RL\\
 
\sidehead{\hspace{0.8em}010036 \hspace{2.3em}- - - - -\hspace{2.65em}01 21 06.778  \hspace{.5em} -09 26 10.13 \hspace{1em} size=  0.60' \hspace{1em}  count:  4  \hspace{1em}comment: t dd?}
010036& 012037&  01 21 04.370&  -09 26 15.74&  SL\\
010036& 012041&  01 21 05.594&  -09 26 12.83&  LOBE\\
010036& 012043&  01 21 06.740&  -09 26 10.64&  CORE\\
010036& 012048&  01 21 07.999&  -09 26 06.92&  LOBE\\
 
\sidehead{\hspace{0.8em}010141 \hspace{2.3em}- - - - -\hspace{2.65em}01 21 56.963  \hspace{.5em} -00 29 29.68 \hspace{1em} size=  1.00' \hspace{1em}  count:  4  \hspace{1em}comment: t? d? sl?}
010141& 012163&  01 21 54.377&  -00 29 35.04&  LOBE\\
010141& 012168&  01 21 57.051&  -00 29 29.72&  A\\
010141& 012174&  01 21 58.145&  -00 29 21.11&  RL\\
010141& 012176&  01 21 58.367&  -00 29 32.90&  RL\\
 
\enddata

\tablecomments{Table \ref{masterlist} is available in its entirety in ASCII FORMAT in the electronic edition of this journal.  A portion is shown here for guidance regarding its form and content.}
\tablecomments{Sizes given in this table are diameters of groups as constructed by grouping software, excluding labeled sidelobes and chance projections.}
\tablenotetext{a}{Group identification number.}
\tablenotetext{b}{Source component identification number.}
\tablenotetext{c}{Source Right Ascension (hr min sec), group or source component.}
\tablenotetext{d}{Source Declination (deg min sec), group or source component.}
\tablenotetext{e}{See Table~\ref{masterkey} for key.}
\end{deluxetable}

\end{appendix}

\clearpage

\clearpage

\begin{figure*}
\centering
\plotone{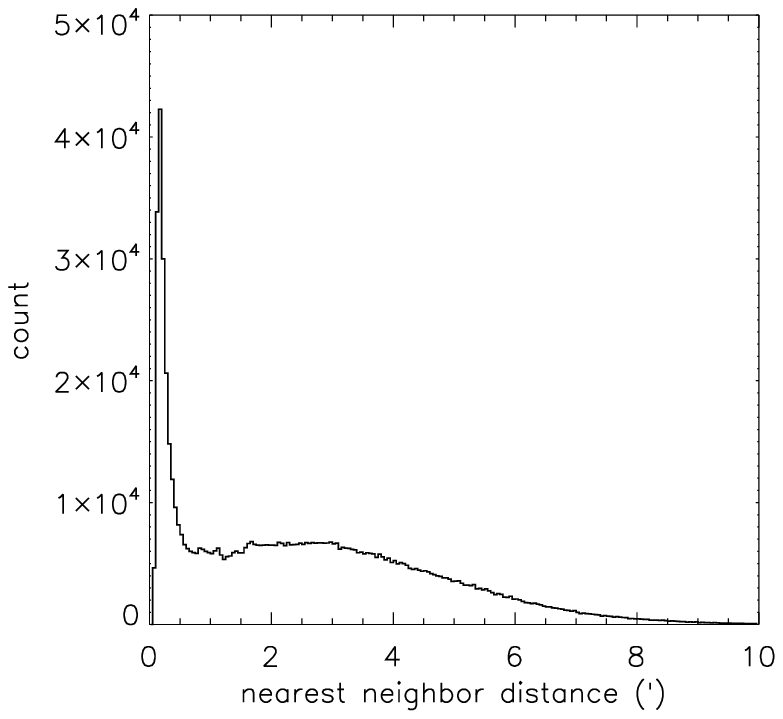}
\caption{Histogram, nearest neighbor distances for 811117 sources in FIRST catalog (April 2003 version).  (There were 1181 sources with nearest neighbor distance greater than 10'.)}\label{nn_hist}
\end{figure*}

\begin{figure*}
  \centering
  \plotone{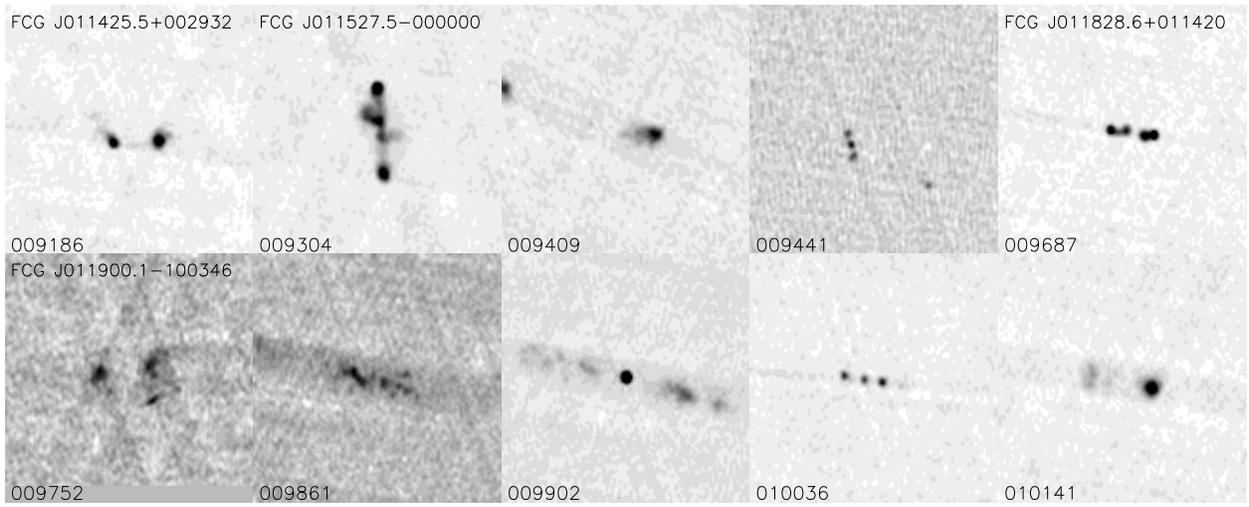}
  \caption{\label{fig:typical_ex}Cutouts associated with Table~\ref{masterlist} examples.  If assigned, the IAU compliant designation is given in the upper portion of the cutout.  The six digit group number is given in the lower portion.} \label{typical_ex}
\end{figure*}

\begin{figure*}
  \centering
  \newcounter{myfigcounter1}   
  \setcounter{myfigcounter1}{\value{figure}}  
  \plotone{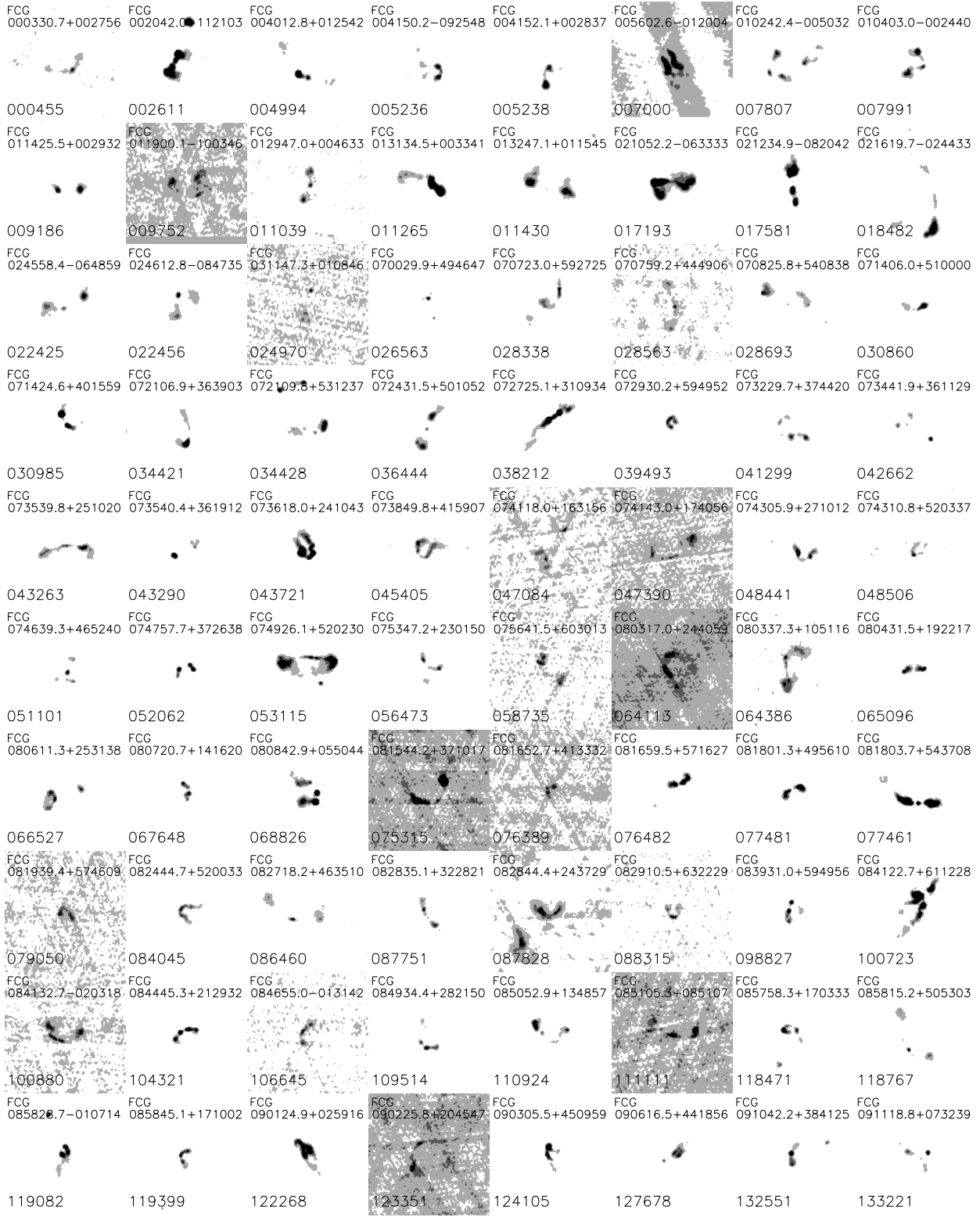}
  \caption{\label{fig:typical_natwat}Typical examples of visually identified WAT and NAT morphology systems, with IAU compliant designation and group number.} \label{typical_natwat}
\end{figure*}

\begin{figure*}
  \centering
  \newcounter{myfigcounter3}   
  \setcounter{myfigcounter3}{\value{figure}}  
  \plotone{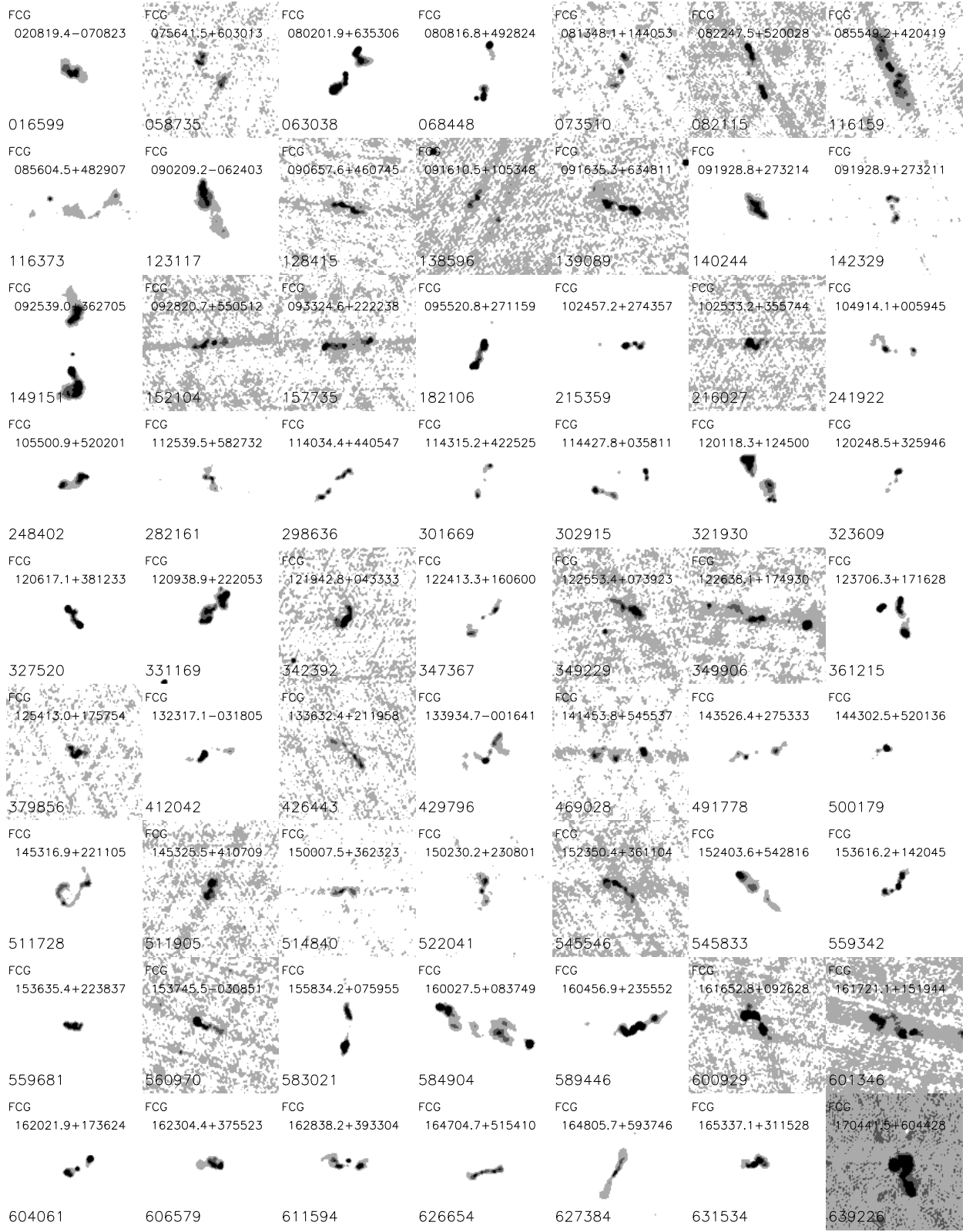}
  \caption{\label{fig:typical_ws}Visually identified W-shaped systems, with IAU compliant designation and group number.}\label{typical_ws}
\end{figure*}

\begin{figure*}
  \centering
  \plotone{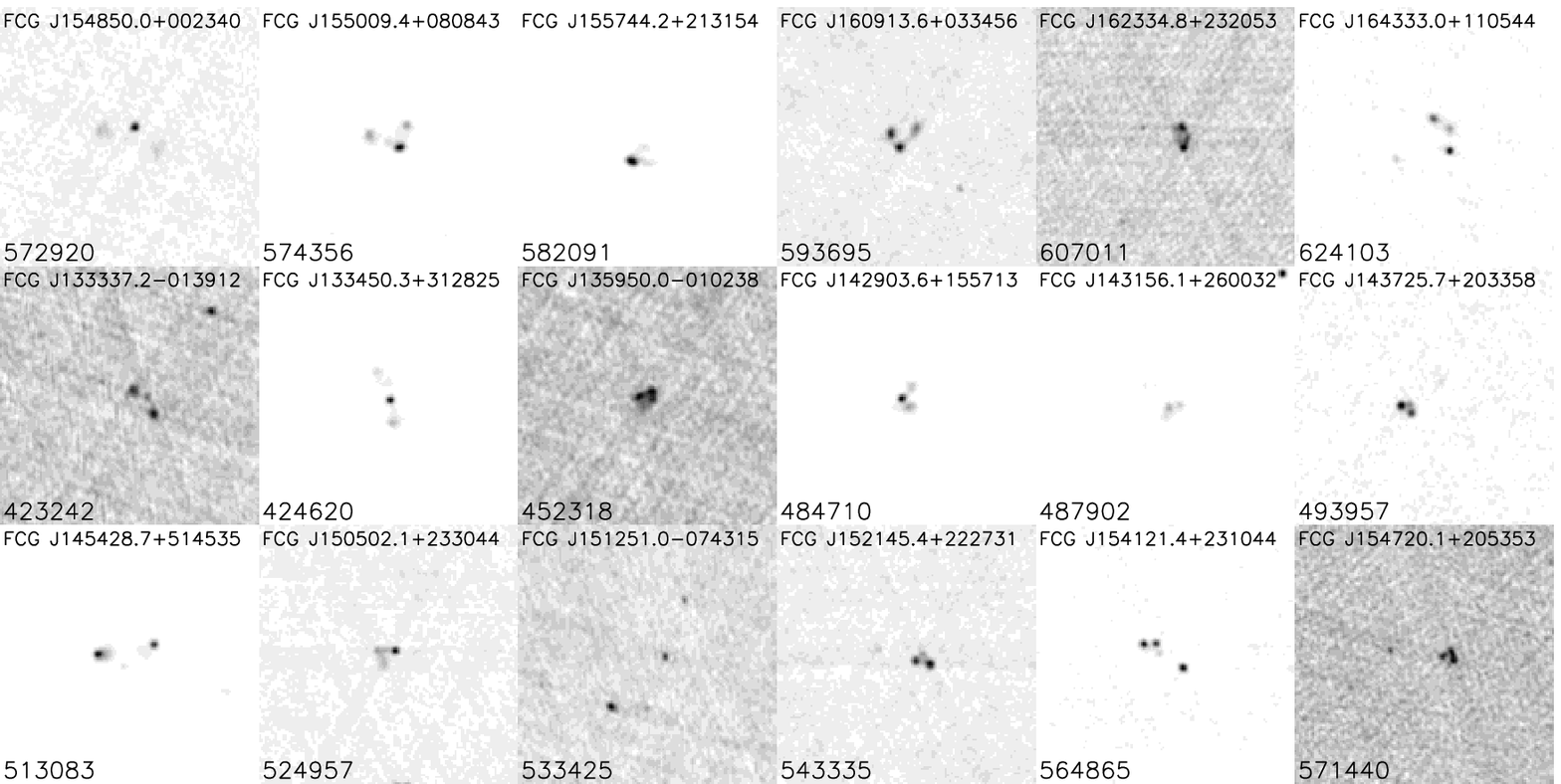}
  \caption{\label{fig:typical_tb}Examples of systems with TB type morphology, with IAU compliant designation and group number.} \label{typical_tb}
\end{figure*}

\begin{figure*}
  \centering
  \plotone{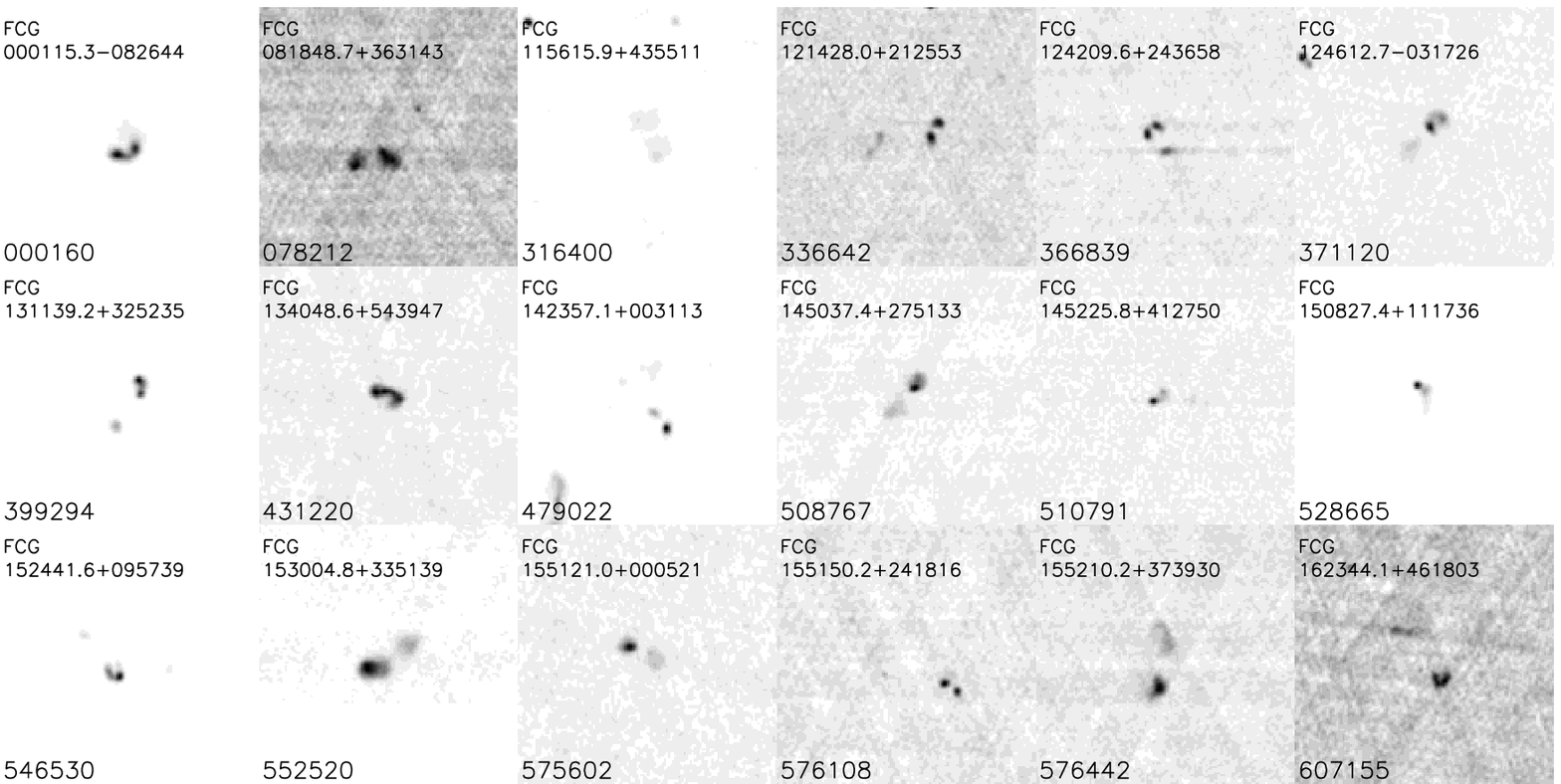}
  \caption{\label{fig:typical_bs}Examples of systems with B type morphology, with IAU compliant designation and group number.} \label{typical_bs}
\end{figure*}

\begin{figure*}
  \centering
  \plotone{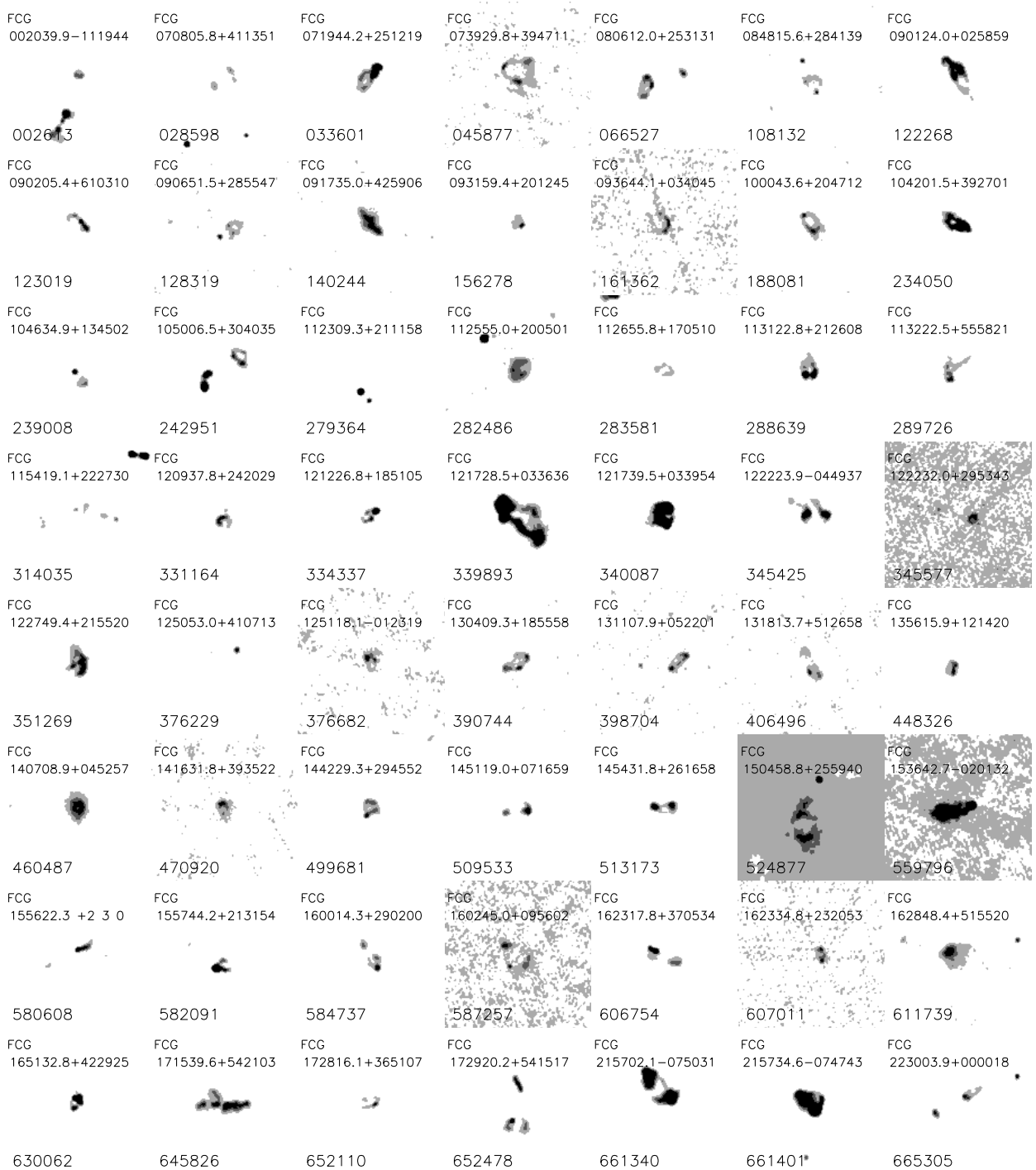}
  \caption{\label{fig:typical_ring}Visually identified ring-like structures, with IAU compliant designation and group number.  FCG J150458.8+255940 (Group 524877) is identified with the lobe of 3C 310.  In FCG J162824.8+230105 (Group 611411) the ring appears transverse to the jet.  A larger image is shown in Fig.~\ref{Cur_ppd_ring}.  While FCG J122749.4+215520 (Group 351269) also has a ring that appears transverse to jets, in this case the ring appears more centered on the system.} \label{typical_ring}
\end{figure*}

\begin{figure*}
  \epsscale{1.}
  \centering
  \plotone{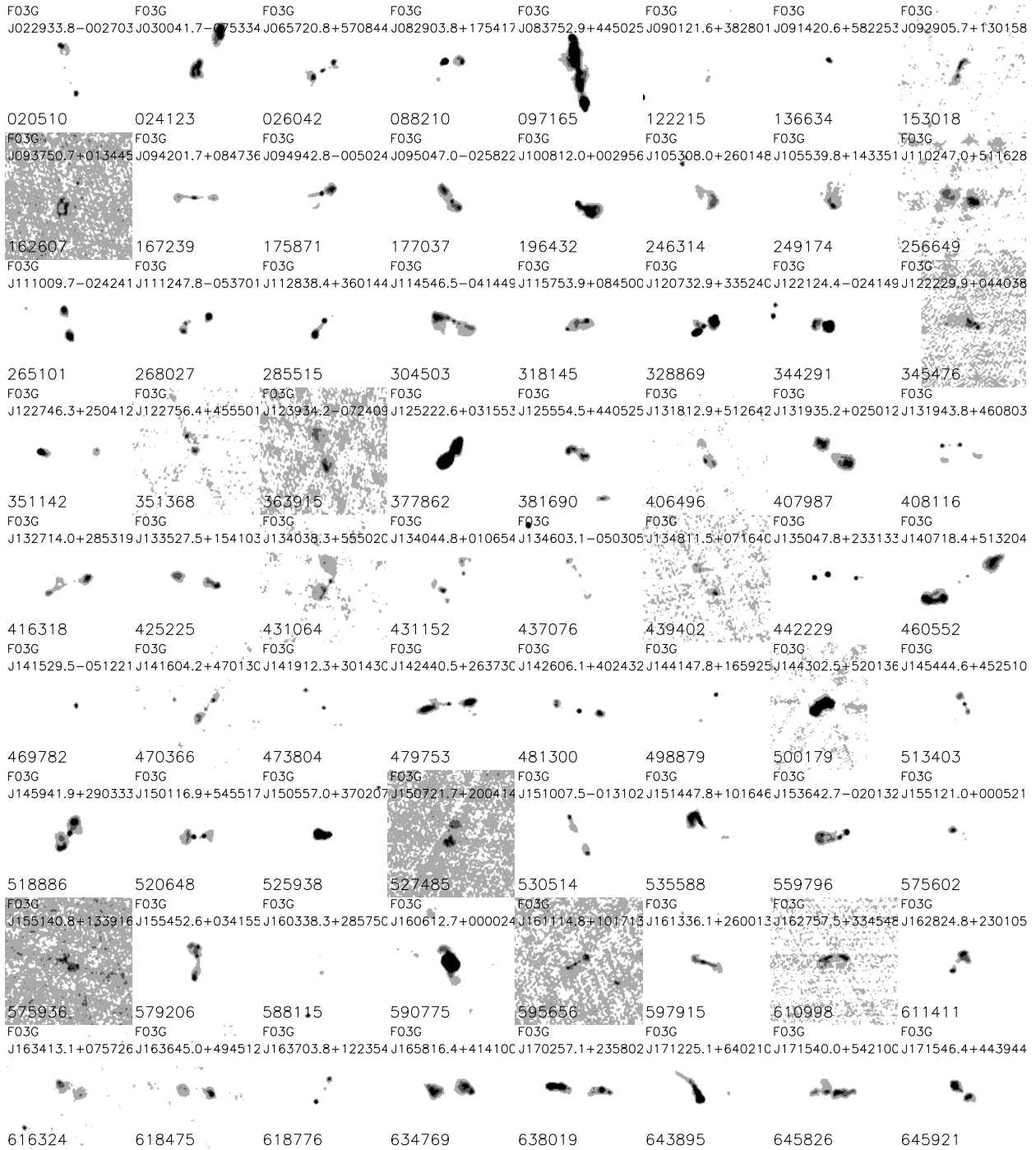}
  \caption{\label{fig:typical_ring-lobes}Visually identified systems having edge-brightened ring-like lobes, with IAU compliant designation and group number.} \label{typical_ring-lobes}
\end{figure*}

\begin{figure*}
  \epsscale{1.}
  \centering
  \plotone{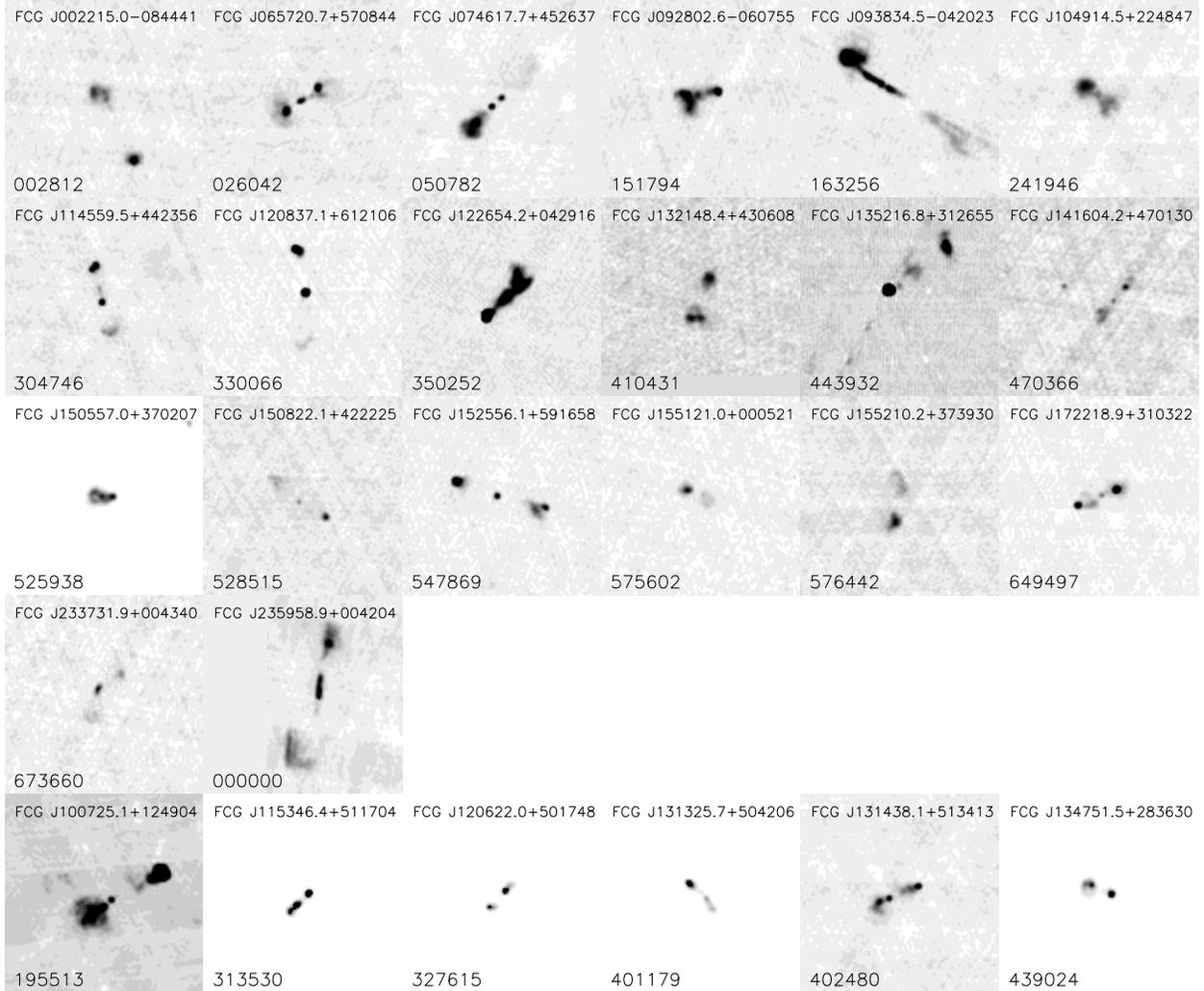}
  \caption{\label{fig:hymors}Visually identified HYMOR candidates, with IAU compliant designation and group number, along with FIRST cutouts of six previously identified HYMORS in the last row for comparison.  FCG J100726.1+124856 (Group 195513, 4C+13.41) was identified by \citet{Gopal-Krishna00} and the remaining five were presented by \citet{Gawronski06}.} \label{hymors}
\end{figure*}

\begin{figure*}
  \epsscale{.5}
  \centering
  \plotone{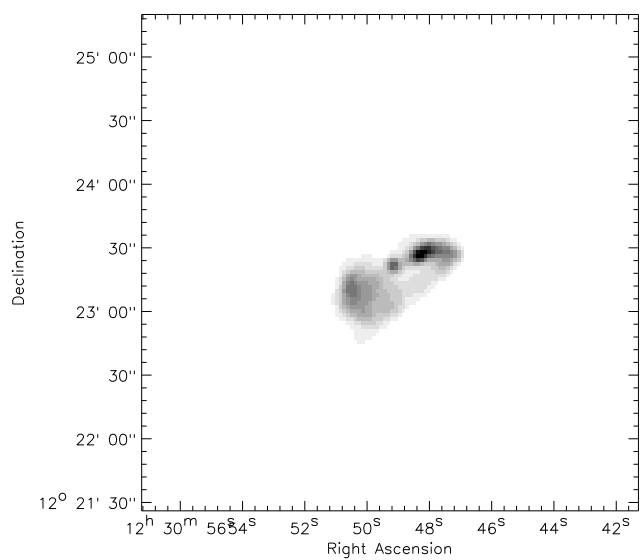}
  \caption{\label{fig:M87} FIRST image of M87.} \label{M87}
\end{figure*}

\begin{figure*}
  \newcounter{myfigcounter2}   
  \setcounter{myfigcounter2}{\value{figure}}  
  \epsscale{1.}
  \centering
  \plotone{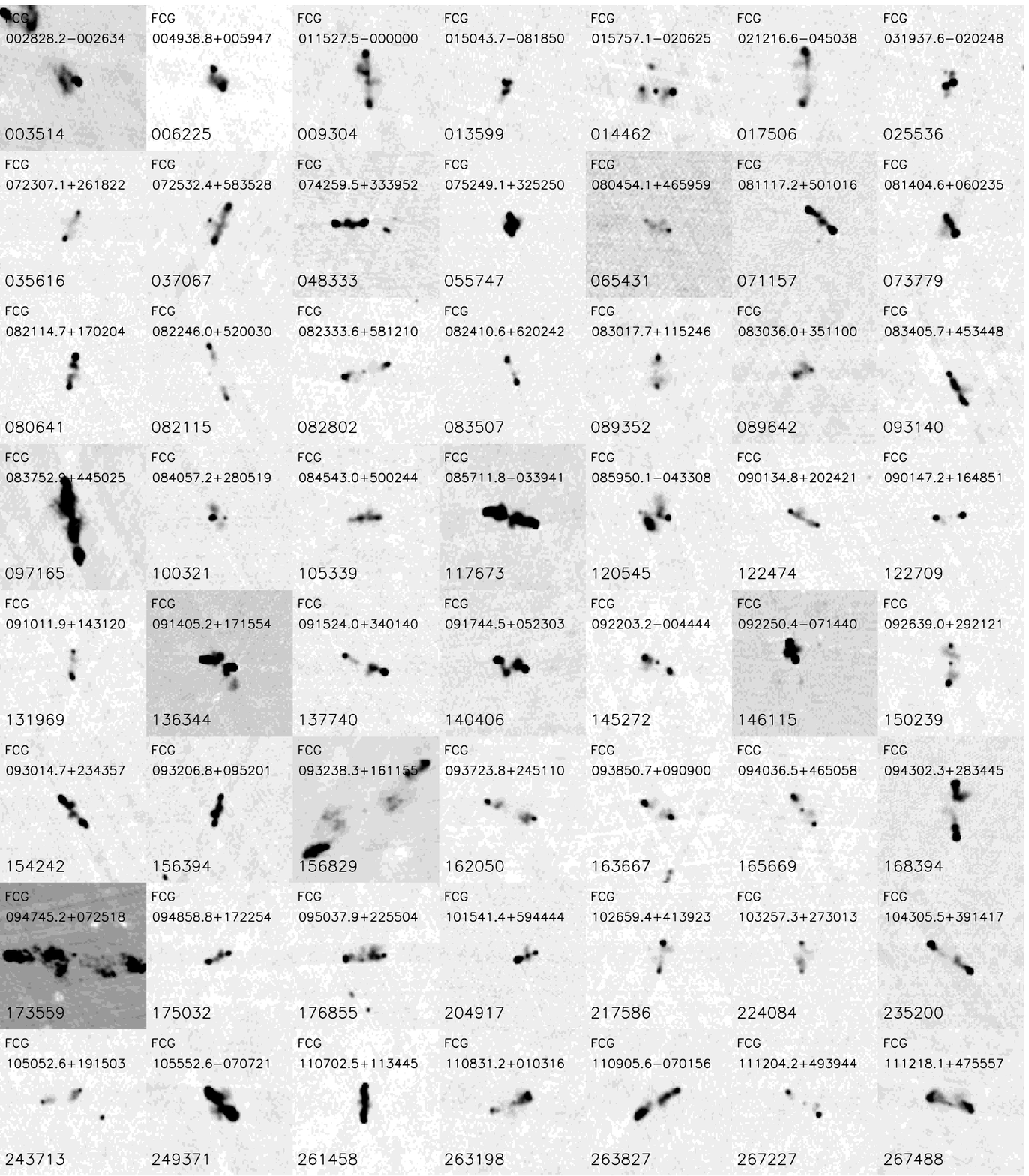}
  \caption{\label{fig:typical_xs}Typical examples of visually identified X-shaped sources, with IAU compliant designation and group number.} \label{typical_xs}
\end{figure*}

\begin{figure*}
  \newcounter{myfigcounter4}   
  \setcounter{myfigcounter4}{\value{figure}}  
  \centering
  \plotone{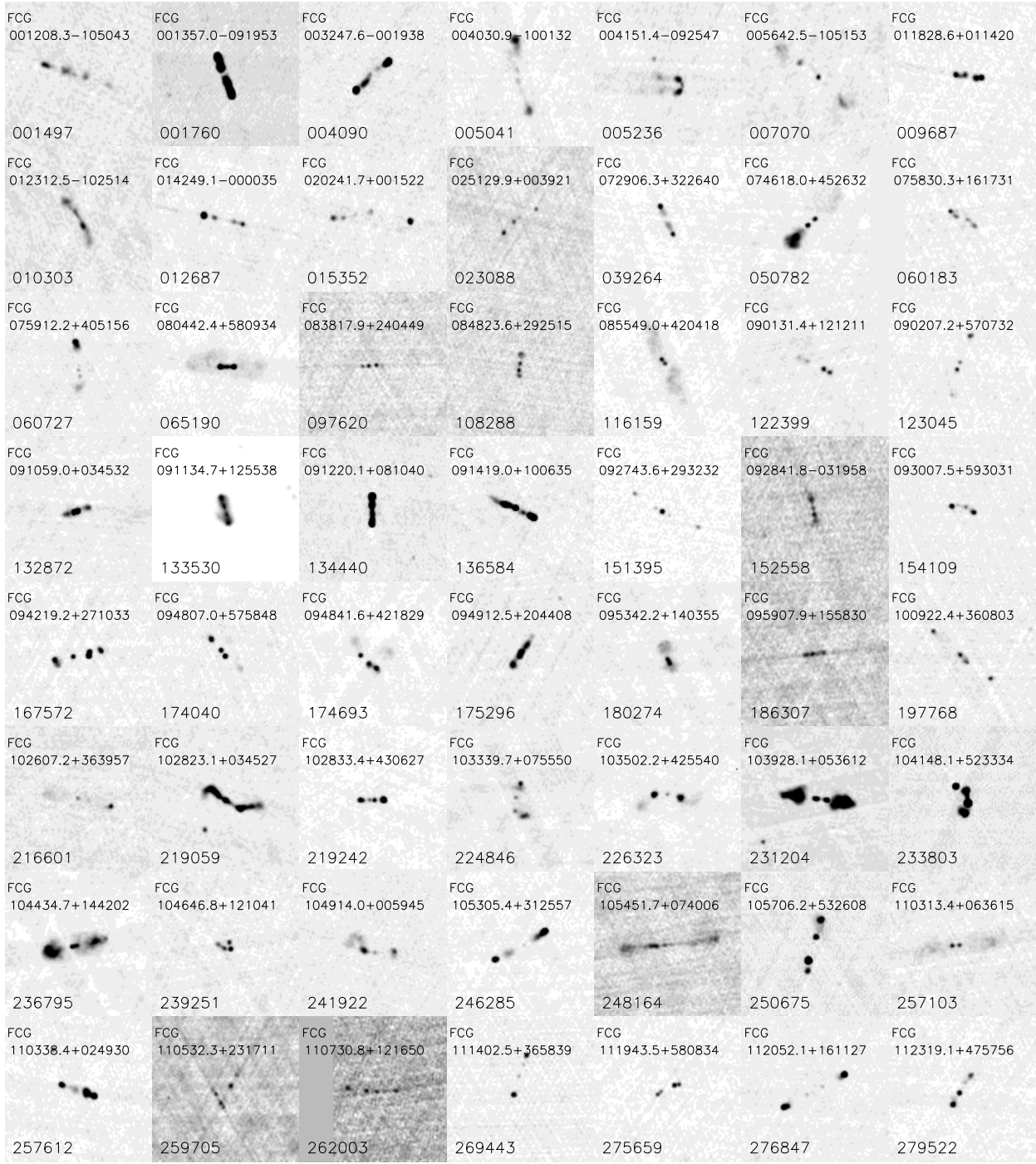}
  \caption{\label{fig:typical_dds} Examples of visually identified DDRGs, with IAU compliant designation and group number.} \label{typical_dds}
\end{figure*}

\begin{figure*}
  \centering
  \plotone{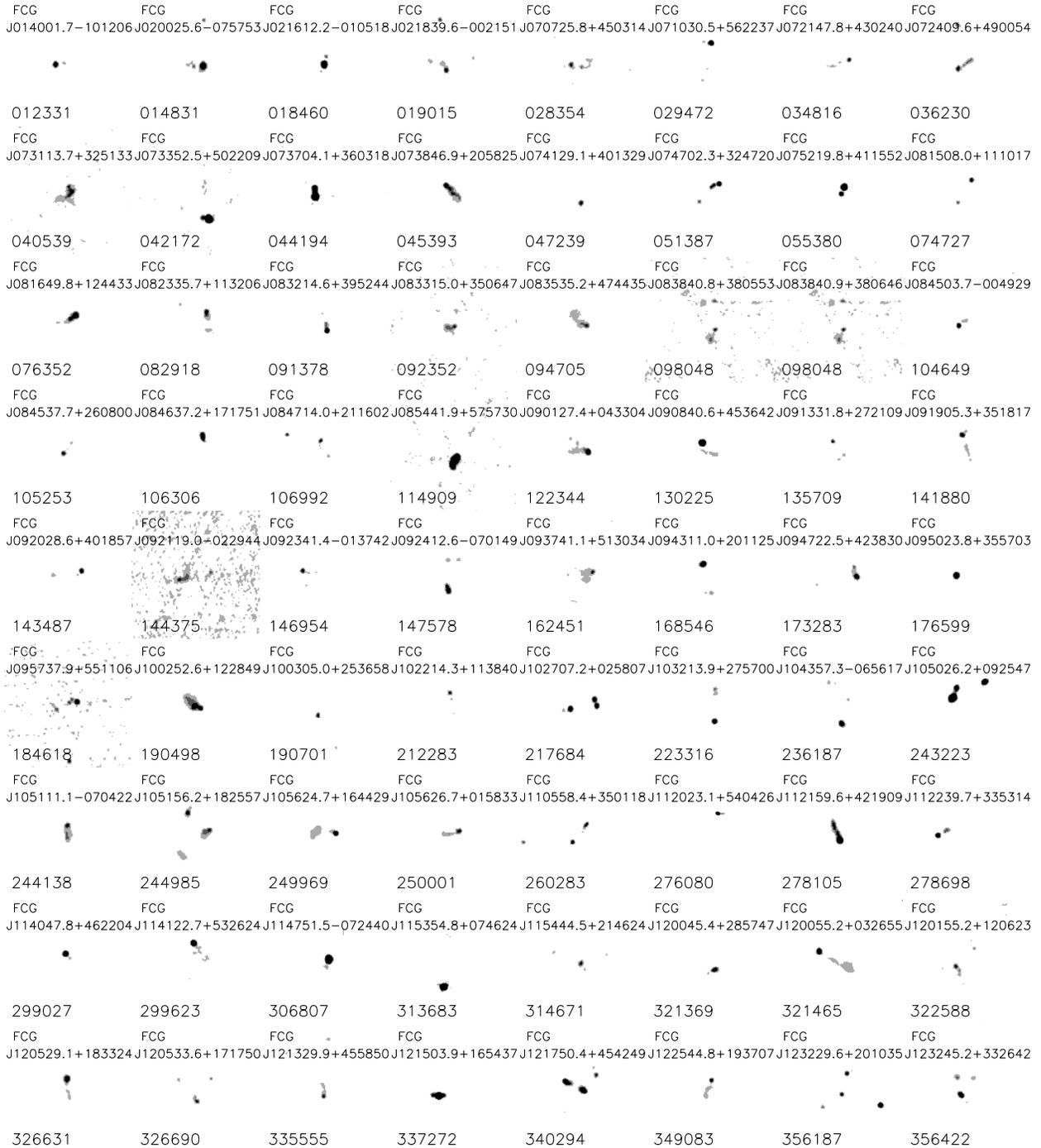}
  \caption{\label{fig:typical_cjs}Examples of visually identified core-jets, with IAU compliant designation and group number.} \label{typical_cjs}
\end{figure*}

\begin{figure*}
  \centering
  \plotone{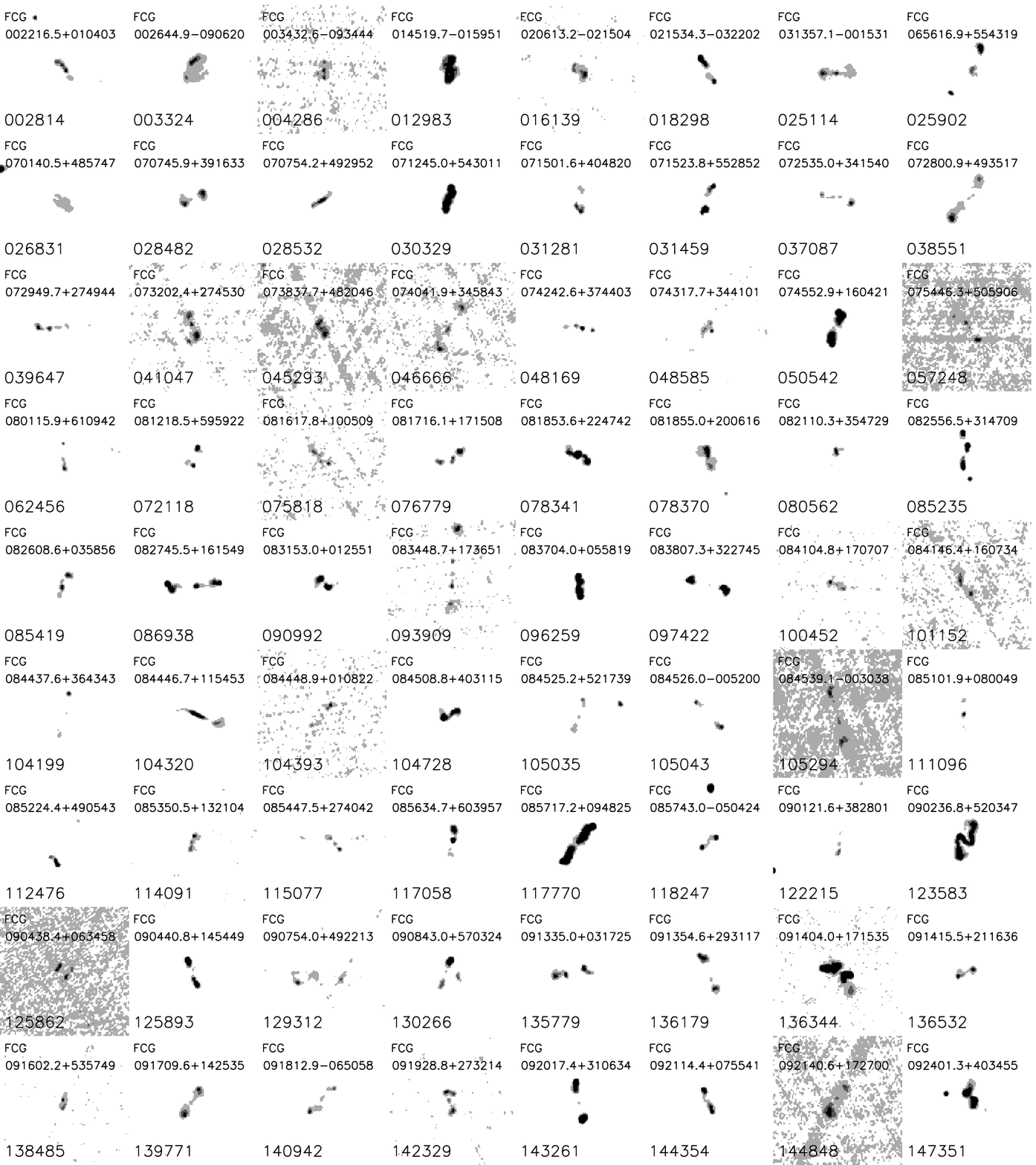}
  \caption{\label{fig:typical_sps}Typical examples of visually identified S or Z-shaped sources, with IAU compliant designation and group number.} \label{typical_sps}
\end{figure*}

\begin{figure*}
  \centering
  \plotone{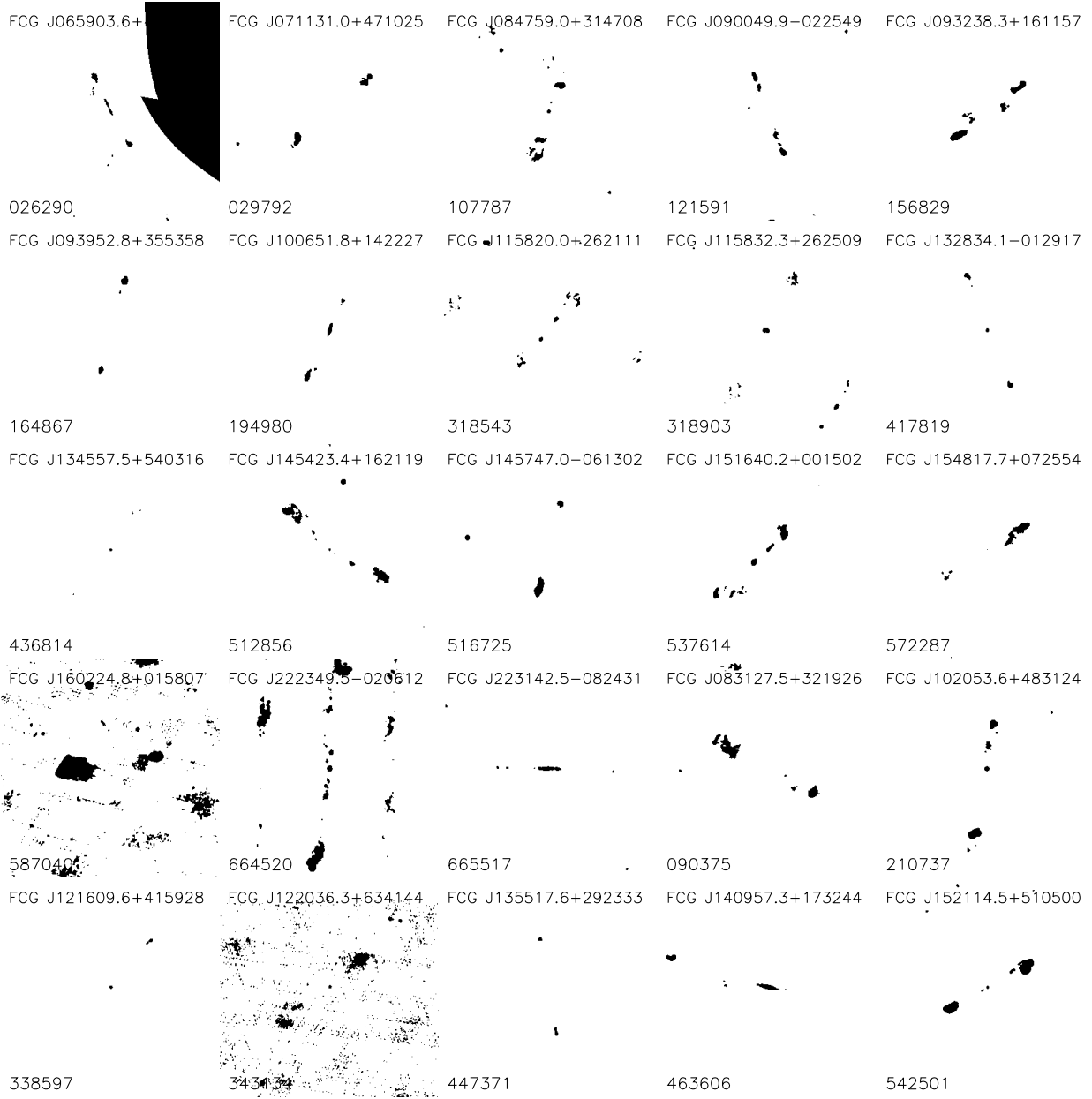}
  \caption{\label{fig:typical_grg}Visually identified GRSs, with IAU compliant designation and group number.  Cutout size is 10'.  The final five images are the FIRST cutouts of known giant radio galaxies.  See Table~\ref{grgtable} for references.} \label{typical_grg}
\end{figure*}

\clearpage


\begin{figure*}
  \epsscale{.5}
  \centering
  \plotone{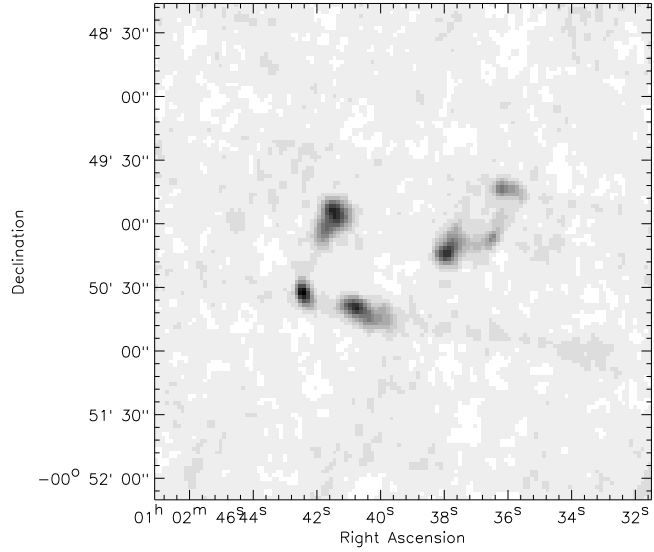}
  \caption{\label{fig:Cur_starwars}FCG J010236.5-005007 and FCG J010242.4-005032 (Group 007807) - An interesting pair of radio galaxies.} \label{Cur_starwars}
\end{figure*}

\begin{figure*}
  \epsscale{.5}
  \centering
  \plotone{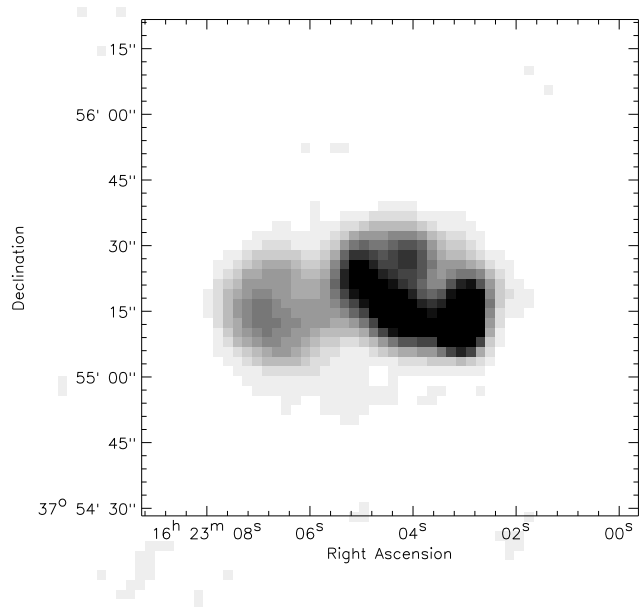}
  \caption{\label{fig:Cur_pretzel}FCG J162304.4+375523 (Group 606579) - An interesting pretzel shaped radio source.} \label{Cur_pretzel}
\end{figure*}

\clearpage

\begin{figure*}
  \epsscale{.5}
  \centering
  \plotone{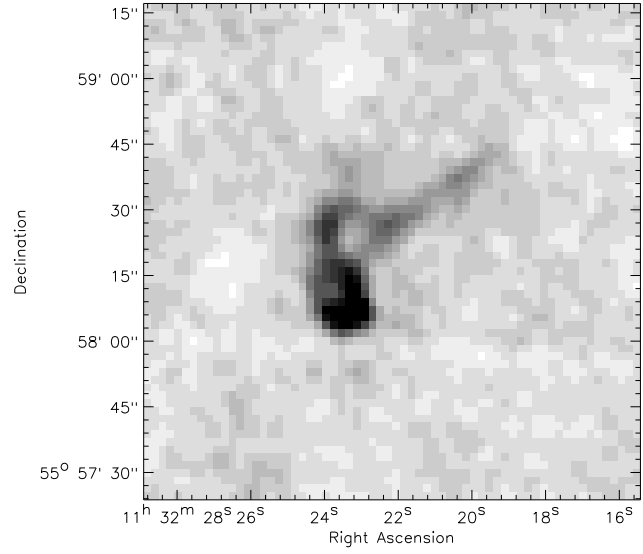}
  \caption{\label{fig:Cur_pretzelwh}FCG J113222.5+555821 (Group 289726) - Another interesting 'pretzel-with-handle' shaped source.} \label{Cur_pretzelwh}
\end{figure*}

\begin{figure*}
  \epsscale{.5}
  \plotone{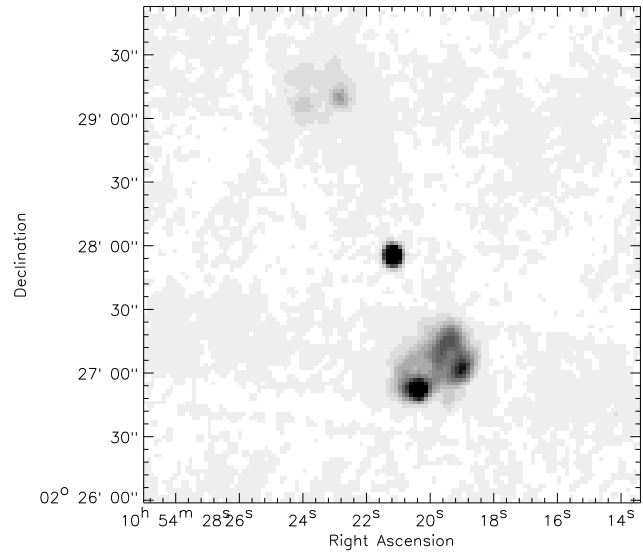}
  \caption{\label{fig:Cur_pretzellobe}FCG J105420.0+022700 (Group 247661) - Another interesting pretzel, possibly a lobe.} \label{Cur_pretzellobe}
\end{figure*}

\clearpage

\begin{figure*}
  \epsscale{.5}
  \centering
  \plotone{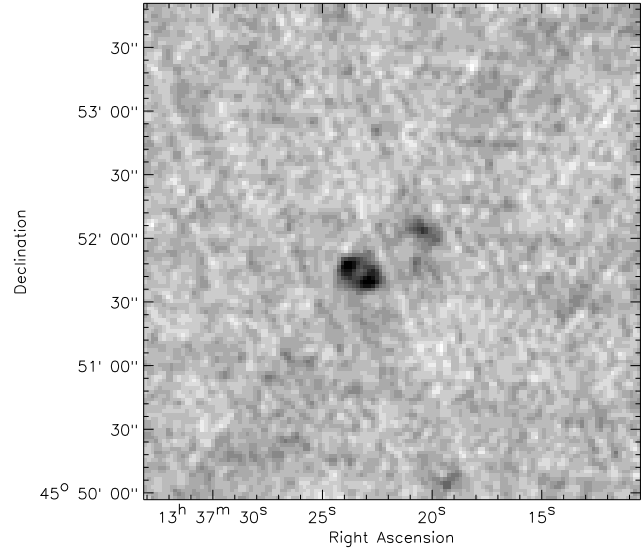}
  \caption{\label{fig:Cur_ring_and_jet}FCG J133724.5+455142 (Group 427348) - An interesting ring with possible plume.} \label{Cur_ring_and_jet}
\end{figure*}

\begin{figure*}
  \centering
  \epsscale{.5}
  \plotone{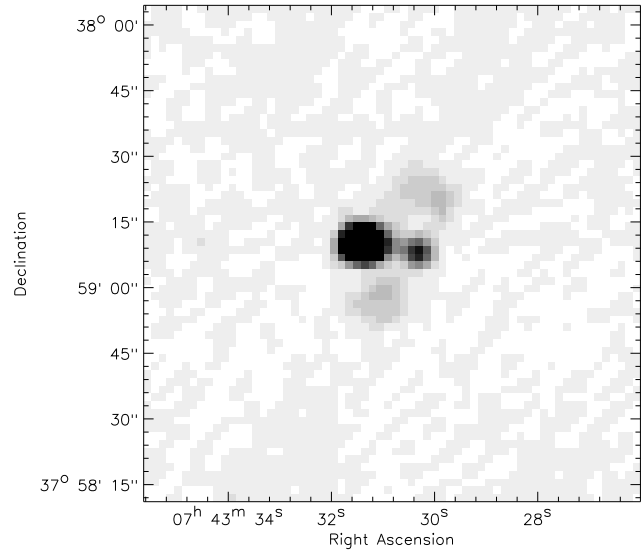}
  \caption{\label{fig:Cur_bug}FCG J074331.0+375908 (Group 048758) - An interesting bug-like morphology source.} \label{Cur_bug}
\end{figure*}

\begin{figure*}
  \epsscale{.5}
  \centering
  \plotone{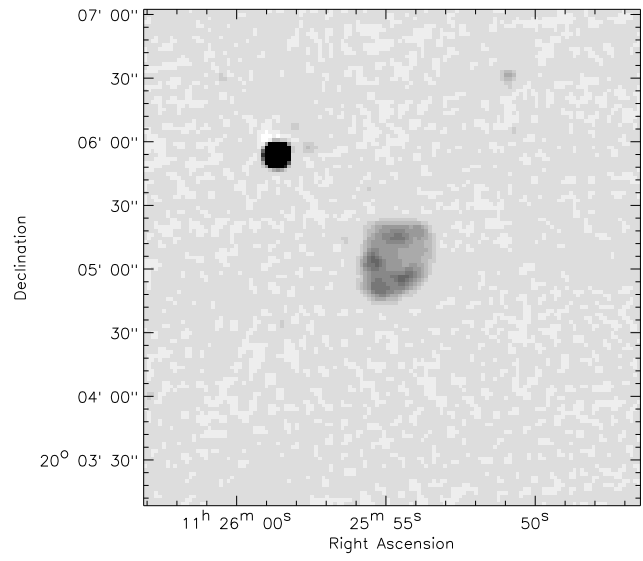}
  \caption{\label{fig:Cur_nova-like}FCG J112555.0+200501 (Group 282486) - An interesting 'nova-like' morphology source.} \label{Cur_nova-like}
\end{figure*}

\clearpage
\begin{figure*}
  \epsscale{.5}
  \centering
  \plotone{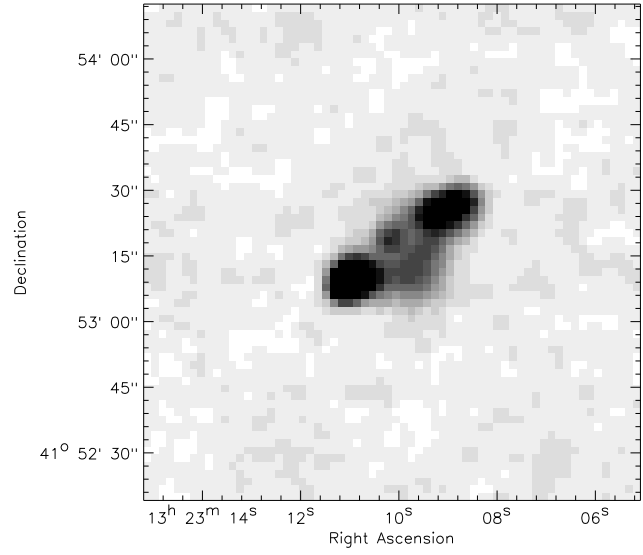}
  \caption{\label{fig:Cur_core_gap}FCG J132309.9+415318 (Group 411941) - An interesting gap around core.} \label{Cur_core_gap}
\end{figure*}

\begin{figure*}
  \epsscale{.5}
  \centering
  \plotone{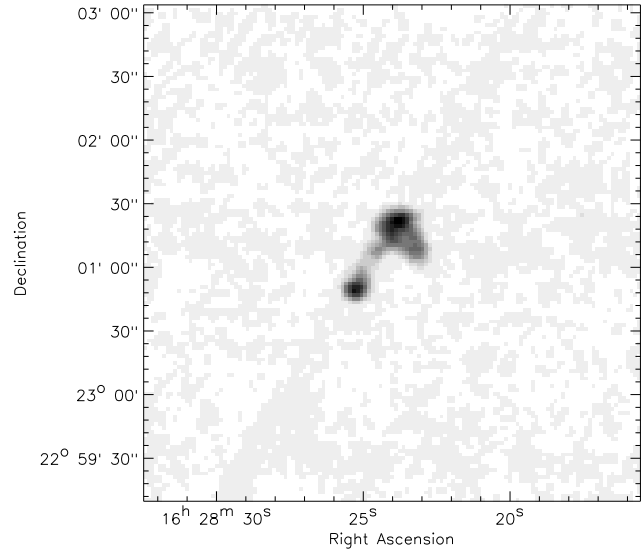}
  \caption{\label{fig:Cur_ppd_ring}FCG J162824.8+230105 (Group 611411) - An interesting ring perpendicular to system axis.} \label{Cur_ppd_ring}
\end{figure*}

\clearpage
\begin{figure*}
 \epsscale{1.}
  \centering
  \plotone{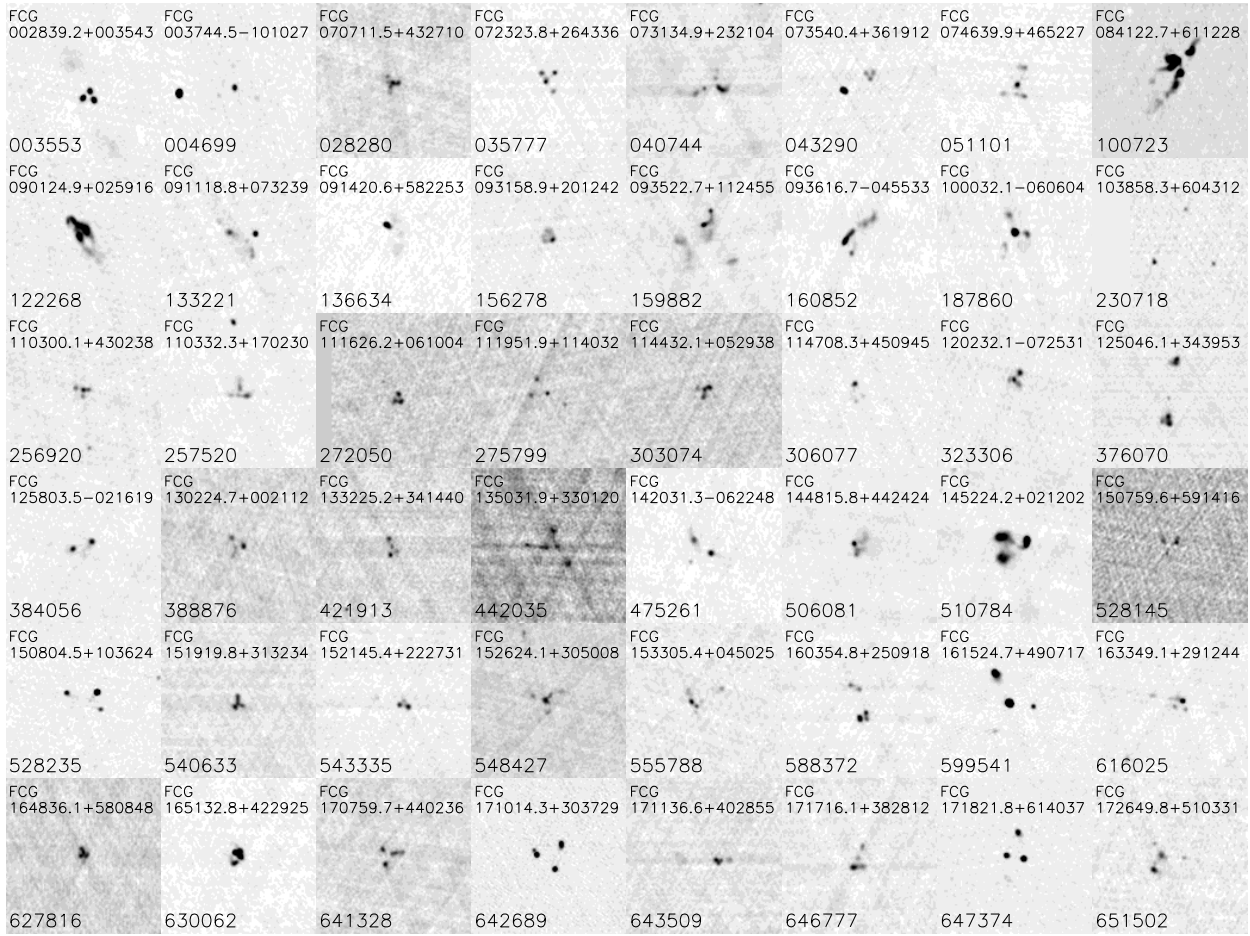}
  \caption{\label{fig:typical_tri}Apparent tri-axial systems, with IAU compliant designation and group number.} \label{typical_tri}
\end{figure*}

\begin{figure*}
  \centering
  \plotone{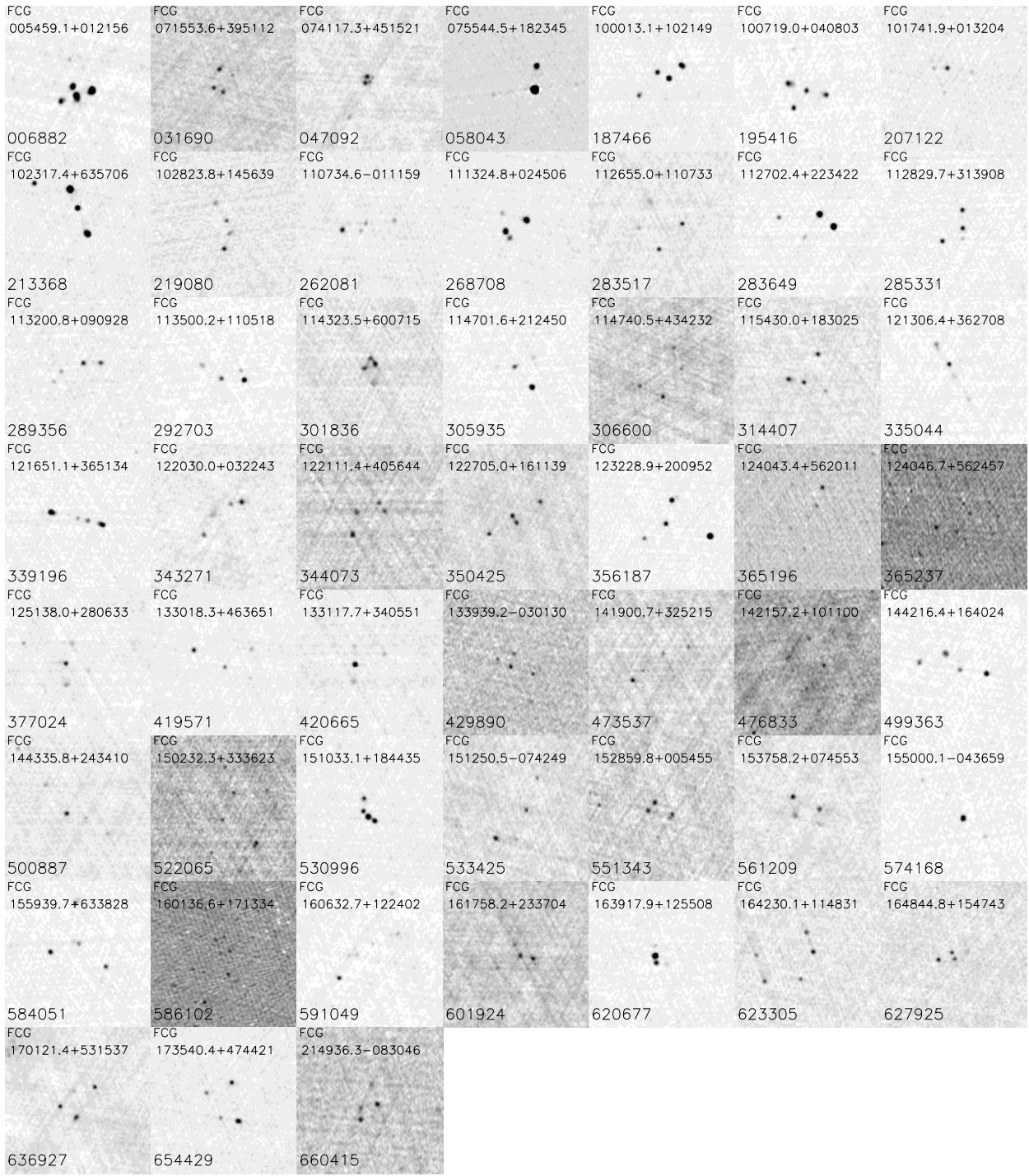}
  \caption{\label{fig:typical_quad}Visually identified quad or quint-morphology groups, with IAU compliant designation and group number.} \label{typical_quad}
\end{figure*}

\begin{figure*}
  \epsscale{1.0}
  \newcounter{myfigcounter5}   
  \setcounter{myfigcounter5}{\value{figure}}  
  \centering
  \plotone{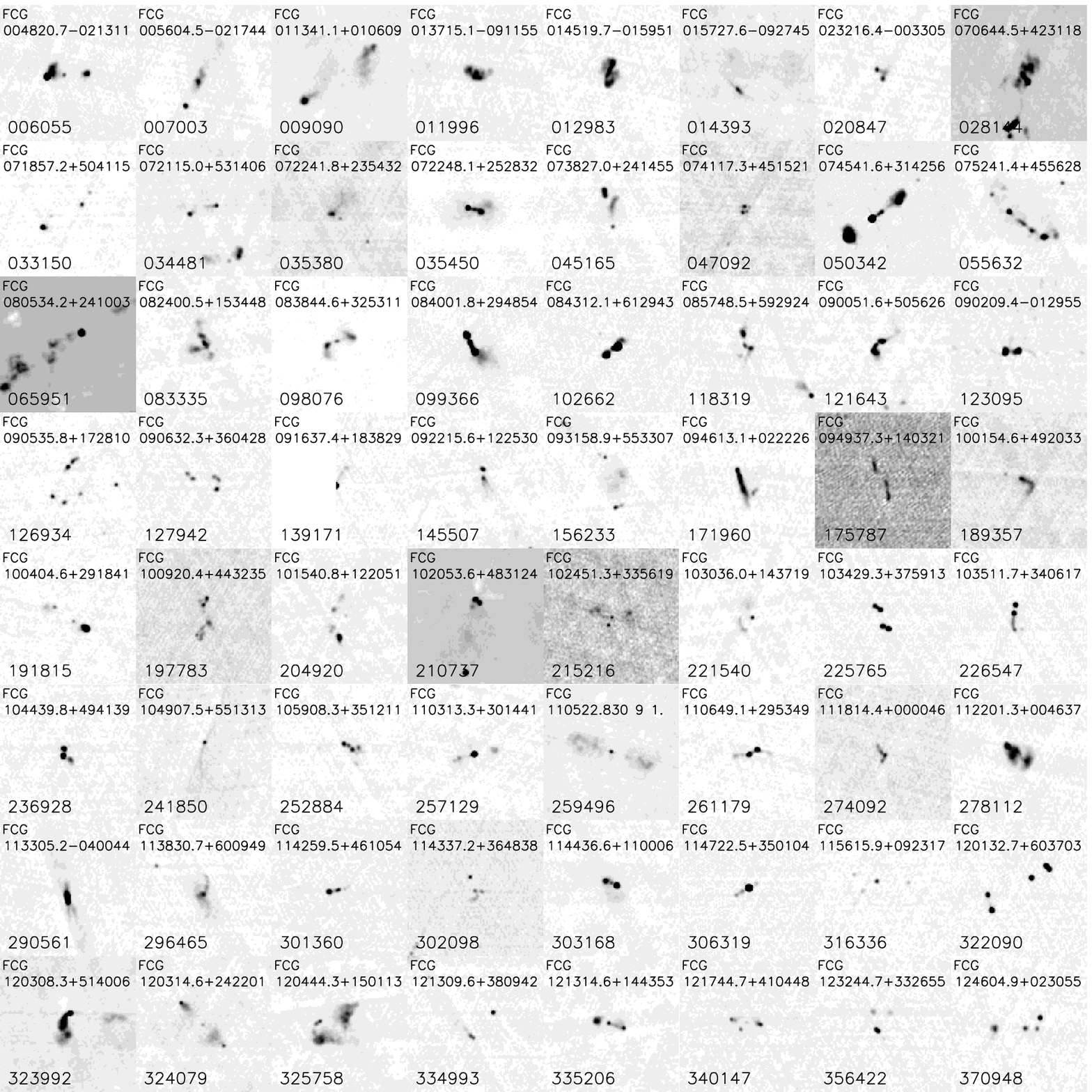}
  \caption{\label{fig:typical_int} Mosaic of other interesting groups with IAU compliant designation and group number.} \label{typical_int}
\end{figure*}

\begin{figure*}
  \centering
  \setcounter{figure}{\value{myfigcounter5}}
  \plotone{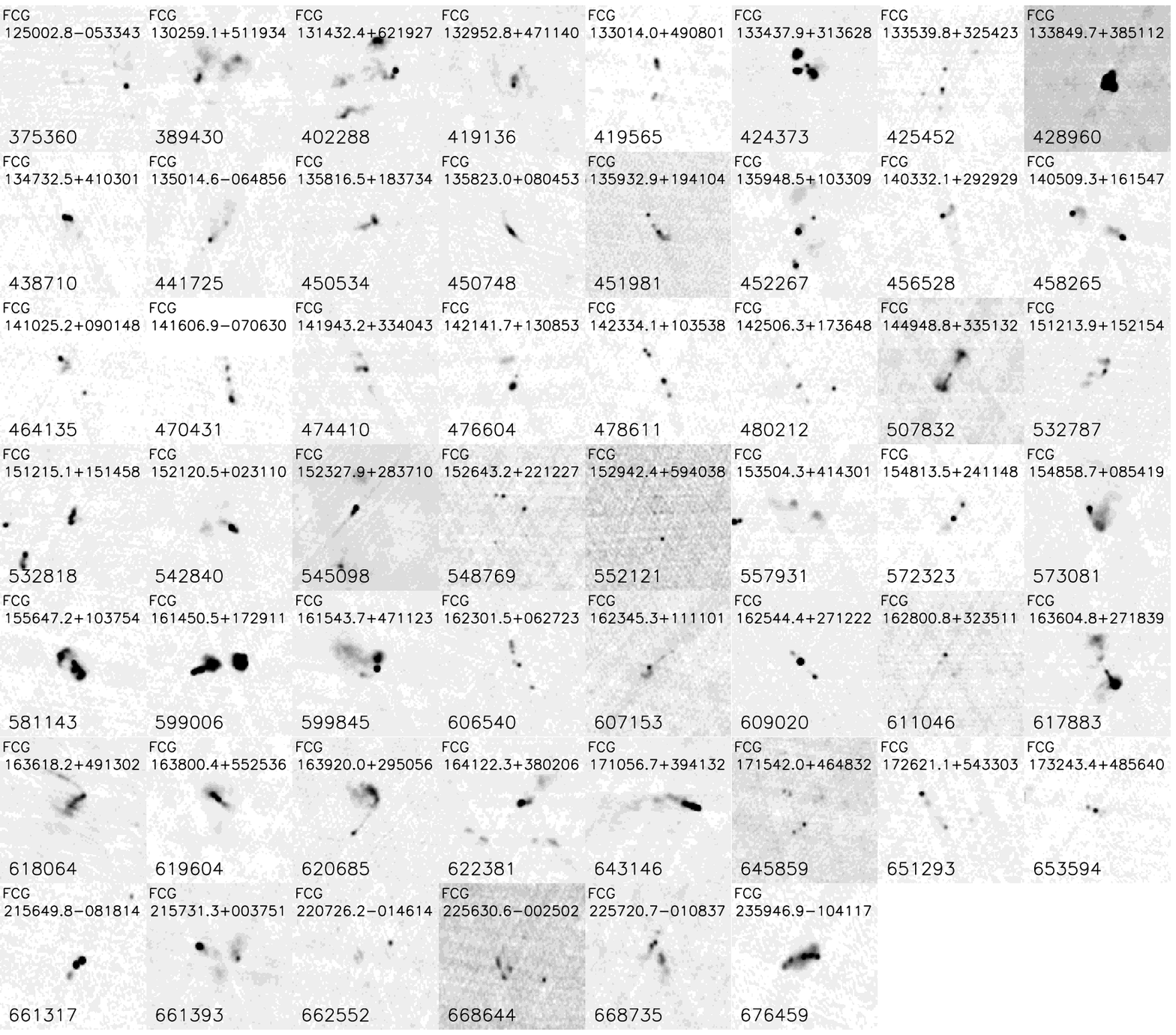}
  \caption{\label{fig:typical_int2} Continued.  Mosaic of other interesting groups with IAU compliant designation and group number.} \label{typical_int2}
\end{figure*}


\clearpage

\begin{figure*}
 \epsscale{1.}
 \centering
  \plotone{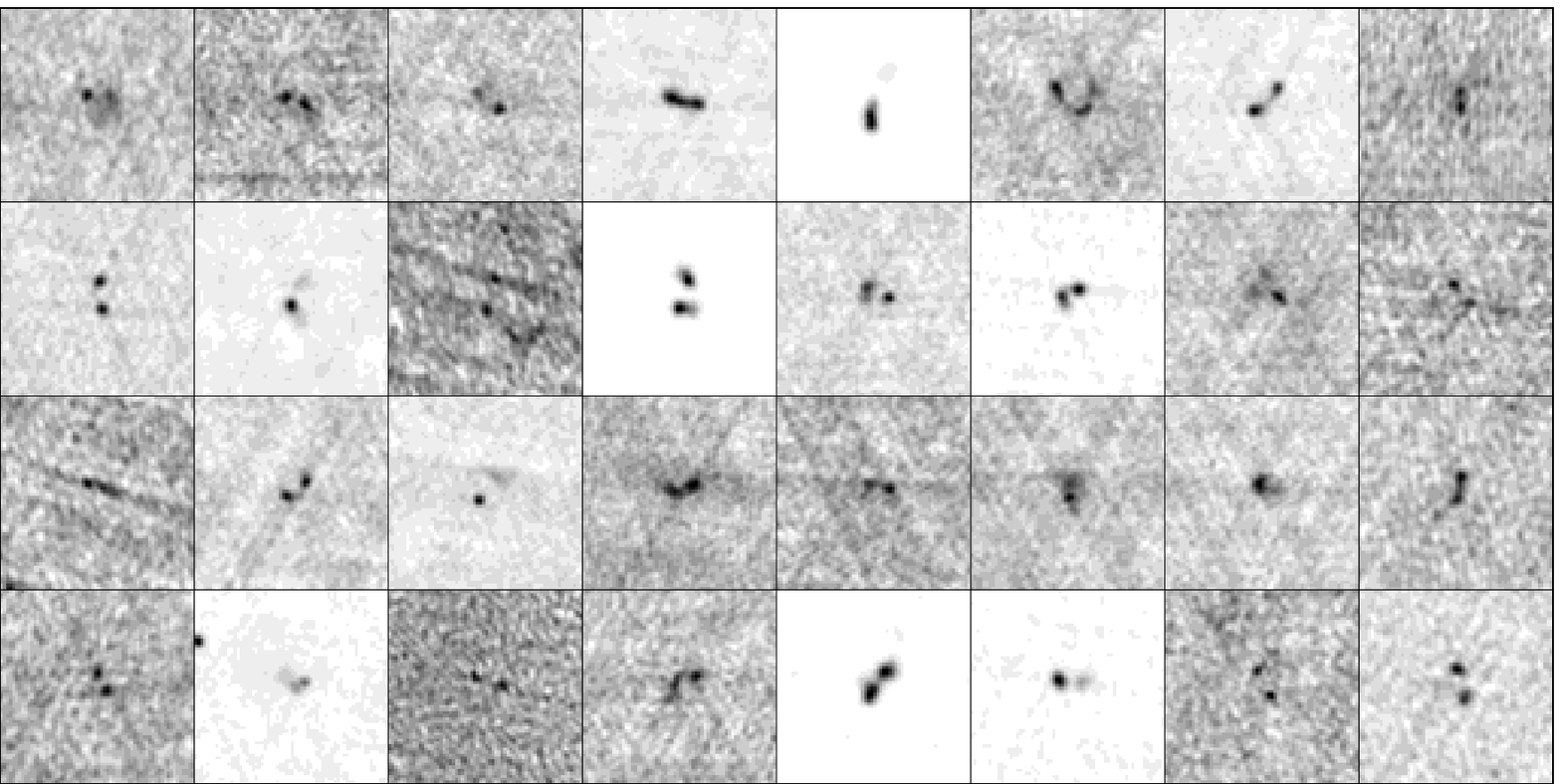}
  \caption{\label{fig:epsarts}A random selection of 32 highest ranked (for being bent) two-component groups.} \label{F_maxvex}
\end{figure*}

\begin{figure*}
  \centering
  \plotone{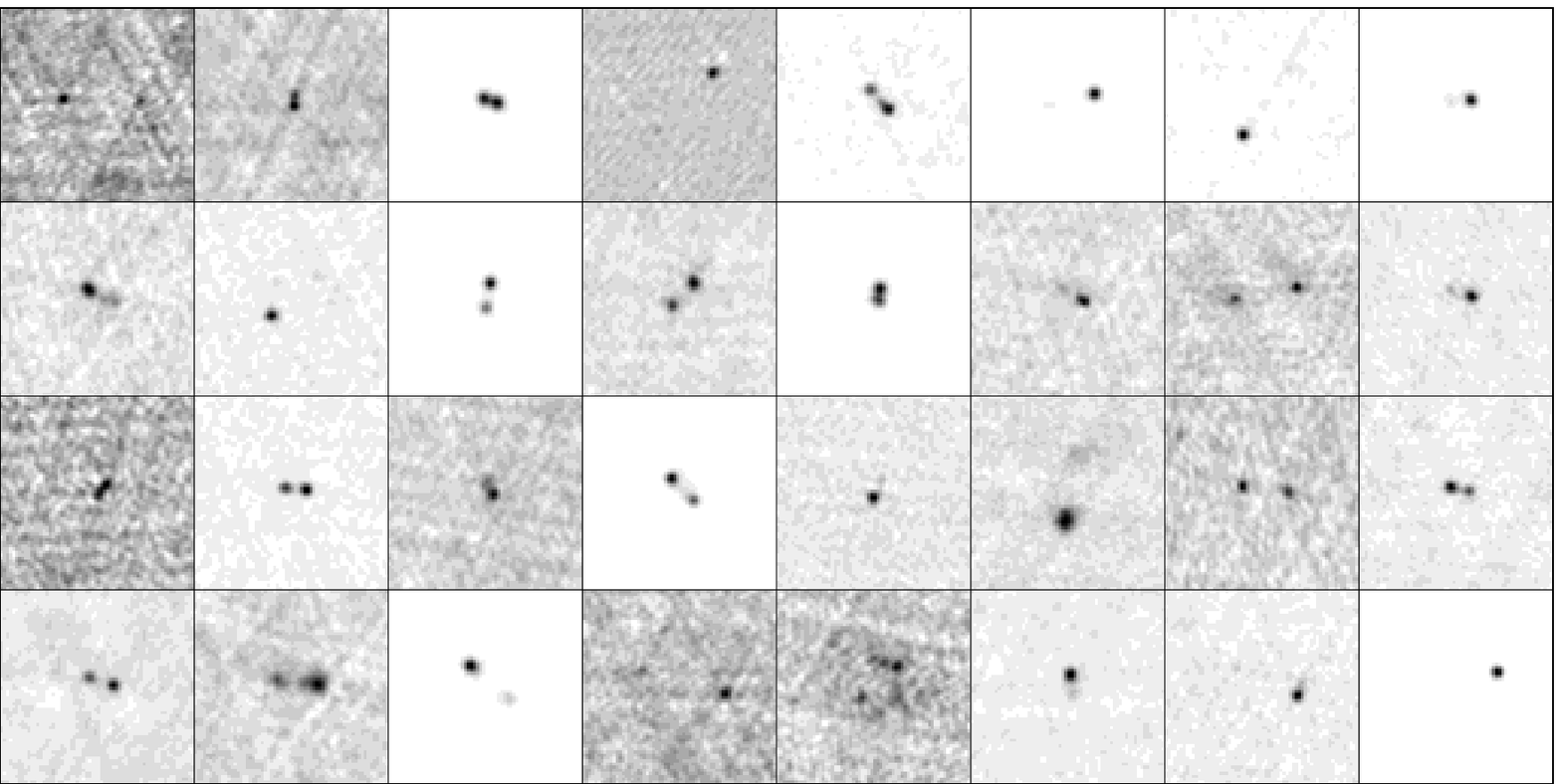}
  \caption{\label{fig:epsartt}Random selection of 32 of lowest ranked (for being bent) two-component groups.} \label{F_minvex}
\end{figure*}

\begin{figure*}
 \centering
 \includegraphics{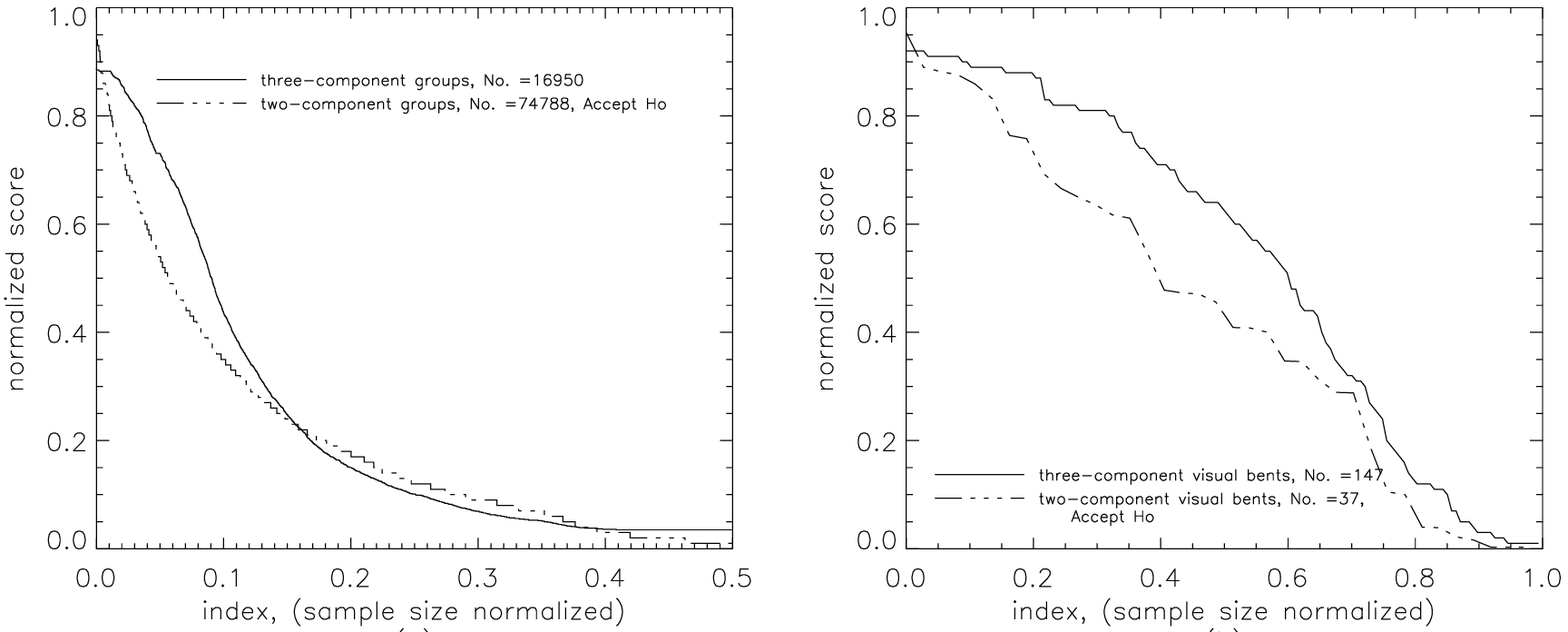}  \caption{\label{fig:epsarte} Vote curve comparisons between three-component group and two-component group pattern recognition, (a) entire populations, (b) visual bents.} \label{F_2-3vcc}
\end{figure*}


\clearpage

\notetoeditor{I couldn't figure out how to get fixed width fonts for the tables so that the designations and coordinates in the tables would all come out the same width instead of ragged allignment.}

\begin{deluxetable}{lccccllll}
\tabletypesize{\scriptsize}
\rotate
\tablecaption{WAT and NAT Sources \label{watnattable}}
\tablewidth{0pt}
\tablehead{
\colhead{\parbox{4em}{Source\\Designation}} & \colhead{\parbox{4em}{system\\R.A.(2000)}} & \colhead{\parbox{4em}{system\\Dec.(2000)}} & \colhead{\parbox{4em}{core\\R.A.(2000)}}  & \colhead{\parbox{4em}{core\\Dec.(2000)}}  & \colhead{\parbox{2em}{core\\ type\tablenotemark{a}}}  & \colhead{size (')}  & \colhead{Group} & \colhead{comments} 
}

\startdata
FCG J000330.7+002756&   00 03 31.218&  +00 28 04.25&   00 03 30.731&  +00 27 56.09& c &  1.12&  000455&  wat\\
FCG J002042.0-112103&   00 20 41.243&  -11 21 07.36&   00 20 42.045&  -11 21 03.42& v &  0.95&  002611&  nat dd?, ring nearby\\
FCG J004012.8+012542&   00 40 13.004&  +01 26 03.05&   00 40 12.891&  +01 25 42.06& v &  1.99&  004994&  wat dd?\\
FCG J004150.2-092548&   00 41 51.461&  -09 25 47.99&   00 41 50.271&  -09 25 48.92& v &  0.64&  005236&  nat dd\\
FCG J004152.1+002837&   00 41 51.586&  +00 28 36.22&   00 41 52.109&  +00 28 37.37& v?&  0.80&  005238&  nat\\
FCG J005602.6-012004&   00 56 01.880&  -01 20 25.26&   00 56 02.666&  -01 20 04.82& v &  0.92&  007000&  nat prototypical 0053-015 in A119\\
FCG J010236.5-005007&   01 02 36.812&  -00 50 00.99&   01 02 36.579&  -00 50 07.69& c &  0.94&  007807&  wat int starwars, first system of group\\
FCG J010242.4-005032&   01 02 41.686&  -00 50 27.71&   01 02 42.429&  -00 50 32.98& c &  1.16&  007807&  wat int starwars, second system of group\\
FCG J010403.0-002440&   01 04 04.046&  -00 24 32.84&   01 04 03.040&  -00 24 40.94& v &  0.76&  007991&  nat\\
FCG J011425.5+002932&   01 14 25.792&  +00 29 36.77&   01 14 25.594&  +00 29 32.59& c &  1.05&  009186&  wat\\
\enddata
\tablecomments{This table is available in its entirety in ASCII FORMAT in the electronic edition of this journal.  A portion is shown here for guidance regarding its form and content.}
\tablenotetext{a}{Core type c indicates (presumed) core coordinate is FIRST catalog entry, v indicates visual estimate from symmetry considerations.}
\end{deluxetable}

\begin{deluxetable}{cccccllll}
\tabletypesize{\scriptsize}
\rotate
\tablecaption{W-Shaped Sources \label{wtable}}
\tablewidth{0pt}
\tablehead{
\colhead{\parbox{4em}{Source\\Designation}} & \colhead{\parbox{4em}{system\\R.A.(2000)}} & \colhead{\parbox{4em}{system\\Dec.(2000)}} & \colhead{\parbox{4em}{core\\R.A.(2000)}}  & \colhead{\parbox{4em}{core\\Dec.(2000)}}   & \colhead{\parbox{2em}{core\\ type\tablenotemark{a}}} & \colhead{size (')}  & \colhead{Group} & \colhead{comments}
}
\startdata
FCG J020819.4-070823&   02 08 19.451&  -07 08 20.44&   02 08 19.461&  -07 08 23.32& v &  0.70&  016599&  w sz?\\
FCG J071006.2+500309&   07 10 05.001&  +50 03 08.27&   07 10 06.281&  +50 03 09.32& c?&  1.12&  029313&  w? sz? dd?\\
FCG J071639.6+413405&   07 16 39.632&  +41 34 05.73&   07 16 39.632&  +41 34 05.73& v &  0.73&  032073&  w? sz??\\
FCG J071735.2+374508&   07 17 34.873&  +37 45 19.85&   07 17 35.290&  +37 45 08.41& c?&  1.53&  032482&  w? irr distorted int\\
FCG J073423.4+350648&   07 34 23.462&  +35 06 48.19&               &              &   &  1.59&  042511&  w? irr x? sz? fan distorted int\\
\enddata
\tablecomments{This table is available in its entirety in ASCII FORMAT in the electronic edition of this journal.  A portion is shown here for guidance regarding its form and content.}
\tablenotetext{a}{Core type c indicates (presumed) core coordinate is FIRST catalog entry, v indicates visual estimate from symmetry considerations.}
\end{deluxetable}

\begin{deluxetable}{ccclll}
\tabletypesize{\scriptsize}
\tablecaption{'TB'-Type Sources \label{tbtable}}
\tablewidth{0pt}
\tablehead{
\colhead{\parbox{4em}{Source\\Designation}} & \colhead{R.A.(2000)} & \colhead{Dec.(2000)} & \colhead{size (')}  & \colhead{Group} & \colhead{comments}
}
\startdata
FCG J002839.2+003543&   00 28 39.260&  +00 35 43.88&     0.28&   003553&  tb int tri triple core with possible lobes to 1.54'\\
FCG J075954.8+262337&   07 59 54.838&  +26 23 37.25&     0.65&   061333&  mg tb d+s? unu\\
FCG J075955.1+445017&   07 59 55.123&  +44 50 17.97&     0.33&   061335&  mg tb d+s? unu\\
FCG J083201.9+395547&   08 32 01.906&  +39 55 47.54&     0.31&   091104&  tb\\
FCG J092239.9-072428&   09 22 39.940&  -07 24 28.71&     0.52&   145955&  wat? tb\\
\enddata
\tablecomments{This table is available in its entirety in ASCII FORMAT in the electronic edition of this journal.  A portion is shown here for guidance regarding its form and content.}
\end{deluxetable}

\begin{deluxetable}{cccllll}
\tabletypesize{\scriptsize}
\tablecaption{'B'-Type Sources \label{btable}}
\tablewidth{0pt}
\tablehead{
\colhead{\parbox{4em}{Source\\Designation}} & \colhead{R.A.(2000)} & \colhead{Dec.(2000)} & \colhead{size (')}  & \colhead{Group} & \colhead{\parbox{8em}{additional\\comments}}
}
\startdata
FCG J000115.3-082644&   00 01 15.303&  -08 26 44.34&     0.40&   000160&  cat missed core?\\
FCG J073232.3+374451&   07 32 32.399&  +37 44 51.32&     0.43&   041299&  with wat interacting?\\
FCG J081848.7+363143&   08 18 48.759&  +36 31 43.91&     0.62&   078212&  core id ambiguous\\
FCG J084851.2+090125&   08 48 51.240&  +09 01 25.86&     0.30&   108783&  sz?\\
FCG J093050.8+202000&   09 30 50.866&  +20 20 00.74&     0.60&   154918&  mg nat/wat?\\
\enddata
\tablecomments{This table is available in its entirety in ASCII FORMAT in the electronic edition of this journal.  A portion is shown here for guidance regarding its form and content.}
\end{deluxetable}

\begin{deluxetable}{cccllll}
\tabletypesize{\scriptsize}
\tablecaption{Ring Type Sources \label{ringtable}}
\tablewidth{0pt}
\tablehead{
\colhead{\parbox{4em}{Source\\Designation}} & \colhead{R.A.(2000)} & \colhead{Dec.(2000)} & \colhead{size (')}  & \colhead{Group} & \colhead{comments}
}
\startdata
FCG J002039.9-111944&   00 20 39.955&  -11 19 44.06&     0.25&   002613&  ring, a two-component group ring\\
FCG J070805.8+411351&   07 08 05.884&  +41 13 51.40&     0.63&   028598&  ring\\
FCG J071944.2+251219&   07 19 44.207&  +25 12 19.44&     0.68&   033601&  ring system center\\
FCG J071944.7+251213&   07 19 44.729&  +25 12 13.17&     0.48&   033601&  ring center\\
FCG J073929.8+394711&   07 39 29.843&  +39 47 11.58&     0.77&   045877&  ring nat?\\
\enddata
\tablecomments{This table is available in its entirety in ASCII FORMAT in the electronic edition of this journal.  A portion is shown here for guidance regarding its form and content.}
\end{deluxetable}

\begin{deluxetable}{cccccllll}
\tabletypesize{\scriptsize}
\rotate
\tablecaption{Ring-lobe Type Sources \label{ringlobetable}}
\tablewidth{0pt}
\tablehead{
\colhead{\parbox{4em}{Source\\Designation}} & \colhead{\parbox{4em}{system\\R.A.(2000)}} & \colhead{\parbox{4em}{system\\Dec.(2000)}} & \colhead{\parbox{4em}{core\\R.A.(2000)}}  & \colhead{\parbox{4em}{core\\Dec.(2000)}}  & \colhead{\parbox{2em}{core\\ type\tablenotemark{a}}} & \colhead{size (')}  & \colhead{Group} & \colhead{comments}
}
\startdata
FCG J022933.8-002703&   02 29 33.846&  -00 27 05.21&   02 29 33.822&  -00 27 03.96& c &  1.54&  020510&  ring-lobe  sz\\
FCG J030041.7-075334&   03 00 41.731&  -07 53 34.40&               &              &   &  1.68&  024123&  ring-lobe\\
FCG J065720.8+570844&   06 57 20.771&  +57 08 44.04&   06 57 20.832&  +57 08 44.42& c &  0.85&  026042&  ring-lobe? t sz?? bifurcation hymor\\
FCG J082903.8+175417&   08 29 03.819&  +17 54 17.41&               &              &   &  0.60&  088210&  ring-lobe? ring nat? int\\
FCG J083752.9+445025&   08 37 52.983&  +44 50 25.79&   08 37 52.983&  +44 50 25.79& v?&  2.44&  097165&  ring-lobe? x dd?\\
\enddata
\tablecomments{This table is available in its entirety in ASCII FORMAT in the electronic edition of this journal.  A portion is shown here for guidance regarding its form and content.}
\tablenotetext{a}{Core type c indicates (presumed) core coordinate is FIRST catalog entry, v indicates visual estimate from symmetry considerations.}
\end{deluxetable}

\begin{deluxetable}{cccllll}
\tabletypesize{\scriptsize}
\tablecaption{HYMOR Candidates - An Incidental List \label{hymortable}}
\tablewidth{0pt}
\tablehead{
\colhead{\parbox{4em}{Source\\Designation}} & \colhead{\parbox{4em}{system\\R.A.(2000)}} & \colhead{\parbox{4em}{system\\Dec.(2000)}} & \colhead{\parbox{2em}{core\\ type\tablenotemark{a}}}  & \colhead{size (')}  & \colhead{Group} & \colhead{comments}
}

\startdata
FCG J002215.0-084441&   00 22 15.010&  -08 44 41.32& v &  1.53&   002812&  hymor? mg lobe of large sz?\\
FCG J065720.7+570844&   06 57 20.771&  +57 08 44.04& c &  1.32&   026042&  hymor  t sz?? bifurcation ring-lobe?\\
FCG J074617.7+452637&   07 46 17.774&  +45 26 37.41& v &  1.91&   050782&  hymor? dd\\
FCG J092802.6-060755&   09 28 02.636&  -06 07 55.52& v &  0.85&   151794&  hymor  rc dd? bifurcation\\
FCG J093834.5-042023&   09 38 34.550&  -04 20 23.28& v? &  3.10&   163256&  hymor? sz? wat? dd? bifurcation int\\
\enddata
\tablecomments{This table is available in its entirety in ASCII FORMAT in the electronic edition of this journal.  A portion is shown here for guidance regarding its form and content.}
\tablenotetext{a}{Core type c indicates (presumed) core coordinate is FIRST catalog entry, v indicates visual estimate from symmetry considerations.}
\end{deluxetable}

\begin{deluxetable}{ccclll}
\tabletypesize{\scriptsize}
\tablecaption{X-Shaped Sources \label{xtable}}
\tablewidth{0pt}
\tablehead{
\colhead{\parbox{4em}{Source\\Designation}} & \colhead{R.A.(2000)} & \colhead{Dec.(2000)} & \colhead{size (')}  & \colhead{Group} & \colhead{\parbox{8em}{additional\\comments}}
}
\startdata
FCG J002828.2-002634&   00 28 28.288&  -00 26 34.65&     2.74&   003514&  mg\\
FCG J004938.8+005947&   00 49 38.814&  +00 59 47.20&     0.57&   006225&  rc sz dl? int\\
FCG J011527.5-000000&   01 15 27.574&  -00 00 00.98&     1.38&   009304&  int\\
FCG J015043.7-081850&   01 50 43.762&  -08 18 50.06&     0.42&   013599&  butterfly\\
FCG J015757.1-020625&   01 57 57.184&  -02 06 25.20&     1.26&   014462&  t atypical\\
\enddata
\tablecomments{This table is available in its entirety in ASCII FORMAT in the electronic edition of this journal.  A portion is shown here for guidance regarding its form and content.}
\end{deluxetable}

\begin{deluxetable}{ccclll}
\tabletypesize{\scriptsize}
\tablecaption{Double-Double Sources \label{ddtable}}
\tablewidth{0pt}
\tablehead{
\colhead{\parbox{4em}{Source\\Designation}} & \colhead{R.A.(2000)} & \colhead{Dec.(2000)} & \colhead{size (')}  & \colhead{Group} & \colhead{\parbox{8em}{additional\\comments}}
}
\startdata
FCG J001208.3-105043&   00 12 08.373&  -10 50 43.35&     1.77&   001497&  \\
FCG J001357.0-091953&   00 13 57.093&  -09 19 53.08&     1.07&   001760&  \\
FCG J003247.6-001938&   00 32 47.626&  -00 19 38.74&     1.16&   004090&  bs\\
FCG J004030.9-100132&   00 40 30.995&  -10 01 32.34&     1.92&   005041&  bs\\
FCG J004151.4-092547&   00 41 51.461&  -09 25 47.99&     0.64&   005236&  nat int\\
\enddata
\tablecomments{This table is available in its entirety in ASCII FORMAT in the electronic edition of this journal.  A portion is shown here for guidance regarding its form and content.}
\end{deluxetable}

\begin{deluxetable}{cccccllll}
\tabletypesize{\scriptsize}
\tablecaption{Core-jet Sources \label{cjtable}}
\tablewidth{0pt}
\tablehead{
\colhead{\parbox{4em}{Source\\Designation}} & \colhead{\parbox{4em}{system\\R.A.(2000)}} & \colhead{\parbox{4em}{system\\Dec.(2000)}} & \colhead{\parbox{4em}{core\\R.A.(2000)}}  & \colhead{\parbox{4em}{core\\Dec.(2000)}} & \colhead{size (')}  & \colhead{Group} & \colhead{\parbox{8em}{additional\\comments}}
}
\startdata
FCG J014001.7-101206&   01 39 59.890&  -10 12 13.15&   01 40 01.763&  -10 12 06.82&     1.09&  012331&  with resolved jet?\\
FCG J020025.6-075753&   02 00 26.361&  -07 57 54.19&   02 00 25.616&  -07 57 53.55&     0.57&  014831&    \\
FCG J021612.2-010518&   02 16 12.209&  -01 05 18.75&   02 16 12.209&  -01 05 18.75&     0.51&  018460&  both sides\\
FCG J021839.6-002151&   02 18 39.991&  -00 21 45.08&   02 18 39.630&  -00 21 51.49&     0.40&  019015&     \\
FCG J070725.8+450314&   07 07 24.389&  +45 03 11.23&   07 07 25.868&  +45 03 14.69&     0.70&  028354&  mg\\
\enddata
\tablecomments{This table is available in its entirety in ASCII FORMAT in the electronic edition of this journal.  A portion is shown here for guidance regarding its form and content.}
\end{deluxetable}

\begin{deluxetable}{cccccllll}
\tabletypesize{\scriptsize}
\rotate
\tablecaption{S or Z-Shaped Sources \label{sptable}}
\tablewidth{0pt}
\tablehead{
\colhead{\parbox{4em}{Source\\Designation}} & \colhead{\parbox{4em}{system\\R.A.(2000)}} & \colhead{\parbox{4em}{system\\Dec.(2000)}} & \colhead{\parbox{4em}{core\\R.A.(2000)}}  & \colhead{\parbox{4em}{core\\Dec.(2000)}}  & \colhead{\parbox{2em}{core\\ type\tablenotemark{a}}} & \colhead{size (')}  & \colhead{Group} & \colhead{\parbox{8em}{additional\\comments}}
}
\startdata
FCG J002216.5+010403&   00 22 16.564&  +01 04 03.27&               &              &   &  0.62&  002814&  dd?\\
FCG J002644.9-090620&   00 26 45.390&  -09 06 27.05&   00 26 44.954&  -09 06 20.22& v &  0.84&  003324&  rc b? pinwheel  SDSS galaxy\\
FCG J003432.6-093444&   00 34 32.769&  -09 34 46.13&   00 34 32.619&  -09 34 44.52& c &  0.44&  004286&  dl\\
FCG J014519.7-015951&   01 45 19.751&  -01 59 51.82&               &              &   &  0.78&  012983&  sz x? hook int pretzel\\
FCG J020613.2-021504&   02 06 13.589&  -02 15 02.89&   02 06 13.276&  -02 15 04.61& c &  0.43&  016139&    \\
\enddata
\tablecomments{This table is available in its entirety in ASCII FORMAT in the electronic edition of this journal.  A portion is shown here for guidance regarding its form and content.}
\tablenotetext{a}{Core type c indicates (presumed) core coordinate is FIRST catalog entry, v indicates visual estimate from symmetry considerations.}
\end{deluxetable}

\begin{deluxetable}{cccllll}
\tabletypesize{\scriptsize}
\rotate
\tablecaption{Giant Radio Sources - An Incidental List\label{grgtable}}
\tablewidth{0pt}
\tablehead{
\colhead{\parbox{4em}{Source\\Designation}} & \colhead{R.A.(2000)} & \colhead{Dec.(2000)} &  \colhead{\parbox{2em}{core\\ type\tablenotemark{a}}}  & \colhead{size (')}  & \colhead{Group} & \colhead{comments}
}
\startdata
FCG J065903.6+490221&   06 59 03.616&  +49 02 21.17& c&   4.0&    026290 026276              &  GRS  core id ambiguous dl\\
FCG J071131.0+471025&   07 11 31.002&  +47 10 25.95& v&   4.8&    029792 029901              &  GRS?\\
FCG J084759.0+314708&   08 47 59.050&  +31 47 08.73& c&   4.4&    107787 107867 107830 108041&  GRS  ring? dd? also nearby GRS?\\
FCG J090049.9-022549&   09 00 49.950&  -02 25 49.06& v&   4.0&    121591 121654 121517       &  GRS \\
FCG J093238.3+161157&   09 32 38.315&  +16 11 57.78& c&   4.0&    156829                     &  GRS  x dd??\\
\enddata
\tablecomments{This table is available in its entirety in ASCII FORMAT in the electronic edition of this journal.  A portion is shown here for guidance regarding its form and content.}
\tablenotetext{a}{Core type c indicates (presumed) core coordinate is FIRST catalog entry, v indicates visual estimate from symmetry considerations.}
\end{deluxetable}

\clearpage

\begin{deluxetable}{cccccllll}
\tabletypesize{\scriptsize}
\rotate
\tablecaption{Tri-Axial Type Morphology Sources \label{tritable}}
\tablewidth{0pt}
\tablehead{
\colhead{\parbox{4em}{Source\\Designation}} & \colhead{\parbox{4em}{system\\R.A.(2000)}} & \colhead{\parbox{4em}{system\\Dec.(2000)}} & \colhead{\parbox{4em}{core\\R.A.(2000)}}  & \colhead{\parbox{4em}{core\\Dec.(2000)}}  & \colhead{\parbox{2em}{core\\ type\tablenotemark{a}}}  & \colhead{size (')}  & \colhead{Group} & \colhead{\makebox[8em][l]{comments}}
}
\startdata
FCG J002839.2+003543&   00 28 39.260&  +00 35 43.88&               &              &   &  2.58&  003553&  tri  dd? triple core with lobe(s) or sl? int\\
FCG J003744.5-101027&   00 37 44.528&  -10 10 27.58&               &              &   &  1.34&  004699&  tri? mg int\\
FCG J070711.5+432710&   07 07 11.597&  +43 27 10.48&               &              &   &  0.51&  028280&  tri  rc? sz? x? unu pinwheel?\\
FCG J072323.8+264336&   07 23 23.896&  +26 43 36.48&               &              &   &  0.57&  035777&  tri? quad unu\\
FCG J073134.9+232104&   07 31 34.935&  +23 21 04.69&               &              &   &  1.15&  040744&  tri  wat? irr\\
FCG J073540.4+361912&   07 35 40.689&  +36 19 17.76&   07 35 40.471&  +36 19 12.76& c &  0.30&  043290&  tri? nat unu\\
\enddata
\tablecomments{This table is available in its entirety in ASCII FORMAT in the electronic edition of this journal.  A portion is shown here for guidance regarding its form and content.}
\tablenotetext{a}{Core type c indicates (presumed) core coordinate is FIRST catalog entry, v indicates visual estimate from symmetry considerations.}
\end{deluxetable}

\begin{deluxetable}{ccclll}
\tabletypesize{\scriptsize}
\tablecaption{Quad-Type Morphology Sources \label{quadtable}}
\tablewidth{0pt}
\tablehead{
\colhead{\parbox{4em}{Source\\Designation}} & \colhead{R.A.(2000)} & \colhead{Dec.(2000)} & \colhead{size (')}  & \colhead{Group} & \colhead{comments}
}
\startdata
FCG J005459.1+012156&   00 54 59.185&  +01 21 56.60&     1.04&   006882&  mg unu quad complex\\
FCG J071553.6+395112&   07 15 53.615&  +39 51 12.27&     0.62&   031690&  mg quad unu\\
FCG J074117.3+451521&   07 41 17.309&  +45 15 21.94&     0.27&   047092&  nat? int quad\\
FCG J075544.5+182345&   07 55 44.564&  +18 23 45.86&     0.00&   058043&  mg quad? d+sl? d? unu\\
FCG J100013.1+102149&   10 00 13.186&  +10 21 49.99&     1.53&   187466&  mg quad d+d sz? unu\\
\enddata
\tablecomments{This table is available in its entirety in ASCII FORMAT in the electronic edition of this journal.  A portion is shown here for guidance regarding its form and content.}
\end{deluxetable}

\begin{deluxetable}{cccllll}
\tabletypesize{\scriptsize}
\tablecaption{Other Interesting Sources \label{inttable}}
\tablewidth{0pt}
\tablehead{
\colhead{\parbox{4em}{Source\\Designation}} & \colhead{\parbox{4em}{system\\R.A.(2000)}} & \colhead{\parbox{4em}{system\\Dec.(2000)}} & \colhead{\parbox{2em}{type\tablenotemark{a}}}   & \colhead{size (')}  & \colhead{Group} & \colhead{additional comments}
}
\startdata
FCG J004820.7-021311&   00 48 20.747&  -02 13 11.21& c&   1.30&   006055&  t lobe with fan dl? x?\\
FCG J005604.5-021744&   00 56 04.599&  -02 17 44.25&  &   2.31&   007003&  mg sz? rc\\
FCG J011341.1+010609&   01 13 41.119&  +01 06 09.19& c&   2.74&   009090&  x? sz? dl?\\
FCG J013715.1-091155&   01 37 15.175&  -09 11 55.76&  &   0.68&   011996&  rc? lobe? arc? beads\\
FCG J014519.7-015951&   01 45 19.751&  -01 59 51.82&  &   0.78&   012983&  sz x? hook pretzel\\
\enddata
\tablecomments{This table is available in its entirety in ASCII FORMAT in the electronic edition of this journal.  A portion is shown here for guidance regarding its form and content.}
\tablenotetext{a}{Type c indicates (presumed) core coordinate is FIRST catalog entry, v indicates visual estimate from symmetry considerations.}
\end{deluxetable}

\begin{deluxetable}{cccccccl}
\tabletypesize{\scriptsize}
\rotate
\tablecaption{Selected Sources from Three Component Groups Training Set \label{threecomptrsetint}}
\tablewidth{0pt}
\tablehead{
\colhead{\parbox{4em}{Source\\Designation}} & \colhead{\parbox{4em}{system\\R.A.(2000)}} & \colhead{\parbox{4em}{system\\Dec.(2000)}} & \colhead{\parbox{4em}{core\\R.A.(2000)}}  & \colhead{\parbox{4em}{core\\Dec.(2000)}}  & \colhead{size (')}  & \colhead{Group} & \colhead{\makebox[8em][l]{comments}}
}
\startdata
FCG J070247.8+500203&  07 02 47.888&  +50 02 03.16&              &              &   0.84&  027113&  x butterfly\\
FCG J070643.8+423111&  07 06 43.827&  +42 31 11.86&              &              &   2.26&  028148&  wat int\\
FCG J071510.1+491052&  07 15 10.114&  +49 10 52.42&              &              &   0.44&  031341&  x\\
FCG J071519.2+482933&  07 15 20.431&  +48 29 21.74&  07 15 19.230&  +48 29 33.34&   0.64&  031427&  nat\\
FCG J072014.7+403747&  07 20 14.784&  +40 37 47.41&              &              &   0.46&  033890&  x butterfly, may be larger\\
\enddata
\tablecomments{This table is available in its entirety in ASCII FORMAT in the electronic edition of this journal.  A portion is shown here for guidance regarding its form and content.}
\tablecomments{The components given are for the individual components of the three-component group.}
\end{deluxetable}

\begin{deluxetable}{rcccccc}
\tabletypesize{\scriptsize}
\tablecaption{Probability Estimates for a Three Component Group Being Bent \label{three_comp_vote}}
\tablewidth{0pt}
\tablehead{
\colhead{vote\tablenotemark{a}} & \colhead{R.A.(2000)} & \colhead{Dec.(2000)} &
                 \colhead{R.A.(2000)} & \colhead{Dec.(2000)} &
                 \colhead{R.A.(2000)} & \colhead{Dec.(2000)}

}
\startdata
0.89&  14 28 21.875&  +23 53 19.83&    14 28 22.806&  +23 53 13.04&     14 28 21.731&  +23 53 34.84\\
0.89&  14 56 56.031&  +50 17 48.69&    14 56 55.185&  +50 17 57.92&     14 56 57.336&  +50 17 48.05\\
0.89&  12 17 18.435&  +32 53 19.18&    12 17 17.808&  +32 53 27.65&     12 17 19.428&  +32 53 18.72\\
0.89&  17 39 00.154&  +44 24 02.72&    17 39 00.521&  +44 24 13.49&     17 39 00.721&  +44 23 51.57\\
0.89&  14 35 21.564&  +11 50 54.66&    14 35 22.332&  +11 51 01.92&     14 35 21.535&  +11 50 40.09\\
\enddata
\tablenotetext{a}{vote (estimated probability of being bent radio galaxy).}
\tablecomments{This table is available in its entirety in ASCII FORMAT in the electronic edition of this journal.  A portion is shown here for guidance regarding its form and content.}
\tablecomments{The components given are for the individual components of the three-component group.}
\end{deluxetable}

\begin{deluxetable}{cccccccl}
\tabletypesize{\scriptsize}
\rotate
\tablecaption{Selected Sources from Two Component Groups Training Set \label{twocomptrsetint}}
\tablewidth{0pt}
\tablehead{
\colhead{\parbox{4em}{Source\\Designation}} & \colhead{\parbox{4em}{system\\R.A.(2000)}} & \colhead{\parbox{4em}{system\\Dec.(2000)}} & \colhead{\parbox{4em}{core\\R.A.(2000)}}  & \colhead{\parbox{4em}{core\\Dec.(2000)}}  & \colhead{size (')}  & \colhead{Group} & \colhead{\makebox[8em][l]{comments}}
}
\startdata
FCG J001620.4-090705&  00 16 20.410&  -09 07 05.23&              &              &   0.30&  002072&  ring?\\
FCG J011643.9-094606&  01 16 43.948&  -09 46 06.93&              &              &   0.68&  009475&  b? nat?\\
FCG J014317.3-011858&  01 43 17.356&  -01 18 58.03&              &              &   0.73&  012759&  bs int low level\\
FCG J031947.1+004526&  03 19 47.121&  +00 45 26.38&              &              &   0.36&  025548&  ring?\\
FCG J072050.7+335853&  07 20 50.748&  +33 58 53.58&              &              &   0.38&  034261&  tb cat missed component\\
FCG J072459.4+335020&  07 24 59.451&  +33 50 20.32&              &              &   0.42&  036741&  b? tri?\\
FCG J073626.0+185725&  07 36 26.094&  +18 57 25.71&              &              &   0.19&  043822&  d int region - may be larger ring\\
FCG J075134.6+292122&  07 51 34.513&  +29 21 15.48&  07 51 34.641&  +29 21 22.54&   0.65&  054862&  tb? cat missed lobe\\
\enddata
\tablecomments{This table is available in its entirety in ASCII FORMAT in the electronic edition of this journal.  A portion is shown here for guidance regarding its form and content.}
\end{deluxetable}

\begin{deluxetable}{ll}
\tabletypesize{\scriptsize}
\tablecaption{List of Features for Two Component Pattern Recognition \label{two_comp_features}} \label{feature_list}
\tablehead{
\colhead{Feature} & \colhead{Description}}
\startdata
 $R_{ss}$.................... & ratio of silhouette sizes, component with smaller integrated flux to component with larger integrated flux. \\
 $d_{int\_l}$................. & distance between position of component with larger integrated flux and intersection of major axes of both. \\
 $d_{int\_s}$................ & distance between position of component with smaller integrated flux and intersection of major axes. \\
 ($d_{mid}$)/$d_{max}$ ... & ratio of distances. \\
 $fpa_{diff}$ ...........  & absolute value of differences in position angles of two sources. \\
 $d_{cent}$.................  & distance between centers components. \\
 $R_{ap}$...................  & ratio minor to major axis for component with larger integrated flux. \\
 $R_{as}$...................  & ratio minor to major axis for component with smaller integrated flux. \\
\enddata
\end{deluxetable}

\begin{deluxetable}{rcccccc}
\tabletypesize{\scriptsize}
\tablecaption{Probability Estimates for a Two Component Group Being Bent \label{two_comp_vote}}
\tablewidth{0pt}
\tablehead{
\colhead{vote\tablenotemark{a}} & \colhead{R.A.(2000)} & \colhead{Dec.(2000)} &
                 \colhead{R.A.(2000)} & \colhead{Dec.(2000)}
}
\startdata
0.94&  22 33 17.746&  -09 07 05.38&   22 33 18.384&  -09 07 06.47\\
0.94&  17 01 05.735&  +50 51 46.29&   17 01 06.837&  +50 51 55.24\\
0.94&  16 13 40.569&  +38 38 06.13&   16 13 41.238&  +38 38 10.82\\
0.94&  16 12 53.579&  +17 59 45.08&   16 12 54.189&  +17 59 56.26\\
0.94&  16 02 59.264&  -00 34 57.85&   16 02 59.569&  -00 35 16.60\\

\enddata
\tablecomments{This table is available in its entirety in ASCII FORMAT in the electronic edition of this journal.  A portion is shown here for guidance regarding its form and content.}
\tablecomments{The coordinates given are for the individual components of the two component group.}
\tablenotetext{a}{vote (estimated probability of being bent radio galaxy).}
\end{deluxetable}

\begin{deluxetable}{cc}
\tabletypesize{\scriptsize}
\tablecaption{Percentages of Morphological Types for FIRST Higher Count Groups \label{MorphTypesTable}}
\tablewidth{0pt}
\tablehead{ \colhead{Range (\%) }  & \colhead{Morphology} }
\startdata
 9.8 - 37.7&  S-shape or Z-shape\\ 
 3.4 - 12.9& DD \\           
 2.7 - 17.7& WAT or NAT \\
 2.5 - 8.0&  X-shape \\                    
 0.9 - 1.7&  W-shape \\   
 0.5 - 1.2& ring \\
 0.4 - 0.5& edge-brightened lobe \\ 
\enddata
\end{deluxetable}



\begin{thebibliography}{}

\bibitem[Beck et al.(2005)] {Beck05} Beck, R., Fletcher, A., Shukurov, A., Snodin, A., Sokoloff, D. D., Ehle, M., Moss, D., Shoutenkov, V. 2005, A\&A, 444, 739

\bibitem[Becker et al.(1995)] {BBWH} Becker, R. H., White, R. L., \& Helfand, D. J. 1995, \apj, 450, 559

\bibitem[Begelman et al.(1980)] {Begelman80} Begelman, M. C., Blanford, R. D., Rees, M. J. 1980, Nature, 287, 307

\bibitem[Belokurov et al.(2007)] {Belokurov07} Belokurov, V., Evans, N. W., Moiseev, A., King, L. J., Hewett, P. C., Pettini, M., Wyrzykowsi, L., McMahon, R. G., Smith, M. C., Gilmore, G., Sanchez, S. F., Udalski, A., Koposov, S., Zucker, D. B., Walcher, C. J. 2007, \apj, 671, L9

\bibitem[Belokurov et al.(2009)] {Belokurov09} Belokurov, V., Evans, N. W., Hewett, P. C., Moiseev, A., McMahon, R. G., Sanchez, S. F., King, L. J. 2009, \mnras, 392, 104


\bibitem[Blanton et al.(2000)] {Blanton00} Blanton, E. L., Gregg, M. D., Helfand, D. J., Becker, R. H., \& White, R. L. 2000, \apj, 531, 118

\bibitem[Blanton et al.(2001)] {Blanton01} Blanton, E. L., Gregg, M. D., Helfand, D. J., Becker, \& Leighly, K. M. 2001, \aj, 121, 2915

\bibitem[Blanton et al.(2003)] {Blanton03} Blanton, E. L., Gregg, M. D., Helfand, D. J., Becker, \& White, R. L. 2003, \aj, 125, 1635

\bibitem[Bolton et al.(2008)]{Bolton08} Bolton, A. S., Burles, S., Koopmans, L. V. E., Treu, T., Gavazzi, R., Moustakas, L. A., Wayth, R., Schlegel, D. J. 2008, \apj, 682, 964

\bibitem[Brocksopp et al.(2007)] {Brocksopp07} Brocksopp, C., Kaiser, C. R., Schoenmakers, A. P. de Bruyn, A. G. 2007, \mnras, 382, 1019

\bibitem[Burns et al.(1983)] {Burns83} Burns, J. O., Schwendeman, E., White, R. A. 1983, \apj, 271, 575
\bibitem[Buta \& Combes(1996)] {Buta96} Buta, R., Combes, F. 1996, Fund. Cosmic Physics, 17, 95

\bibitem[Carilli et al.(2003)] {Carilli03} Carilli, C. L., Lewis, G. F., Djorgovski, S. G., Mahabal, A., Cox, P., Bertoldi, F., Omont, A. 2003, Science, 300, 773

\bibitem[Chen \& Liu(2007)] {Chen07} Chen, X., Liu, F. 2007, \mnras, Proceedings of IAU Symposium \#238, Cambridge, UK: Cambridge University Press, pp 341-342

\bibitem[Cheung(2007)]{Cheung07} Cheung, C. C. 2007, \aj, 133, 2007

\bibitem[Chyzy et al.(2008)] {Chyzy08} Chyzy, K. T., Buta, R. J. 2008, \apj, 677, L17

\bibitem[Combes et al.(1992)]{Combes92} Combes, F., Gerin, M., Nakai, N., Kawabe, R., Shaw, M. A. 1992, A\&A, 259, L27 

\bibitem[Condon \& Mitchell (1984)] {Condon84} Condon, J. J., Mitchell, K. J. 1984, \apj, 276, 472

\bibitem[Condon et al. (1998)] {Condon98} Condon, J. J., Cotton, W. D., Greisen, E. W., et al. 1998, \aj, 115, 1693

\bibitem[Conner et al.(1998)] {Conner98} Conner, S. R., Cooray, A. R., Fletcher, A. B., Burke, B. F., Leh\'{a}r, J., Garnavich, P. M., Muxlow, T. W. B., Thomasson, P., Blakeslee, J. P. 1998, \aj, 115, 37

\bibitem[Conover(1967)] {BCo} Conover, W. J. 1967, Annals of Mathematical Statistics, 38, 1208   

\bibitem[Cress et al.(1996)] {Cress96} Cress, C.M., Helfand, D. J., Becker, R. H., \&  White R. L. 1996, \apj, 473, 7

\bibitem[de Vries et al.(2006)] {deVries06} De Vries, W. H., Becker, R. H., White, R. L. 2006, \aj, 131, 666

\bibitem[Douglass et al.(2008)] {Douglass08} Douglass, E. M., Blanton, E. L., Clarke, T. E., Sarazin, C. L., Wise, M. 2008, \apj, 673, 763

\bibitem[Duric et.al(1983)] {Duric83} Duric, N., Seaquist, E. R., Crane, P. C., Bignell, R. C. 1983, \apj, 273, L11

\bibitem[Fanaroff \& Riley(1974)] {Fanaroff74} Fanaroff, B. L., Riley, J. M. 1974, \mnras, 167, 31

\bibitem[Florido et al.(1990)] {Florido90} Florido, E., Battaner, E., Sanchez-Saavedra, M. L. 1990, Ap\&SS, 164, 131


\bibitem[Forman et al.(2007)] {Forman07} Forman, W., Jones, C., Churazov, E., Markevitch, M., Nulsen, P., Vikhlinin, A., Begelman, M., Bahringer, H., Eilek, J., Heinz, S., Kraft, R., Owen, F., Pahre, M. 2007, \apj, 665, 1057

\bibitem[Garcia-Barreto et al.(1991)]{Garcia91} Garcia-Barreto, J. A., Downes, D., Combes, F., Gerin, M., Magri, C., Carrasco, L., Cruz-Gonazles, I. 1991, A\&A, 244, 257

\bibitem[Gawro\'{n}ski et al.(2006)] {Gawronski06} Gawro\'{n}ski, M. P., Marecki, A., Kunert-Bajraszewska, M, Kus, A. J. 2006, A\&A, 447, 63

\bibitem[Ghosh(2009)]{Ghosh09} Ghosh, K. K. 2009, \apj, 692, 694

\bibitem[Gizani et al.(2002)] {Gizani02} Gizani, Nectaria A. B., Garrett, M. A., Leahy, J. P. 2002, J. Astrophys. Astr., 23, 89

\bibitem[Gopal-Krishna \& Wiita (2000)] {Gopal-Krishna00} Gopal-Krishna, Wiita, P. J. 2000, A\&A, 363, 507

\bibitem[Hines et al.(1989)] {Hines89} Hines, D., Eilek, J., Owens, F. 1989, \apj, 347, 713

\bibitem[Ishwara-Chandra \& Saikia(1999)] {Ishwara99} Ishwara-Chandra, C. H., Saikia, D. J. 1999, \mnras, 309, 100

\bibitem[Kaiser et al.(2000)] {Kaiser00} Kaiser, C. R., Schoenmakers, A. P., R\"{o}ttgering, H. J. A. 2000, \mnras, 315, 381

\bibitem[Kempner et al.(2003)] {Kempner03} Kempner, J. C., Blanton, E. L., Clarke, T. C., En{\ss}lin, T. A., Johnston-Hollitt, M., Rudnick, L. Proceedings of the Riddle of Cooling Flows in Galaxies and Clusters of Galaxies: E25 May 31-June4, 2003, Charlottesville, Virginia Ed. T. H. Reiprich, J. C. Kempner, \& N. Soker

\bibitem[Kigure et al.(2004)] {Kigure04} Kigure, H., Uchida, Y., Nakamura, M., Hirose, S., Cameron, R. 2004, \apj, 608, 119

\bibitem[Kong et al.(2004)] {Kong04} Kong, A. K. H., Sjouwerman, L. O., Williams, B. F. 2004, \aj, 128, 2783

\bibitem[Kundt \& Saripalli (1987)] {Kundt87} Kundt, W., Saripalli, L. 1987, Journal of Astrophysics and Astronomy, 8, 211

\bibitem[Kurosawa \& Proga(2008)] {Kurosawa08} Kurosawa, R., Proga, D. 2008, \apj, 674, 97

\bibitem[Laine et al.(2006)] {Laine06} Laine, S., Kotilainen, J. K., Reunanen, J., Ryder, S. D., Beck, R. 2006, \aj, 131, 701

\bibitem[Lal et al.(2008)] {Lal08} Lal, D. V., Hardcastle, M. J., Kraft, R. P. 2008, \mnras, 390, 1105

\bibitem[Lal \& Rao(2007)] {Lal07} Lal, D.V, Rao, A.P. 2007, \mnras, 374, 1085

\bibitem[Lara et al.(2001)] {Lara01} Lara, L., Cotton, W. D., Feretti, L., Giovannini, G., Marcaide, J. M., M\'{a}rquez, I., Venturi, T. 2001, A\&A, 370, 409

\bibitem[Lara et al.(2004)] {Lara04} Lara, L., Giovannini, G., Cotton, W. D., Feretti, L., Marcaide, J. M., M\'{a}rquez, I., Venturi, T. 2004, A\&A, 421, 899

\bibitem[Liu et al.(2003)] {Liu03} Liu, F. K., Wu, X., Cao, S. L. 2003, \mnras, 340, 411

\bibitem[Liu(2004)] {Liu04} Liu, F. K. 2004, \mnras, 347, 1357

\bibitem[Lu(1990)] {Lu90} Lu, J. F. 1990, A\&A, 229, 424

\bibitem[Lu \& Zhou(2005)] {Lu05} Lu, J.-F., Zhou, B.-Y. 2005, \apj, 635, L17

\bibitem[Machalski \& Condon(1985)] {Machalski85} Machalski, J., Condon, J. J., 1985, \aj, 90, 5

\bibitem[Machalski et al.(2001)] {Machalski01} Machalski, J., Jamrozy, M., Zola, S. 2001, A\&A, 371, 445


\bibitem[Machalski \& Jamrozy(2006)] {Machalski06b} Machalski, J., Jamrozy, M. 2006, A\&A, 454, 95

\bibitem[Mackay(1971)] {Mackay71} Mackay, C. D. 1971, \mnras, 154, 209

\bibitem[Mao et al.(2009)] {Mao09} Mao, M. Y., Johnston-Hollitt, M., Stevens, J. B., Wotherspoon, S. J. 2009, \mnras, 392, 1070

\bibitem[Marecki et al.(2006)] {Marecki06} Marecki, A., Thomasson, P., Mack, K.-H. 2006, A\&A, 448, 479

\bibitem[Ostle(1964)] {BOS} Ostle, B. 1964, Statistics in Research, (2nd ed.; Ames: Iowa State University Press)

\bibitem[Owen \& Laing(1989)] {Owen89} Owen, F. N., Laing, R. A. 1989, \mnras, 238, 357

\bibitem[Owen \& White(1991)] {Owen91} Owen, F. N., White, R. A. 1991, \mnras, 249, 164

\bibitem[Proctor(2002)] {Proctor02} Proctor, D. D. 2002, J. Electron. Imaging, 12, 398

\bibitem[Proctor(2006)] {Proctor06} Proctor, D. D. 2006, \apjs, 165,95

\bibitem[Riechers et al.(2008)]{Riechers08} Riechers, D. A., Walter, F., Brewer, B. J., Carilli, C. L., Lewis, G. F., Bertoldi, F., Cox, P. 2008, \apj, 686, 851

\bibitem[Sadun \& Morrison (2002)] {Sadun02} Sadun, A. C., Morrison, P. 2002, \aj, 123, 2312

\bibitem[Saikia(2006)] {Saikia06} Saikia, D. J., Konar, C., Kulkarni, V. K. 2006, \mnras, 366, 1391



\bibitem[Saripalli(2005)] {Saripalli05} Saripalli, L., Hunstead, R. W., Subrahmanyan, R., Boyce, E. 2005, \aj, 130, 896

\bibitem[Saunders et al. (1987)] {Saunders87} Saunders, R., Baldwin, J. E., Warner, P. J. 1987, \mnras, 225, 713

\bibitem[Saxton, et al.(2002)] {Saxton02} Saxton, C. J., Bicknell, G. V., Sutherland, R. S. 2002, \apj, 579, 176



\bibitem[Schoenmakers et al.(2000a)] {Schoenmakers00a} Schoenmakers, A. P., deBruyn, A. G., R\"{o}ttgering, H. J. A., van der Laan, H., Kaiser, C. R. 2000, \mnras, 315, 371

\bibitem[Schoenmakers et al.(2000b)] {Schoenmakers00b} Schoenmakers, A. P., de Bruyn, A. G., R\"{o}ttgering, H. J. A., van der Laan, H. 2000, \mnras, 315, 395

\bibitem[Schoenmakers et al.(2000)] {Schoenmakers00} Schoenmakers, A. P., Mack, K.-H.,  de Bruyn, A. G., R\"{o}ttgering, H. J. A., Klein, U., van der Laan, H. 2000, Astron. Astrophys. Suppl. Ser., 146, 293

\bibitem[Schoenmakers et al.(2001)] {Schoenmakers01} Schoenmakers, A. P., de Bruyn, A. G., R\"{o}ttgering, H. J. A., van der Laan, H. 2001, A\&A, 374, 861


\bibitem[Sirothia et al.(2009)] {Sirothia09} Sirothia, S. K., Saikia, D. J., Ishwara-Chandra, C. H., Kantharia, N. G. 2009, \mnras, 392, 1403


\bibitem[Sofue(1991)] {Sofue91} Sofue, Y. 1991, Publ. Astron. Soc. Japan, 43, 671

\bibitem[Sun et al.(2002)] {Sun02} Sun, X-H, Han, J-L, Qiao,G-J 2002, Chin. J. Astron. Astrophys., 2, 133

\bibitem[Taylor et al.(1994)]{Taylor94} Taylor, G. B., Barton, E. J., Ge, J.-P. 1994, \aj, 107, 1942
\bibitem[Taylor et al.(2002)]{Taylor02} Taylor, G. B., Fabian, A. C., Allen, S. W. 2002, \mnras, 334, 769

\bibitem[Tsao(1954)] {BTs} Tsao, C. K. 1954, Ann. Math. Statist., 25, 587

\bibitem[Turner \& Ho(1994)] {Turner94} Turner, J. L., Ho, P. T. 1994, \apj, 421, 122


\bibitem[van Breugel et al.(1983)]{vanBreugel83} van Breugel, W., Balick, B., Heckman, T., Miley, G., Helfand, D. 1983, \aj, 88, 1

\bibitem[van Breugel\& Fomalont(1984)]{vanBreugel84} van Breugel, W., Fomalont, E. B. 1984, \apj, 282, L55

\bibitem[Waller et al. (2001)]{Waller01} Waller, H. W., et al. 2001, \aj, 121, 1395


\bibitem[White, et al.(1997)] {BWBHG} White, R. L., Becker, R. H., \& Helfand, D. J., Gregg, M. D 1997, \apj, 475, 479



\bibitem[Wong \& Blitz (2000)]{Wong00} Wong, T., Blitz, L. 2000, \apj, 540, 771

\bibitem[Zhang et al.(2007)] {Zhang07} Zhang, X-G, Dultzin-Hacyan, D., Wang, T-G 2007, \mnras, 377, 1215

\end{thebibliography}
\end{document}